\documentclass[aps,twocolumn,nofootinbib,preprintnumbers,superscriptaddress]{revtex4-1}
\usepackage{amssymb,amsmath,amstext,amsfonts,textcomp,soul}
\usepackage{graphicx,multirow}
\usepackage{physics}
\usepackage{bbold}
\usepackage{mathrsfs}  
\usepackage{hyperref}
\usepackage[capitalise]{cleveref}
\usepackage{color,xcolor,colortbl}
\usepackage{orcidlink}

\crefformat{equation}{Eq.~(#2#1#3)}
\crefformat{table}{Tab.~(#2#1#3)}
\crefformat{figure}{Fig.~(#2#1#3)}
\crefformat{appendix}{App.~(#2#1#3)}
\crefformat{section}{Sec.~(#2#1#3)}
\crefformat{subsection}{Subsec.~(#2#1#3)}

\begin{document}
\preprint{YITP-25-53, IPMU25-0016}

\title{Ghostly interactions in (1+1) dimensional classical field theory}

\author{C\'edric~Deffayet\,\orcidlink{0000-0002-1907-5606}}
\email{cedric.deffayet AT ens.fr}
\affiliation{Laboratoire de Physique de l’\'Ecole normale sup\'erieure, ENS, Universit\'e PSL, CNRS, Sorbonne Universit\'e, Universit\'e Paris Cit\'e, F-75005 Paris, France}

\author{Aaron Held\,\orcidlink{0000-0003-2701-9361}}
\email{aaron.held@phys.ens.fr}
\affiliation{
Institut de Physique Théorique Philippe Meyer, Laboratoire de Physique de l’\'Ecole normale sup\'erieure (ENS), Universit\'e PSL, CNRS, Sorbonne Universit\'e, Universit\'e Paris Cit\'e, F-75005 Paris, France
}

\author{Shinji Mukohyama\,\orcidlink{0000-0002-9934-2785}} 
\email{shinji.mukohyama@yukawa.kyoto-u.ac.jp}
\affiliation{Center for Gravitational Physics and Quantum Information, Yukawa Institute for Theoretical Physics, Kyoto University, 606-8502, Kyoto, Japan}
\affiliation{Kavli Institute for the Physics and Mathematics of the Universe (WPI), The University of Tokyo Institutes for Advanced Study, The University of Tokyo, Kashiwa, Chiba 277-8583, Japan}

\author{Alexander Vikman\,\orcidlink{0000-0003-3957-2068}} 
\email{vikman@fzu.cz}
\affiliation{CEICO--Central European Institute for Cosmology and Fundamental Physics, FZU--Institute of Physics of the Czech Academy of Sciences, Na Slovance 1999/2, 182 00 Prague 8, Czech Republic}

\begin{abstract}
We investigate the classical stability of two coupled scalar fields with opposite-sign kinetic terms evolving in 1+1  dimensional Minkowski spacetime.
In the first part, we characterise unquenched ghostly interactions and present numerical solutions that support the following statements. First, the classical instability is not instantaneous and can even be benign, i.e., free of finite-time singularities. Second, while the classical instability can cascade towards higher frequency excitations, it is not driven by high frequency modes: At fixed amplitude, high-frequency modes are more stable than low-frequency modes.
In the second part, we demonstrate that the classical instability can be quenched by mass terms. In particular, we exemplify that heavy high-frequency ghost fields seem to not violate the decoupling theorem and can be integrated out classically.
In the third part, we demonstrate how self-interactions can quench the instability, for instance, by postponing its onset to parametrically large times. Extrapolating numerical results at large but finite evolution time to infinite evolution time, we demonstrate that classical fluctuations around trivial and nontrivial field-theory vacua are increasingly long-lived with (i) smaller initial amplitude of fluctuations, (ii) higher initial frequency of fluctuations, (iii) larger masses of the fields, or (iv) weaker interaction coupling. Moreover, our numerical simulations for field-theoretical generalisations of some globally-stable ghostly mechanical models do not feature any instability.
\end{abstract} 

\maketitle

\section{Introduction}
\label{sec:intro}
Positivity of kinetic energies for all dynamical degrees of freedom -- the absence of ghosts -- is often considered to be a fundamental requirement for a physically consistent theory. Yet such ghosts with negative kinetic terms frequently appear in cosmology and modified gravity. In particular, very recent observations \cite{DESI:2025zgx,Lodha:2025qbg} seem to prefer the so-called Phantom fields \cite{Caldwell:1999ew}, and even systems crossing the so-called Phantom divide \cite{Vikman:2004dc,Hu:2004kh,Caldwell:2005ai,Kunz:2006wc,Nesseris:2006er}, as the dynamical entity behind Dark Energy. Restricting the level of nonlinearity to only quadratic kinetic terms in the action implies that the simplest naive realization of the Phantom is a ghost. On the other hand, the crossing of the Phantom divide, in the same naive fashion, can be arranged in the so-called Quintom which is a combination of one ghostly and one usual scalar fields \cite{Guo:2004fq,Cai:2009zp}. However, allowing for specific higher-derivative kinetic terms results in a stable crossing of the Phantom divide with a single degree of freedom and without ghostly perturbations \cite{Deffayet:2010qz}.
Yet, generic higher-derivative theories are accompanied by ghosts, due to the Ostrogradski theorem~\cite{Ostrogradsky:1850fid,Woodard:2015zca}. Indeed, the original formalism by Ostrogradski yields the Hamiltonian for a non-degenerate mechanical (point-particles) Lagrangian involving arbitrarily high time derivatives. The thereby constructed Hamiltonian is necessarily unbounded from above and, crucially, also from below, which invokes ghosts. This formalism can be straightforwardly generalised to classical relativistic field theory~\cite{Pais:1950za,deUrries:1998obu}.

Since higher derivatives appear naturally in the context of quantum corrections and effective field theory (EFT), the above considerations have wide-ranging implications in quantum field theory~\cite{Pais:1950za,Lee:1970iw,Garriga:2012pk}, for modifications~\cite{Lovelock:1971yv} and for the quantisation~\cite{Stelle:1976gc,Hawking:2001yt} of gravity, as well as for cosmology beyond dark energy~\cite{Linde:1988ws,Kaplan:2005rr,Cline:2003gs,Cline:2023cwm,Cline:2023hfw,Brandenberger:2016vhg}.
Thus, the absence of higher-order derivatives has been elevated to a construction principle in the search for new fundamental physics and is prominently used as a viability criterion, see, e.g.,~\cite{Woodard:2006nt,Copeland:2006wr,Sotiriou:2008rp,Clifton:2011jh,Joyce:2014kja,deRham:2014zqa,Berti:2015itd} for respective reviews.
\\

While the unboundedness of the Ostrogradski Hamiltonian is a mathematical fact, conclusions about an inevitable catastrophic instability are not. To our best knowledge, said conclusions instead rest on various physical expectations concerning (i) point particles, (ii) classical field theory, and (iii) quantisation, see~\cite{Woodard:2015zca,Woodard:2006nt} for pedagogical reviews. 
\\

In the following, we will not address any issues related to quantisation. Instead, we take the viewpoint that it is pertinent to clarify the classical dynamics first. Indeed, we are motivated by two recent developments which underscore that, even at the classical level, the dynamical evolution is not understood sufficiently well to conclude that a catastrophic instability is inevitable.

The first concerns classical point particles, see~\cite{ErrastiDiez:2024hfq} for a classification of stability results. In particular, a presence of an additional integral of motion has allowed for proofs of global stability in a class of models~\cite{Deffayet:2021nnt,Deffayet:2023wdg}, i.e., to establish bounded phase-space motion for any choice of initial data, hence providing the first transparent physical counterexamples to the above expectations with a rigorous proof of global stability. For a mathematically inclined discussion of integrals of motion in this context, see~\cite{Kaparulin:2014vpa}. 
For previous investigations of local stability for specific regions of initial data (``islands of stability''), see~\cite{Pagani:1987ue,Smilga:2004cy,Carroll:2003st,Ilhan:2013xe,Pavsic:2016ykq, Smilga:2017arl,Pavsic:2013noa, Pavsic:2020aqi, Boulanger:2018tue,Damour:2021fva}.

The second concerns classical field theory and, in particular, the field of numerical relativity. Recent progress on a locally well-posed initial value formulation for effective field theories of gravity~\cite{Figueras:2024bba} (see also~\cite{Noakes:1983,Held:2021pht,Held:2023aap} for prior, less-general results) explicitly relies on higher-derivative terms and thus involves ghostly interactions. In this context, it is crucial that first numerical solutions suggest that sufficiently heavy ghost fields effectively decouple~\cite{Held:2023aap} such that longlived time evolution becomes possible~\cite{Held:2025ckb}.
\\

The combination of global stability of point-particle systems and local well-posedness in gravitational field theories motivates us to revisit physical conclusions about the time evolution in classical field theories with opposite-sign kinetic terms, see~\cite{Gross:2020tph,Damour:2021fva,Fring:2024xhd,Fring:2023pww,Fring:2024brg,Fring:2023ijk} for related work. We focus on the simplest possible setup, i.e., on two scalar fields coupled via non-derivative interactions and evolving in $(1+1)$-dimensional Minkowski spacetime. Even in this simple setup, analytical solutions cannot generally be obtained. In the following, we thus resort to numerical solutions to gain physical insights. 
By nature, numerical solutions are limited to finite evolution time and exemplary families of initial data. As such, they cannot provide proof of global stability. We can nevertheless confidently demonstrate several important physical conclusions which dispel the inevitability of a catastrophic instability and suggest that classical ghostly field theories can exhibit arbitrarily longlived evolution from generic initial data. 
\\

We detail our numerical setup along with three suitable families of initial data in~\cref{sec:setup}. The three following sections present our results: \cref{sec:characterising-instabilities} characterises the unquenched instability, demonstrating, in particular, that higher-frequency initial data is more stable rather than less stable; \cref{sec:massive} concerns mass terms, demonstrating, in particular, that heavy ghost fields decouple and can apparently be integrated out; \cref{sec:longlived} introduces self-interactions to demonstrate that these can quench the instability and can lead to apparently longlived motion. In particular, as we discuss in \cref{sec:longlived:PRL-model}, our numerical simulations do not reveal any instability for field-theoretical extension of the mechanical model introduced in \cite{Deffayet:2021nnt}.
We provide our conclusions and a discussion of open questions in~\cref{sec:conclusions}.

\section{Setup}
\label{sec:setup}

We consider (1+1) dimensional field theories of two scalar fields $\phi(t,x)$ and $\chi(t,x)$, coupled by an  interaction potential $V[\phi,\chi]$. Depending on a parameter $\sigma=\pm1$, the two scalar fields can be chosen to have opposite-sign kinetic terms. To be specific, the associated Lagrangian density
\begin{align}
\label{eq:field-theory-Lagrangian}
	\mathcal{L} = 
	- \frac{1}{2}\phi\left[\Box + m_\phi^2\right]\phi
	-  \frac{\sigma}{2}\chi\left[\Box + m_\chi^2\right]\chi
	- V[\phi,\,\chi]\;,
\end{align}
and equivalently the Hamiltonian density 
\begin{align}
\label{eq:field-theory-Hamiltonian}
	\mathcal{H} &= 
    \frac{1}{2}\Big[\dot{\phi}^2 + \phi'^2 + m_\phi^2\phi^2\Big]
    + \,\frac{\sigma}{2}\Big[\dot{\chi}^2 + \chi'^2+ m_\chi^2\chi^2\Big]
    + V\;,
\end{align}
exhibit a ghost whenever $\sigma=-1$. Here, we denote time derivatives by overdots and spatial derivatives by primes. Moreover, the (flat-space) d'Alembertian reads 
\begin{equation}
\label{eq:dAlembert}
\Box\phi=\partial_t^2\phi - \partial_x^2\phi \equiv \ddot{\phi} - \phi''\,,
\end{equation}
and equivalently for the field $\chi$.
The respective 2nd-order field equations can be written as
\begin{align}
	\left[\Box + m_\phi^2\right]\phi &= 
    - \partial_{\phi} V
    \;,
	\notag\\
	\left[\Box + m_\chi^2\right]\chi &= 
    - \sigma \,\partial_{\chi} V
    \;,
	\label{eq:eoms}
\end{align}
from which one can immediately see that the presence/absence of a ghost field shows up only in interactions without derivatives. In particular, $\sigma$ does not change the principal part of the system of partial differential equations (PDEs), see~\cref{eq:eoms}. Consequently the presence/absence of a ghost does not change the characteristics of the system and preserves its hyperbolicity. Therefore, the Cauchy problem is well-posed, despite the presence of a ghost. 
In particular, the well-posedness of the Cauchy problem means that, at least for some finite time interval, any growth of the solutions is universally bounded~\cite{Sarbach:2012pr}.

Nevertheless, $\sigma=-1$ changes the structure of the field equations. In particular, we expect that it is impossible to obtain the ghostly field equations from any modified action without ghost, even when arbitrarily modifying the interaction potential. (For the exception of fully decoupled positive- and negative-energy, see below.) 
\\

As for the point-particle systems in~\cite{Deffayet:2023wdg}, it will be useful to split the potential into three parts, i.e.,
\begin{align}
    V_\phi[\phi] &= V[\phi,\,\chi=0]\;,
    \notag\\
    V_\chi[\chi] &= V[\phi=0,\,\chi]\;,
    \notag\\
    V_\text{int}[\phi,\,\chi] &= V[\phi,\,\chi] - V_\phi[\phi] - V_\chi[\chi]\;.
\end{align}
We will refer to $V_\phi$ and $V_\chi$ as the self-interaction potentials and to $V_\text{int}$ as the ghostly interaction potential.

Similarly, we may split the Hamiltonian densities, i.e.,
\begin{align}
    \mathcal{H}_\phi &= 
    \frac{1}{2}\Big[\dot{\phi}^2 + \phi'^2 + m_\phi^2\phi^2\Big]
    + V_\phi[\phi]
    \;,
    \notag\\
    \mathcal{H}_\chi &= 
    \sigma\,\frac{1}{2}\Big[\dot{\chi}^2 + \chi'^2+ m_\chi^2\phi^2\Big]
    + V_\chi[\chi]
    \;,
    \notag\\[0.5em]
    \mathcal{H}_\text{int} &= 
    \mathcal{H} - \mathcal{H}_\phi - \mathcal{H}_\chi
    = 
    V_\text{int}[\phi,\,\chi]
    \;,
\end{align}
and integrate in the spatial domain to obtain the respective component energies, i.e.,
\begin{align}
    H_\phi &= \int_x\,\mathcal{H}_\phi\;,
    \notag\\
    H_\chi &= \int_x\,\mathcal{H}_\chi\;,
    \notag\\
    H_\text{int} &= \int_x\,\mathcal{H}_\text{int} \;,
    \notag\\
    H &= \int_x\,\mathcal{H} \;.
    \label{eq:component-energies}
\end{align}
We emphasize that $H$ corresponds to the total energy and, hence, is always be conserved, irrespective of the sign of $\sigma$. This conservation is guaranteed for all self-interactions $V_\chi[\chi]$, $V_\phi[\phi]$ as well as the ghostly interactions or couplings between $\chi$ and $\phi$, given by $V_\text{int}[\phi,\,\chi]$ above. In contrast, $H_\phi$, and $H_\chi$ are only conserved if the two modes decouple, i.e., if $\mathcal{H}_\text{int} \equiv V_\text{int} \equiv 0$. In the absence of masses $m_{\chi}=m_{\phi}=0$ and self-interactions, i.e., for $V_\phi=0$ and $V_\chi=0$, the component energies $H_\phi$ and $H_\chi$ correspond to the sum of kinetic and gradient energies of the individual fields. 
Whenever self-interactions do not contribute to $H_\phi$ and $H_\chi$, we loosely refer to them as ``kinetic energy''.

As we will see below, the unquenched ghost instability is accompanied by $|H_\phi|$ and $|H_\chi|$ growing with opposite signs. To characterise the instability, it will thus be useful to monitor how these component energies evolve with time.
\\

In~\cref{sec:characterising-instabilities,sec:massive,sec:longlived}, we investigate different potentials $V[\phi,\,\chi]$ and study the (in)stability of the resulting motion. Before we present these results, we describe the physics of our numerical setup in~\cref{sec:setup:numerics} (see also~\cref{app:numerics} for technical details) and several exemplary families of initial data, see~\cref{sec:setup:ID}, which we will in turn use to characterise the time evolution.

\subsection{Numerical Setup}
\label{sec:setup:numerics}

We numerically solve the initial value problem defined by \cref{eq:eoms} and suitable initial data (see below) for different potentials using the \texttt{julia} package \texttt{DifferentialEquations.jl}~\cite{rackauckas2017differentialequations}. The latter has been extensively tested and is a flexible but efficient tool for solving partial differential equations\footnote{
    For reproducibility, the complete \texttt{julia} code is available on \href{https://github.com/aaron-hd/ghostlyPDE_1D}{\texttt{github.com/aaron-hd/ghostlyPDE\_1D}}.
}. 

In the following, we describe the physical setup of this evolution and then present our results.
The interested reader can find details of our numerical setup as well as convergence tests in~\cref{app:numerics}. The latter firmly establish that all presented numerical results converge with the expected convergence rate and thus approximate the continuum field theory up to an error which can straightforwardly be obtained from the indicated numerical precision. Crucially, since we establish convergence rates, there is no reason to expect that a further refinement of the numerical resolution would lead to different conclusions about the continuum field theory. It is, in the above sense, appropriate to say that we have solved the initial value problem in the continuum field theory itself. The well-posedness of the initial value problem guaranties the convergence of these discretised approximations.
\\

We simulate in a spatial domain of $[0,L]$. Throughout the rest of this paper, we work in a natural unit system and express all dimensionful quantities in units of~$L$. We have confirmed that the physics is determined only by these dimensionless ratios. Hence, the reader can also rescale our results to different spatial domains by rescaling all other dimensionful quantities accordingly. 

Further, we implement periodic boundary conditions. In particular, this causes (many) repeated interactions of the field configurations as soon as the evolution time $T/L\gtrsim1$ ($T/L\gg1$). It will be interesting to investigate the impact of other boundary conditions and we intend to do so in future work.

We discretize the spatial domain by means of central 4th-order finite differencing. For the time evolution, we chose a 4th-order Runge-Kutta (\texttt{RK4}) scheme~\cite{runge1895numerische, kutta1901beitrag}, see also~\cite{press1988numerical}). 
We find that this standard choice is suitable to solve our problem efficiently and with sufficient precision. We have also verified (for exemplary evolutions) that other time-stepping routines reproduce the same conclusions about the continuum solution, see~\cref{app:numerics}.
Since we want to rigorously establish numerical convergence rates, we have turned off adaptive step-size control algorithms (e.g., with an \texttt{AutoTsit5} algorithm~\cite{tsitouras2011runge}) for all presented results, despite the fact that such algorithms significantly accelerate the computation time. For all results in this paper, we have implemented automatic self-convergence tests by running three neighbouring resolutions and verifying that the discrete $L^2$~norm of the state vector of all evolution variables converges at the expected 4th-order rate $\mathcal{C}=4$. To be specific, we ensure that $3<\mathcal{C}<5$ (although for the majority of the evolution time it remains much closer to $\mathcal{C}=4$) at all times. A detailed discussion of these convergence rates is presented in~\cref{app:numerics}.

\subsection{Initial data}
\label{sec:setup:ID}

We setup three different families of initial data.
For the first family of initial data~$\Phi^{(Gauss)}(k,\,A)$, we construct Gaussian wave packets, corresponding to localized and initially separate configurations of both fields. 
For the second family of initial data~$\Phi^{(wave)}(k,\,A)$, we construct plane waves which correspond to delocalized field configurations filling the whole spatial domain. 
For the third family of initial data~$\Phi^{(rand)}(k,\,A)$, we construct a stochastic superposition of the previous plane waves, corresponding to (pseudo)random initial data.

Each family of initial data is constructed such that it exhibits a characteristic amplitude~$A$ and wave number~$k$. We recall that, while $k$ is measured in units of $L^{-1}$, the fields are already dimensionless in $(1+1)$ dimensions, hence the characteristic amplitude $A$ is also dimensionless to begin with. In terms of dimensionless ratios, we thus expect that the physics of each initial-data family is fully determined by the characteristic dimensionless quantities $A$ and $k\times L/(2\pi)$.

\subsubsection{Gaussian wave packets}
\label{sec:setup:ID:wave-packets}

The first initial data family is constructed from Gaussian wave-packets, i.e.,
\begin{align}
\label{eq:Init_Data_Gaussian}
    \phi_0(x) &=A_{\phi}\exp\left(-\frac{\left(x-x_{\phi}\right)^{2}}{2\ell_{\phi}^{2}}\right)
    \;,
    \notag\\
    \dot\phi_0(x) &=A_{\phi}\frac{c_{\phi}\left(x-x_{\phi}\right)}{\ell_{\phi}^{2}}\exp\left(-\frac{\left(x-x_{\phi}\right)^{2}}{2\ell_{\phi}^{2}}\right)
    \;,
    \notag\\    
    \chi_0(x) &=
    A_{\chi}\exp\left(-\frac{\left(x-x_{\chi}\right)^{2}}{2\ell_{\chi}^{2}}\right)
    \;,
    \notag\\
    \dot\chi_0(x) &=
    A_{\chi}\frac{c_{\chi}\left(x-x_{\chi}\right)}{\ell_{\chi}^{2}}\exp\left(-\frac{\left(x-x_{\chi}\right)^{2}}{2\ell_{\chi}^{2}}\right)
    \;,
\end{align}
where the initial velocities $c_{\phi}=1$ and $c_{\chi}=-1$ are set such that the Gaussian wave packets move in opposite directions. Their initial position is set by $x_{\phi}$ and $x_{\chi}$, for which we pick $x_{\phi}=0.3\,L$ and $x_{\chi}=0.7\,L$.
For concreteness, we also work with equal amplitudes $A\equiv A_{\phi}\equiv A_{\chi}$ and equal width (standard deviation) $\ell\equiv\ell_{\phi}\equiv\ell_{\chi}$.  
This leaves a characteristic dimensionless amplitude $A$ and the dimensionful standard deviation $\ell$ as the only remaining free parameters that characterise the Gaussian initial data family. 
To unify notation with the other two initial data families (see below) and since the Gaussian is an eigenfunction of the Fourier transform, we can equivalently characterise the Gaussian initial data family by a characteristic momentum scale $k=1/(4\ell)$ such that
$\Phi^{(Gauss)}(k,\,A)$ is fully determined by the dimensionless characteristic amplitude $A$ and the dimensionless characteristic wave number $k\times L/(2\pi)$. We note that $k\times L/(2\pi) \geqslant 1$ is required to guarantee that the Gaussian is sufficiently suppressed at the boundary. Indeed, this is the reason for the inclusion of the otherwise arbitrary factor of four in the identification $k=1/(4\ell)$ which guarantees that $k\times L/(2\pi) = 1$ corresponds to the broadest admissible Gaussian initial data.

\subsubsection{Plane waves}
\label{sec:setup:ID:waves}

As a second family of initial data we consider plane waves with initial frequencies determined by the wave vectors $k_{\phi}$ and $k_{\chi}$. For completeness, the explicit initial conditions are given by
\begin{align}
\label{eq:Init_Data_Plane}
    \phi_0(x) &=
    A_\phi\,\sin(k_{\phi}(x-x_{\phi}))
    \;,
    \notag\\
    \dot\phi_0(x) &=
    -c_\phi\,k_{\phi}\,A_\phi\,\cos(k_{\phi}(x-x_{\phi}))
    \;,
    \notag\\    
    \chi_0(x) &=
    A_\chi\,\sin(k_{\chi}(x-x_{\chi}))
    \;,
    \notag\\
    \dot\chi_0(x) &=
    -c_\chi\,k_{\chi}\,A_\chi\,\cos(k_{\chi}(x-x_{\chi}))
    \;.
\end{align}
Here we have introduced initial phase velocities
\begin{align}
    c_\phi &=
    \pm \sqrt{k_{\phi}^2 + m_\phi^2}/k_{\phi}
    \;,
    \notag\\    
    c_\chi &=
    \pm \sqrt{k_{\chi}^2 + m_\chi^2}/k_{\chi}
    \;,
\end{align}
where the positive sign corresponds to the motion towards larger values of $x$. Further $A_{\phi}$, $k_{\phi}$, and $x_{\phi}$ denote amplitude, wave number, and initial position of the $\phi$-wave and analogously for the $\chi$-wave.  The above initial dispersion relations ensure plane-wave initial data for free massive fields. Without interactions (that is neither self-interactions nor ghostly interactions, i.e., $V_\phi\equiv V_\chi\equiv V_\text{int}\equiv0$) these plane waves freely propagate as 
\begin{align}
    \phi(x) &=
    A_\phi\,\sin(k_{\phi}(x-x_{\phi}-c_\phi t))
    \;,
    \notag\\ 
    \chi(x) &=
    A_{\chi}\,\sin(k_{\chi}(x-x_{\chi}-c_{\chi} t))
    \;.
\end{align}
For massless fields, $c_{\phi}=\pm1$ and/or $c_{\chi}=\pm1$.
\\

To reduce the complexity and obtain a family of initial data $\Phi^{(wave)}(k,A)$ with a single characteristic wave number~$k$ and a single characteristic amplitude~$A$, we fix $x_\phi=0$, $x_\chi=L/3$, $A \equiv A_\phi = A_\chi$, $k \equiv k_\phi = k_\chi/2$, and chose the two plane waves to move in opposite direction. We do not expect that generic modifications of these choices alter any of our conclusions. 

\subsubsection{Stochastic initial conditions}
\label{sec:setup:ID:stochastic}

For the third family of initial data, we combine plane waves (both left-moving and right-moving) at different wave numbers $k_{\phi/\chi}$ into stochastic initial conditions by randomly drawing their initial location $x_{\phi/\chi}/L$ from a flat distribution and their initial amplitude from a normal distribution with width $A_{\phi/\chi}$. For simplicity, we pick $A \equiv A_\phi = A_\chi$.
For each wave number $k_{\phi/\chi}$, we draw initial amplitudes for both a left-moving and a right-moving wave from a normal distribution with width (standard deviation) $\text{SD} = A_{\phi/\chi}/\sqrt{2\,k_{\phi/\chi}\,L}$. Summing up all such plane waves with wave number $k_N=2\pi N / L$ and $N\in\mathbb{N}$ results in combined initial data for the fields and their derivatives.
For reproducibility and consistency among numerics at different resolution, we initialise respective pseudo-random number generator with persistent seeds\footnote{The specific random seeds for all evolutions are available on \href{https://github.com/aaron-hd/ghostlyPDE_1D}{\texttt{github.com/aaron-hd/ghostlyPDE\_1D}.}. 
}. In particular, fixing the random seed allows us to study the same instance of (pseudo-)random initial data at various amplitudes.
We choose to sum up frequencies between $k=k_N$ and $k_{N+4}$ and characterise the family of initial data by referring to the lowest wave number $k\equiv k_N$ of the four neighbouring frequencies that compose the stochastic initial data.
In summary, the respective stochastic family of initial data $\Phi^{(rand)}(k,A)$ is characterised by a single (minimum) wave number $k\equiv k_N$ and a single characteristic amplitude $A$.

\begin{figure*}
        \includegraphics[width=\linewidth]{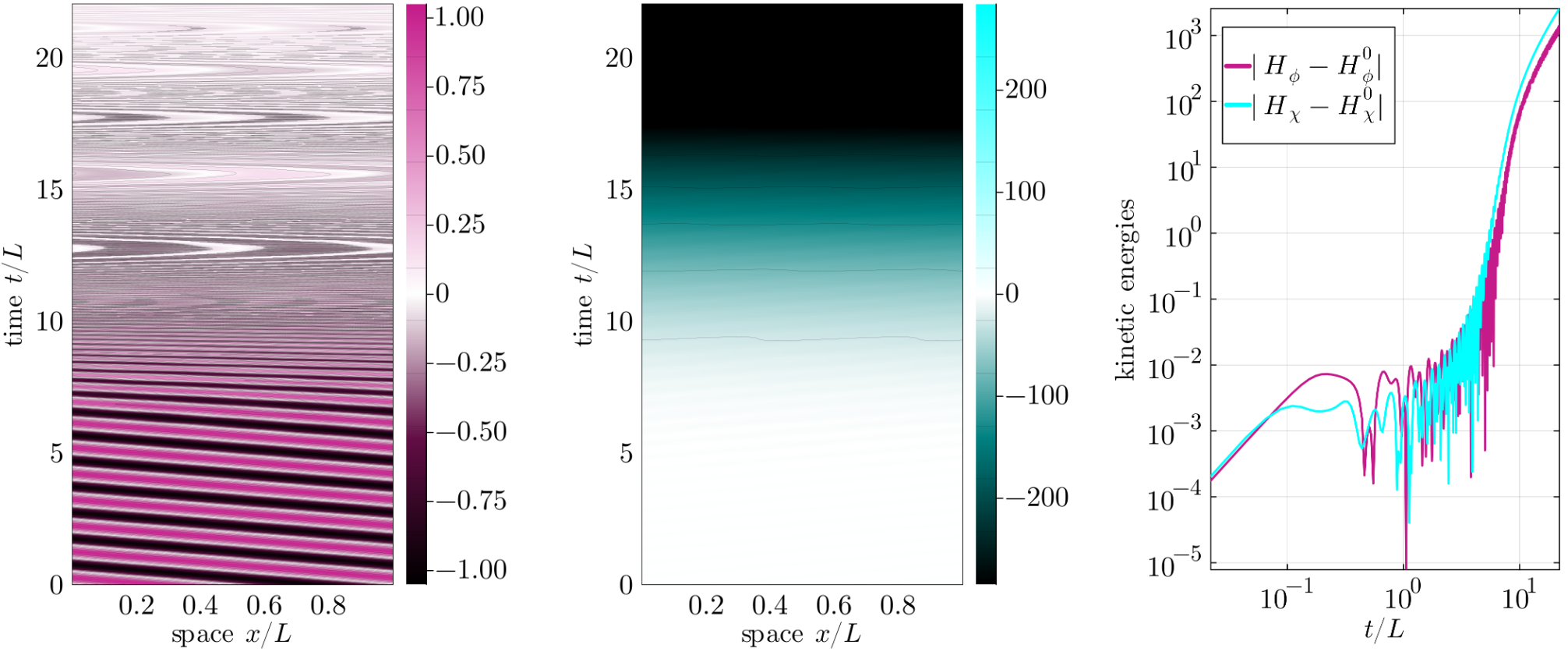}
    \caption{
        \label{fig:benign}
        Exemplary evolution of the unquenched benign ghost, i.e., with potential $V_\text{int}^{(22)}=\lambda_{22}\,\phi^2\chi^2$, choosing $\lambda_{22}\times L^2=1$, starting from plane-wave initial data $\Phi^{(wave)}(k,\,A)$ (see~\cref{sec:setup:ID}) with $k\times L/(2\pi)=1$ and amplitude $A=1$. The left and central panels show the evolution of $\phi$ and $\chi$, respectively. 
        The right panel shows the growth of component (``kinetic'') energies $H_\phi$ and $H_\chi$ compared to their initial values $H_\phi^{0}$ and $H_\chi^{0}$. Note that the latter is shown as a log-log plot such that straight lines correspond to polynomial growth.
        For clarity, we have dropped the explicit mass terms, i.e., have set $m_\phi^2 = m_\chi^2 =0$. 
        \href{https://zenodo.org/records/15209689}{Animations available online.}
    }
\end{figure*}
\begin{figure}
        \includegraphics[width=\linewidth]{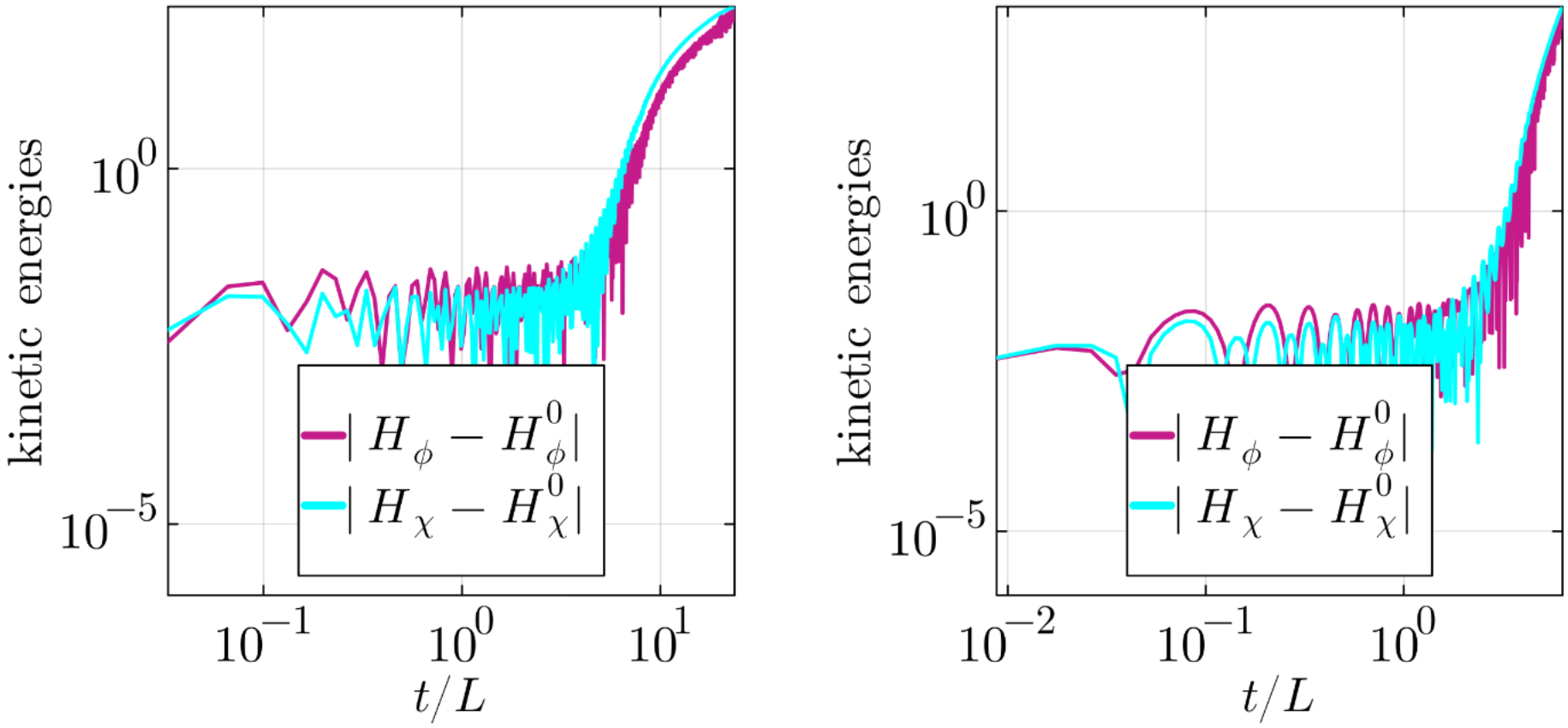}
    \caption{
        \label{fig:benign_ghost_general-class}
        As in the right-hand panel of~\cref{fig:benign} but for a ghostly interaction mediated by $V_\text{int}^{(42)}$ (left panel) and $V_\text{int}^{(44)}$ (right panel). 
        \href{https://zenodo.org/records/15209689}{Animations available online.}
    }
\end{figure}
%

\section{Characterising the instability}
\label{sec:characterising-instabilities}

Before we proceed to investigate a potential quenching of the instability, we characterise the nature of the unquenched instability.
\\

We recall that the field equations in~\cref{eq:eoms} explicitly show that the ghost instability (i.e., $\sigma=-1$) is mediated solely by non-derivative interactions. 

From the perspective of mathematical analysis of partial differential equations, the sign of $\sigma$ in~\cref{eq:eoms} cannot change the principal part. Irrespective of the absence or presence of a ghost, the above constitutes a system of coupled nonlinear wave equations. As we have already mentioned, the respective initial value problem is thus manifestly well-posed, at least locally well-posed with respect to a suitable norm. We omit a more formal discussion of local well-posedness. For an introduction we refer the reader to, e.g.,~\cite{Sarbach:2012pr}. Instead, we emphasise the physical importance of well-posedness. 

From the perspective of a physicist, well-posedness of an initial value problem defines whether the respective field equations can describe a uniquely defined time evolution of physical initial conditions from some instance in time to some later instance in time. In the context of classical continuum field theories, local well-posedness guarantees that this time evolution exists, is unique, and depends continuously on the initial data (in some suitable norm), at least for some time. Local well-posedness, therefore, constitutes our most rigorous mathematical definition of time evolution in classical field theories. The ghost field equations ($\sigma=-1$) are just as well-posed as the non-ghost field equations ($\sigma=+1$). In particular, a well-posed initial value problem does not tolerate an instantaneous decay of any solution.

As we will explicitly demonstrate below, the local onset of the ghost instability seems to be equivalent to a non-ghostly tachyonic (or higher-order potential) instability.  For non-ghostly tachyons, we know that the local instability can be quenched by suitable self-interactions that lead to a sufficiently steep non-trivial shape of the potential. A key physical question is whether ghost instabilities induced by ghostly interactions can also be quenched by self-interactions. To the best of our knowledge this question has not been systematically investigated and thus remains open. We will take a first step to address it in~\cref{sec:longlived}, focusing exclusively on non-derivative potential interactions. Before doing so, the remainder of this section characterises the unquenched instability in~\cref{sec:characterising-instabilities:benign-vs-catastrophic} and its frequency dependence in~\cref{sec:characterising-instabilities:frequency-dependence}. The following~\cref{sec:massive} will clarify the effect of making one or both fields more and more massive and is of particular relevance in the context of EFT.

\subsection{Unquenched ghost instabilities}
\label{sec:characterising-instabilities:benign-vs-catastrophic}

In this section, we characterise the unquenched instability  in polynomial ghostly interaction potentials 
\begin{align}
	V_\text{int}^{(nm)}[\phi,\chi]=\lambda_{nm}\,\phi^n\chi^m\;.
\end{align}
with $2\leqslant n,m\in\mathbb{N}$. The term unquenched refers to the absence of self-interactions, i.e., throughout this section we set $V_\phi=0$ and $V_\chi=0$.

We distinguish, in particular, between benign runaways and catastrophic runaways. A runaway is referred to as benign if the respective physical divergence occurs at infinite time. In contrast, we refer to a runaway as catastrophic if the divergence occurs at finite time. The terminology of ``benign'' ghostly interactions was introduced in~\cite{Robert:2006nj}, see~\cite{Smilga:2017arl} for review. Subsequently such benign behaviour was analysed in field theory~\cite{Damour:2021fva} where a modified version of the $(1+1)$ dimensional Korteweg-de Vries (KdV) equation is treated as a higher-derivative initial value problem by exchanging time and space, see also \cite{Fring:2024brg,Fring:2024xhd}. Our numerical solutions suggest that benign runaway behaviour could be generic for a large class of simple Lorentz-invariant systems with ghostly interactions. 

\begin{figure*}
        \includegraphics[width=\linewidth]{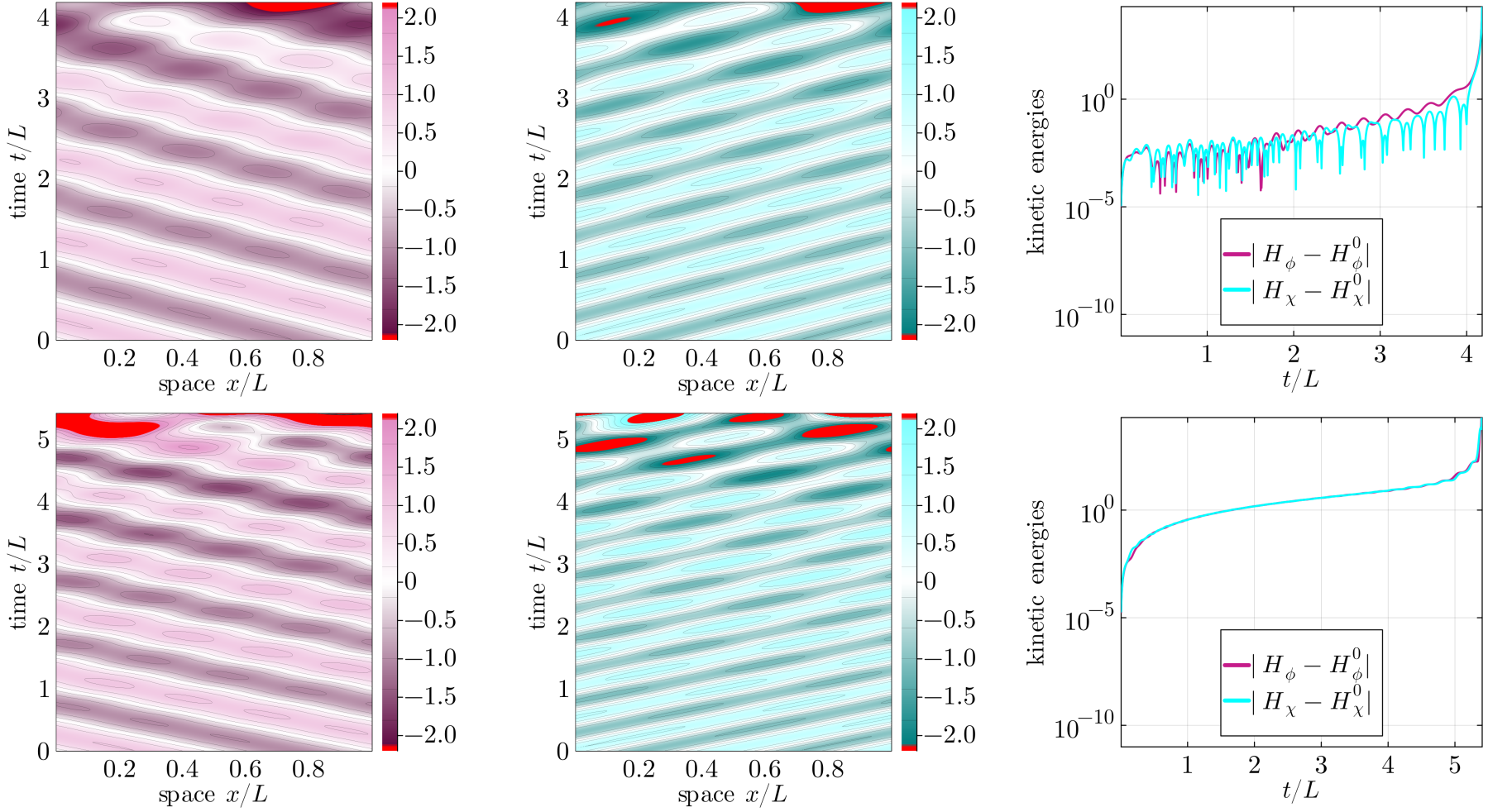}
    \caption{
        \label{fig:catastropic-example}
        Evolution in the unquenched interaction potential $V[\phi,\chi]=\lambda_{33}\,\phi^3\chi^3$ with $\lambda_{33}\times L^2=1$ and starting from plane-wave initial data $\Phi^{(wave)}(k,\,A)$ (see~\cref{sec:setup:ID}) with $k\times L/(2\pi)=1$ and amplitude $A=1$. Both, the non-ghostly case ($\sigma=+1$ and $\lambda_{33}\times L^2=-1$, shown in the upper panels) and the ghostly case ($\sigma=-1$ and $\lambda_{33}\times L^2=1$, shown in the lower panels) develop finite-time singularities with diverging field values. In the left panels, we show density plots of the evolution of $\phi$; in the middle panels, we show density plots of the evolution of $\chi$; and, in the right panels, we show the evolution of the component (``kinetic'') energies. The latter clearly exhibits super-exponential growth, highlighting the approach to a finite-time singularity. Note that, the latter is shown as a log-linear plot such that straight lines correspond to exponential growth. We also highlight the approach to the singularity, colouring out-of-bounds values in the density plots as red regions. 
        \href{https://zenodo.org/records/15209689}{Animations available online.}
    }
\end{figure*}
%

\subsubsection{Benign ghost instabilities}
\label{sec:characterising-instabilities:benign-vs-catastrophic:benign}

It is instructive to start from the field equations of the $(1+1)$ dimensional field theory example in~\cref{eq:field-theory-Lagrangian}. 
We specify to the potential $V_\text{int}^{(22)}=\lambda_{22}\,\phi^2\chi^2$. 
The resulting field equations can be written as
\begin{align}
	\Box\phi &= 
    - \left(m_\phi^2 + 2\,\lambda_{22}\,\chi^2\right)\phi
    \equiv
    - m_{\phi,\text{eff}}^2\,\phi
    \;,
	\notag\\
	\Box\chi &= 
    - \left(m_\chi^2 + 2\,\sigma\,\lambda_{22}\,\phi^2\right)\chi
    \equiv
    - m_{\chi,\text{eff}}^2\,\chi
    \;,
	\label{eq:field-theory-wave-equations-phi2chi2}
\end{align}
where, for clarity, we have defined effective mass terms $m_{\phi,\text{eff}}(\chi)$ and $m_{\chi,\text{eff}}(\phi)$, respectively depending on the other field.  
The close relation between a benign ghost and a tachyonic instability is now fully apparent.
A tachyonic instability occurs whenever $m_\phi^2<0$ or $m_\chi^2<0$ dominate on the right-hand side. In the absence of ghosts ($\sigma=+1$) a negative $\lambda_{22}<0$ leads to negative effective masses and thus to an effective tachyonic instability in both fields while for positive $\lambda_{22}>0$ no such effective tachyonic instability can occur. In the presence of ghosts ($\sigma=-1$), and even for $m_\phi^2>0$ or $m_\chi^2>0$, the key difference is that either sign of $\lambda_{22}$ can lead to an effective tachyonic instability in regions where, respectively, $|\lambda_{22}|\chi^2 > m_\phi^2/2$ or $|\lambda_{22}|\phi^2 > m_\chi^2/2$. At least for this specific potential, the onset of a ghost instability thus seems locally resemble to an effective tachyonic instability in one of the fields. 

Focusing on the ghostly case ($\sigma=-1$) and picking (without loss of generality) $\lambda_{22}>0$, the field $\chi$ will be tachyonic if $|\lambda_{22}|\phi^2 > m_\chi^2/2$. The field~$\chi$ thus grows to larger and larger field values. At the same time, the field~$\phi$ will always remain non-tachyonic and thus its amplitude cannot be amplified as efficiently, as for $\chi$.
Its effective mass, however, will grow as $\chi$ grows to larger field values, thus leading to faster and faster oscillations in time. Without further interactions, this unquenched instability feeds itself and proceeds for all future time, i.e., a runaway occurs, see left and middle panel in~\cref{fig:benign}. To provide a concrete example, the figure shows the evolution of Gaussian initial conditions $\Phi^{(wave)}(k,\,A)$ with $k\times L/(2\pi)=1$ and $A=1$ but we also observe similar late-time behaviour for the other initial-data families.

For the given interaction mediated by $V_\text{int}^{(22)}$, the runway is characterised by polynomially growing component energies $H_\phi$ and $H_\chi$, see right panel in~\cref{fig:benign}. We do not include self-interactions and mass terms, and thus $H_\phi$ and $H_\chi$ correspond to the component (``kinetic'') energies of the individual fields. Since the growth rate is polynomial all physical quantities remain finite at finite time and only diverge at infinitely late time. It is thus appropriate to refer to the instability as ``benign''~\cite{Damour:2021fva}.
\\

For general $n$ and $m$, the field equations generalise to
\begin{align}
	\Box\phi &= 
    - \left(n\,\lambda_{nm}\,\phi^{n-2}\,\chi^{m}\right)\phi
    \equiv
    - \Lambda_{\phi,\text{eff}}\,\phi
    \;,
	\notag\\
	\Box\chi &= 
    - \left(m\,\sigma\,\lambda_{nm}\,\phi^{n}\,\chi^{m-2}\right)\chi
    \equiv
    - \Lambda_{\chi,\text{eff}}\,\chi
    \;,
	\label{eq:field-equations-generalised-benign}
\end{align} 
where, for clarity, we have now dropped the explicit mass terms, i.e., have set $m_\phi^2 = m_\chi^2 =0$, as we do in all explicit evolutions throughout this section. The effective masses in~\cref{eq:field-theory-wave-equations-phi2chi2} are replaced by generalised interaction terms $\Lambda_{\phi,\text{eff}}[\phi,\chi]$ and $\Lambda_{\chi,\text{eff}}[\phi,\chi]$ that now depend on both field values.
We note that, as long as $n$ and $m$ are both even,  it remains true that $\Lambda_{\phi,\text{eff}}[\phi,\chi]$ cannot change sign, irrespective of the local values of either of the fields. The interactions can thus be expected to drive the fields to the same qualitative behaviour as for the benign case in~\cref{eq:field-theory-wave-equations-phi2chi2} and, indeed, this is what we observe when solving the exemplary IVPs. This suggests that all interactions with even $n$ and even $m$ have a qualitatively similar behaviour than the $n=m=2$ case. 

Indeed, our numerical results seem to confirm that all interactions with even $n$ and even $m$ result only in benign runaway behaviour, see~\cref{fig:benign_ghost_general-class}. Again, we use the plane-wave initial data family $\Phi^{(wave)}(k,\,A)$ with $k\times L/(2\pi)=1$ and $A=1$ as exemplary initial data.
With growing $n$ and $m$ it becomes increasingly difficult to confidently distinguish between higher-order polynomial runaway behaviour and exponential runaway behaviour. However, we can confidently assert that the instability remains benign since the evolution can always be extended to later and later evolution time by refining the resolution. The latter becomes impossible when there is a catastrophic finite-time singularity forming, as we will demonstrate in the next section.

\subsubsection{Catastrophic ghost instabilities}
\label{sec:characterising-instabilities:benign-vs-catastrophic:catastrophic}

%
\begin{figure*}
        \includegraphics[width=\linewidth]{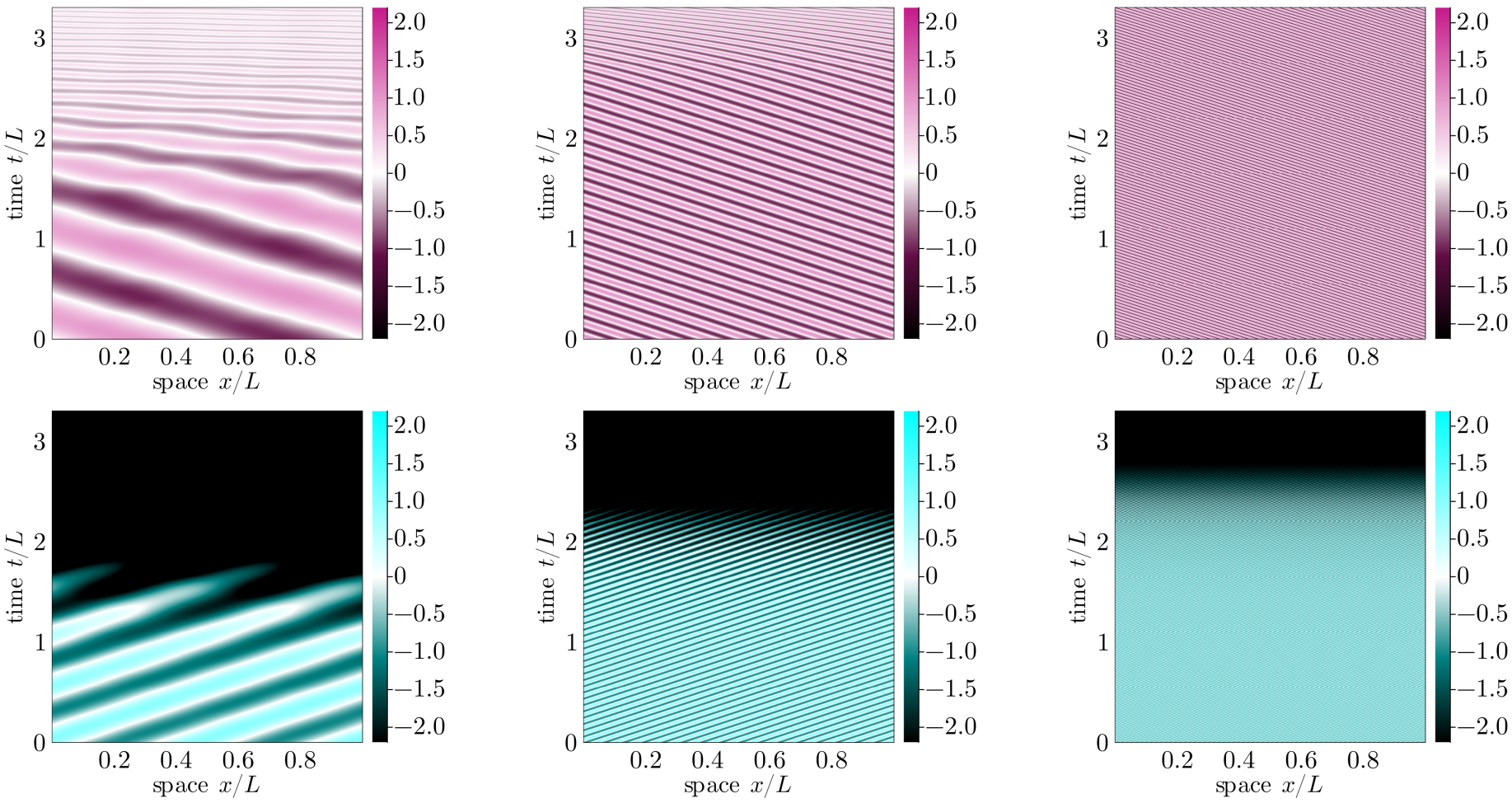}
    \caption{
        \label{fig:IR-instability}
        We exemplify, for the family of plane-wave initial data $\Phi^{(wave)}(k,A)$, that, at fixed amplitude $A=1$, higher frequencies $k$ are more stable not less stable. 
        We show the time evolution of $\phi$ (upper panels) and $\chi$ (lower panels) as propagated by the massless ($m_\phi=m_\chi=0$) field equations in~\cref{eq:field-theory-wave-equations-phi2chi2} with $\sigma=-1$, i.e., such that the $\chi$ field is a ghost with respect to the $\phi$ field, and $\lambda_{22}\times L^2=10$, i.e., such that the $\phi$ and $\chi$ are coupled with a $\phi^2\chi^2$ interaction.
        For the increasing characteristic wave number $k\times L/(2\pi)=1;\,8;\,32$ from left to right, the instability develops less quickly. 
        \href{https://zenodo.org/records/15209689}{Animations available online.}
    }
\end{figure*}

In the previous section, we have argued that all $Z_2\times Z_2$-symmetric polynomial ghostly interactions, i.e., those mediated by a potential $V_\text{int}^{(nm)}[\phi,\chi]=\lambda_{nm}\,\phi^n\chi^m$ with even $2\leqslant m,n\in\mathbb{N}$ lead to a runaway behaviour which is benign, i.e., lead to a divergence at infinitely late time only. In particular, we have presented explicit numerical evidence that, for the case of the lowest-order polynomial and $Z_2\times Z_2$-symmetric ghostly interaction potentials $V_\text{int}^{(22)}$, $V_\text{int}^{(24)}$, $V_\text{int}^{(42)}$, and $V_\text{int}^{(44)}$, the ghost instability leads to a benign runaway with polynomial growth rates in the component energies, at least for the given evolution time.

In contrast, we find that polynomial ghostly interactions with odd $2\leqslant m,n\in\mathbb{N}$ seem to lead to more rapid runaway behaviour. In particular, the latter leads to finite-time singularities in the field values, their first-order time derivatives, and hence the component energies. It is thus appropriate to refer to the respective runaway as ``catastrophic.''~\footnote{In passing, we note that it may be possible to enforce specific conditions to evolve the fields even past such singularities.} Once more, the same behaviour can occur if a non-ghostly field rolls down a sufficiently steep (i.e., growing faster than quadratically) and unbounded potential.
In~\cref{fig:catastropic-example}, we demonstrate this for the exemplary case $V[\phi,\chi]=\lambda_{33}\,\phi^3\chi^3$, where we evolve the same initial conditions, both, for the non-ghostly case ($\sigma=+1$ and $\lambda_{33}\times L^2=-1$, see upper panels) and the ghostly case ($\sigma=-1$ and $\lambda_{33}\times L^2=1$, see lower panels). As exemplary initial conditions, we have chosen, once more, the plane-wave initial data family $\Phi^{(wave)}(k,\,A)$ with $k\times L/(2\pi)=1$ and $A=1$. 
In contrast to the benign cases in the previous section, both fields now grow in amplitude. We highlight this by marking large field values beyond the range of the colour legend as red regions. 
The corresponding finite-time singularity is also evident in the super-exponential growth of the component (``kinetic'') energies, see right panels in~\cref{fig:catastropic-example}. 
Crucially, in the catastrophic case, refining the numerical resolution does not allow us to extend the solution to larger times. We thus conclude that the continuum field theory develops a finite-time singularity. 

We emphasize that such finite-time singularities can occur, both, for non-ghostly ($\sigma=+1$) and for ghostly ($\sigma=-1$) interactions.
The main physical question remains whether or not such a catastrophic instability can be quenched by other interactions, in particular, by higher-order self-interactions. For the non-ghostly case, yet higher-order polynomial interactions can lead to a non-trivial vacuum which, if sufficiently deep (in comparison to the initial amplitude), can quench the instability. For the ghostly case, it is an open question in how far one can quench such instabilities. We will address this question in~\cref{sec:longlived}.
\\

For now, the key message remains the following. The ghost instability is effectively a potential instability. While it can drive (one or both) fields to higher frequencies, crucially, there is no instantaneous decay. We elucidate the physics of this point further by discussing the frequency dependence of the instability.

\subsection{Higher frequencies are more not less stable}
\label{sec:characterising-instabilities:frequency-dependence}

%
\begin{figure*}
        \includegraphics[width=\linewidth]{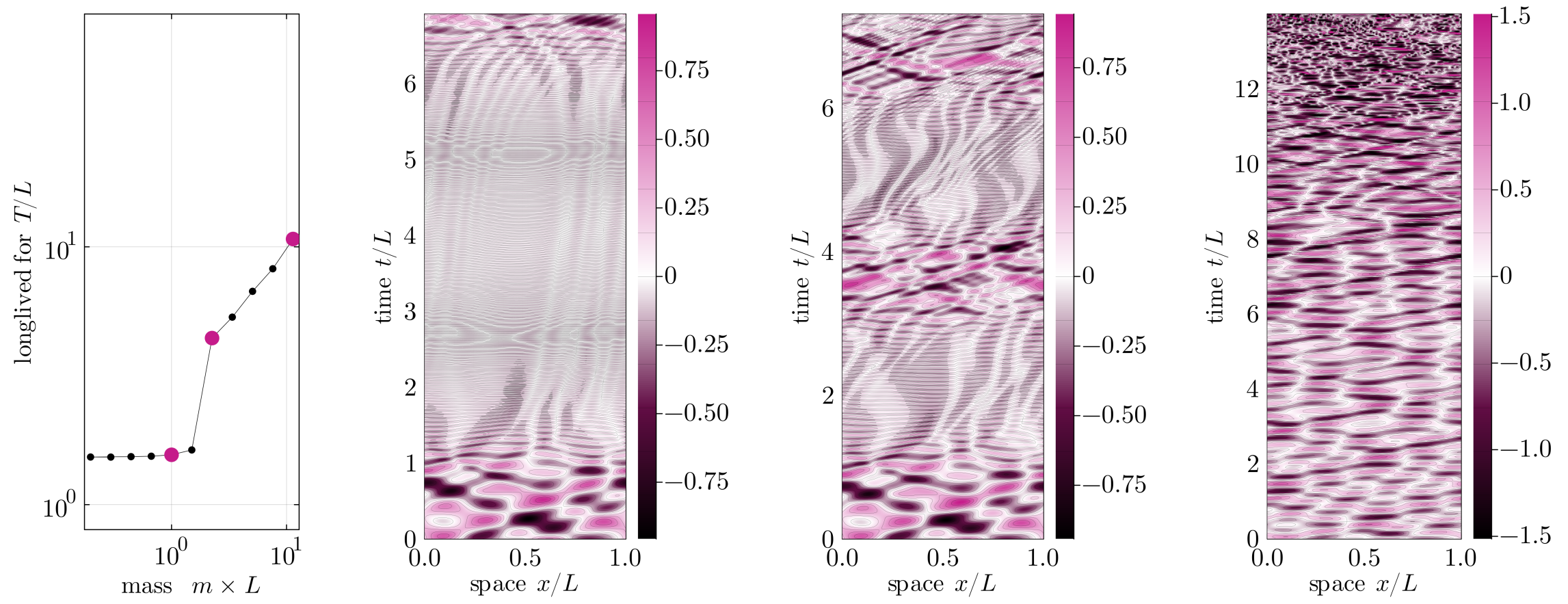}
    \caption{
        \label{fig:benign:mass-quenching}
        For the benign model with $V_\text{int}^{(22)}=\lambda_{22}\,\phi^2\chi^2$, $m_\phi=m_\chi=m$, and $\lambda_{22}\times L^2=100$ and stochastic initial data $\Phi^{(rand)}(k,\,A)$ (see~\cref{sec:setup:ID:stochastic}) with $A=1$ and $k\times L/(2\pi)=1$, we demonstrate that the instability can be quenched by increasingly large masses $m$. The left-most panel shows the time at which one of the component energies has grown $e$-fold as a function of growing dimensionless mass $m\times L$. The other three panels (from left to right) show examples of the evolution of $\phi$, corresponding to the three larger pink points (from left to right) in the left-most panel, with $m\times L=1,\,2.25,\,11.4$, respectively.  
        \href{https://zenodo.org/records/15209689}{Animations available online.}
    }
\end{figure*}
\begin{figure*}
    \begin{centering}
        \includegraphics[width=\linewidth]{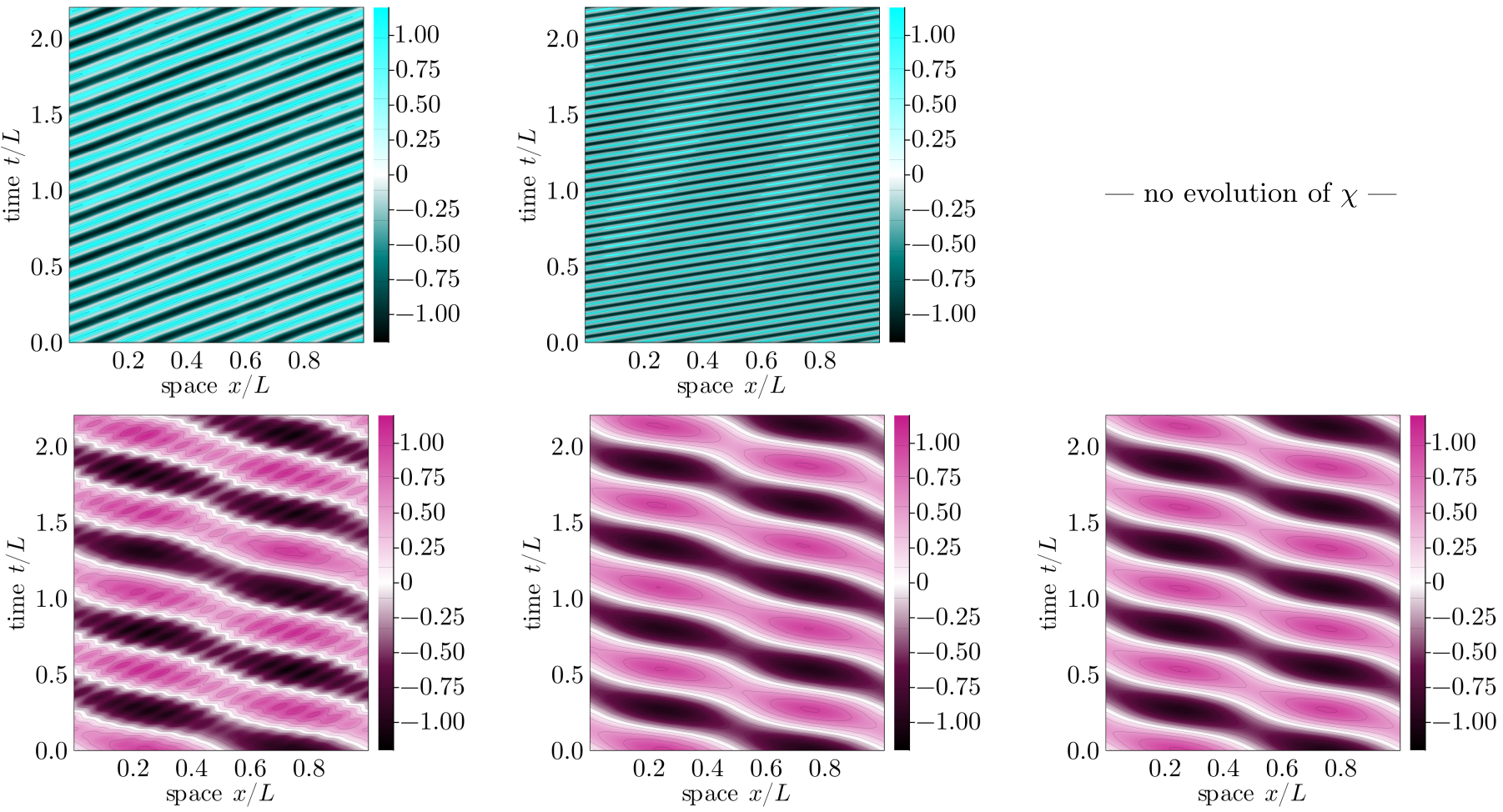}
    \end{centering}
    \caption{
        \label{fig:decoupling}
        We demonstrate that a heavy ghost field $\chi$ (upper panels) can be integrated out and generates an effective mass term for a coupled light non-ghost field $\phi$ (lower panels).
        In the two left-hand panels, we show the time evolution of the field equations in~\cref{eq:field-theory-wave-equations-phi2chi2} with $\sigma=-1$, i.e., such that the $\chi$ field is a ghost with respect to the $\phi$ field, and $\lambda_{22}\times L^2=100$, i.e., such that the $\phi$ and $\chi$ are coupled with a $\phi^2\chi^2$ interaction.
        In both simulations $k_\phi\times L/(2\pi)=1$ and $m_\phi=0$ such that $\phi$ acts a low-energy field. In contrast, we pick $k_\chi\times L/(2\pi)=5$ and increasingly heavy masses $m_\chi^2/\lambda=10$ (left) and $m_\chi^2/\lambda=100$ (middle) to demonstrate that a heavy ghost can effectively be integrated out.
        The lower right-hand panel shows a reference simulations propagating only $\phi$ as per~\cref{eq:field-equation_effective-mass-term} with an effective mass term $m_{\phi,\text{eff}}=1$. (Note that we keep initial data for a massless scalar field $\phi$ throughout. Hence, $\phi$ does not simply propagate as a plane wave but rather oscillates.)        
        We use the plane-wave initial data (see~\cref{sec:setup:ID:waves}) at fixed initial amplitude $A_\phi=A_\chi=1$. 
        \href{https://zenodo.org/records/15209689}{Animations available online.}
    }
\end{figure*}

To demonstrate the dependence on frequency, we focus on the family of plane-wave initial data (see~\cref{sec:setup:ID:waves}) and on the benign case to highlight that, at fixed amplitude, modes with higher frequency become more stable rather than less stable. For the plane-wave initial data in~\cref{eq:Init_Data_Plane}, the field equations in~\cref{eq:field-theory-wave-equations-phi2chi2} reduce to~\footnote{These expressions hold exactly at the initial time and only approximate the system before different Fourier modes are mixed due to the nonlinear interactions.}
\begin{align}
    \partial_t^2\phi &= 
    - \left(k_\phi^2 + m_\phi^2 + \lambda\,\chi^2\right)\phi
    \;,
	\notag\\
	\partial_t^2\chi &= 
    - \left(k_\chi^2 + m_\chi^2 + \sigma\,\lambda\,\phi^2\right)\chi
    \;.
\end{align}
The larger the wave number $k_\phi$ in comparison to all potential terms (including masses), the closer the local time evolution approximates the local time evolution of a free field. Once more, this statement is independent of the sign of~$\sigma\pm1$, i.e., independent of the presence/absence of a ghost.

From the above, one expects that plane wave initial data with increasing frequency is increasingly stable against all types of potential instabilities. This includes, both, tachyonic instabilities (driven by a wrong sign of the mass terms) and ghost instabilities (driven by a wrong sign of the effective mass terms due to the ghostly interaction). 
This expectation can indeed be confirmed by numerical evolution, see~\cref{fig:IR-instability}. The same argument generalises to (and the same numerical results also persist for) the non-benign case. 
\\

We conclude that, while the ghost instability can populate higher frequencies, it is not driven by high frequency fluctuations.

\section{Massive ghost fields}
\label{sec:massive}

We now include mass terms. We first demonstrate that the ghost instability can be quenched if both, or even just one of the fields, is chosen to be heavy (see~\cref{sec:massive:longlived}). Further, we exemplify that this can be used to effectively decouple a heavy ghost field from the remaining dynamics (see~\cref{sec:massive:decoupling}).
\\

By nature, numerical methods cannot provide proof of stability as they are limited to (i) a finite evolution time $T$ and (ii) a fixed family of initial data $\Phi_0$. In spite of these caveats, we can confidently extract critical values of the physical parameters and/or extract finite-time behaviour on a given family of initial data and, from that, obtain scaling relations that suggest how our finite-time results extrapolate to infinite time. 
Throughout this paper, we operationally define the onset of an instability as the evolution time $T$ at which either of the magnitude of the component energies $H_\phi$ and $H_\chi$ (see~\cref{eq:component-energies}) increases $e$-fold for the first time in comparison to the maximum of its magnitude up to $T/10$.

\subsection{Massive benign ghosts}
\label{sec:massive:longlived}

In the following, we demonstrate that a benign ghostly interactions can be quenched by mass terms. To do so, we choose the lowest-order benign ghostly interaction, i.e., $V_\text{int}^{(22)}=\lambda_{22}\,\phi^2\chi^2$ (while keeping $V_\phi=V_\chi=0$; see~\cref{sec:longlived} for the effect of self-interactions). For simplicity, we identify the to masses $m=m_\phi=m_\chi$ such that the model comes with two dimensionless parameters, $\lambda_{22}\times L^2$ and $m\times L$. Focusing on the stochastic initial data family $\Phi^{(rand)}(k,A)$ with $k\times L/(2\pi)=1$ and $A=1$, we further choose $\lambda_{22}\times L^2$ sufficiently large such that the benign instability in the massless model is triggered sufficiently fast. We then demonstrate how the onset of the instability recedes with growing mass $m$, see~\cref{fig:benign:mass-quenching}.
\\

We note that for the massless case (cf.~\cref{fig:benign}), the zero-mode of $\chi$ becomes tachyonic and drives the instability. As zero-mode of $\chi$ grows in amplitude, the amplitude of $\phi$ decreases. 
For non-vanishing mass $m$, the growth of the zero-mode of $\chi$ seems to eventually be quenched and the zero-mode of $\chi$ thus undergoes oscillations in time. 
This can be understood in terms of the effective mass squared $m_{\chi,\text{eff}}^2$ in~\cref{eq:field-theory-wave-equations-phi2chi2}. If $\phi$ is non-vanishing and if $\chi$ is small, and, in particular for the given initial data, $m_{\chi,\text{eff}}^2<0$ throughout most of the domain since it is dominated by the ghostly interaction. The field $\chi$ is thus tachyonic and the zero-mode of $\chi$ is unstable. As $\chi$ grows in amplitude, $\phi$ decreases in amplitude until eventually $m_{\chi,\text{eff}}^2$ is dominated by the explicit mass term $m^2$. Once $m_{\chi,\text{eff}}^2>0$, the zero-mode of $\chi$ is no longer unstable and the growth of $\chi$ slows down and eventually leads to an oscillation of the zero-mode of $\chi$ in time.
For small masses, the timescale (or equivalently the frequency) of these oscillations depends on the mass $m$. The larger the mass, the faster the oscillation of the zero-mode of $\chi$.

For sufficiently large $m=m_\text{crit}(\Phi_0)$ and at fixed initial data $\Phi_0$, the effective mass squared is positive $m_{\chi,\text{eff}}^2>0$ \emph{everywhere}. (Recall that $m_{\phi,\text{eff}}^2>0$ anyways.) Hence, for all $m>m_\text{crit}(\Phi_0)$ there is no remaining effective tachyonic instability. It remains an interesting open question whether there is a distinct mechanism by which the ghostly interaction $V_\text{int}^{(22)}=\lambda_{22}\,\phi^2\chi^2$ can nevertheless lead to a growth of the component energies or whether sufficiently large masses (i.e., dominant in comparison to a fixed choice of initial data) can fully quench the instability.
We will investigate this question, both, analytically and numerically, in future work.

\subsection{Decoupling heavy ghosts}
\label{sec:massive:decoupling}

Given the absence of instantaneous decay and the above understanding that the higher frequencies are more stable not less stable, we now turn to investigate whether one can dynamically decouple ghosts by making them heavy.

To probe this question, we consider again the numerical evolution of the field equations in~\cref{eq:field-theory-wave-equations-phi2chi2} for which the two fields $\phi$ and $\chi$ are coupled with each other by a benign $\phi^2\chi^2$ interaction, focusing on $\sigma=-1$ such that a ghost is present. Once more, we consider plane-wave initial data with wave numbers $k_\phi$ and $k_\chi$, respectively.

In the following, $\phi$ is considered as a low-energy field. It is thus initialised at low frequency, i.e., $k_\phi=1$, and kept massless, i.e., $m_\phi=0$. 
In contrast, $\chi$ is considered as a high-energy field. It is thus initialised at comparatively high frequency, i.e., $k_\chi=10$ and increasingly heavy mass $m_\chi$. The numerical evolution, cf.~\cref{fig:decoupling}, shows how the heavy field $\chi$ decouples from the dynamics of the light field $\phi$: With growing $m_\chi$, the time evolution of $\phi$ converges to the respective time evolution of a free field with an effective mass term. 
The time evolution for $m_\chi=10$ and $m_\chi=100$ is shown in the left and middle panel of~\cref{fig:decoupling}, respectively.
For comparison, we also evolve the decoupled field equation, i.e.,
\begin{align}
	\Box\phi &= 
    - m_{\phi,\text{eff}}^2\,\phi
    \;,
    \label{eq:field-equation_effective-mass-term}
\end{align}
with an effective mass $m_{\phi,\text{eff}}$. The respective time evolution for $m_{\phi,\text{eff}}=1$ is shown in the right panel of~\cref{fig:decoupling}.

The above example clearly demonstrates that the presence of the massive high-frequency ghost amounts to an effective mass term. In the absence of more rigorous results, it remains to be explored whether such a decoupling generalises to other ghostly interactions and other initial conditions.
\\

We expect that the above has a crucial implication in the context of effective field theory: At least in classical field theory, ghosts do not seem to invalidate the decoupling theorem. Rather, the above example suggests that a sufficiently heavy (and high-frequency) ghost effectively decouples from other light fields and can thus be integrated out.
\\

\section{Longlived ghost fields}
\label{sec:longlived}

In this section, we investigate in how far the ghost instability can be quenched by non-derivative self-interactions. Overall, we demonstrate that sufficiently dominant self-interactions can indeed lead to longlived motion.
More specifically, in the following subsections, we present numerical evidence for the following key claims:
\begin{itemize}
    \item 
    First (see~\cref{sec:longlived:Liouville}), we demonstrate that benign ghostly interactions can be quenched by sufficiently dominant self-interactions and that, for suitable initial data in a compact phase-space region, the resulting motion can become longlived. To do so, we work with the field-theory generalisation of an integrable polynomial point-particle system, see~\cite{Deffayet:2023wdg}.
    \item 
    Second (see~\cref{sec:longlived:catastrophic}), we demonstrate that even catastrophic ghostly interactions can be quenched by sufficiently strong self-interactions. In particular, for suitable initial data in a compact phase-space region, the remaining instability is benign and the resulting motion can, once more, become exponentially longlived. 
    \item 
    Third (see~\cref{sec:longlived:PRL-model}), we demonstrate that ghostly interactions can become even more long-lived if the ghostly interactions are localised to small field values. To do so, we numerically solve the field-theory equivalent of the non-polynomial model introduced in~\cite{Deffayet:2021nnt}. 
\end{itemize}
As in the previous section, we operationally define the onset of instability as the evolution time $T$ at which either of the magnitude of the component energies $H_\phi$ and $H_\chi$ (see~\cref{eq:component-energies}) increases $e$-fold for the first time in comparison to the maximum of its magnitude up to $T/10$.

\subsection{Quenching benign ghostly interactions}
\label{sec:longlived:Liouville}

To demonstrate that benign ghostly interactions can be quenched, we specify to a polynomial potential first investigated in~\cite{Deffayet:2023wdg}, in the context of integrable point-particle systems. For the case of point-particles, we have proven global stability, i.e., bounded phase-space motion for all initial conditions~\cite{Deffayet:2021nnt,Deffayet:2023wdg}. Here, we generalise the same potential from point-particles to (1+1) dimensional field theory.

The field theories at hand differ from the integrable point-particle systems in two crucial respects: First, in contrast to point-particle systems with a finite-dimensional phase space, continuum field theories have an infinite-dimensional phase space. Second, in contrast to point-particle systems, the potentials extended from the latter to field theories do not exhibit integrability.
In the absence of integrability, we do not attempt to obtain analytical insights into global stability.
Instead, we resort to numerical methods. Of course, the latter cannot provide a proof of stability. Nevertheless, we will be able to demonstrate that suitable limits of the dimensionless parameters, which characterise the model and the initial data, can lead to exponentially long-lived motion.
\\

The polynomial Liouville model is defined by a Lagrangian/Hamiltonian density as in~\cref{eq:field-theory-Lagrangian} and its potential, i.e.,
we investigate the polynomial potential obtained in~\cite{Deffayet:2023wdg}, i.e.,
\begin{align}
    V_\text{LV}^{(4)}(x,y) 
    &= 
    \frac{1}{\tilde{c}} \left( 
        \frac{m_\phi^2}{2}
        - \frac{m_\chi^2}{2}
    \right)(\phi^2 - \chi^2)^2
    \notag\\&\quad
    + \tilde{c}\;  \mathcal{C}_4 (\phi^4 - \chi^4)
    + \mathcal{C}_4 (\phi^2 - \chi^2)^3 \;.
    \label{eq:potential_polynomial-Liouville}
\end{align}
We note that, in comparison to~\cite{Deffayet:2023wdg}, we do not include the mass terms as part of the potential, since these are already written out in~\cref{eq:field-theory-Lagrangian}. The connection between mass terms and quartic interactions (see first line in~\cref{eq:potential_polynomial-Liouville}) is related to the integrability of the point-particle model. We are not aware that integrability extends to the field-theory case at hand.

In turn, the polynomial Liouville model is a special case of the generic polynomial interactions investigated in~\cref{sec:characterising-instabilities}, with
$\lambda_{22}= - (m_\phi^2-m_\chi^2)$,
$\lambda_{40} = (m_\phi^2-m_\chi^2) + \mathcal{C}_4$,
$\lambda_{04}= (m_\phi^2-m_\chi^2) - \mathcal{C}_4$,
$\lambda_{24}=3\mathcal{C}_4=-\lambda_{42}$,
$\lambda_{60}=\mathcal{C}_4=-\lambda_{06}$, 
and all other $\lambda_{nm}=0$.
\\

In $(1+1)$ dimensions, the fields are dimensionless and, hence, the dimensionality (in natural units of energy/momentum, i.e., in units of inverse length $L$) are
\begin{align}
    [\tilde{c}] &= 0\;,
    \notag\\
    [m_\phi] = [m_\chi] &= 1\;,
    \notag\\
    [\mathcal{C}_4] &= 2\;.
\end{align}
For simplicity, we set $\tilde{c}=1$ from here on such that the potential is fully characterised by the choice of the three dimensionful parameters $m_\phi$, $m_\chi$, and $\mathcal{C}_4$.
For the point-particle case, we have proven global stability~\cite{Deffayet:2023wdg} for $\mathcal{C}_{4}\geqslant0$ and $\tilde{c}\geqslant0$.
\\

\begin{figure*}
        \includegraphics[width=\linewidth]{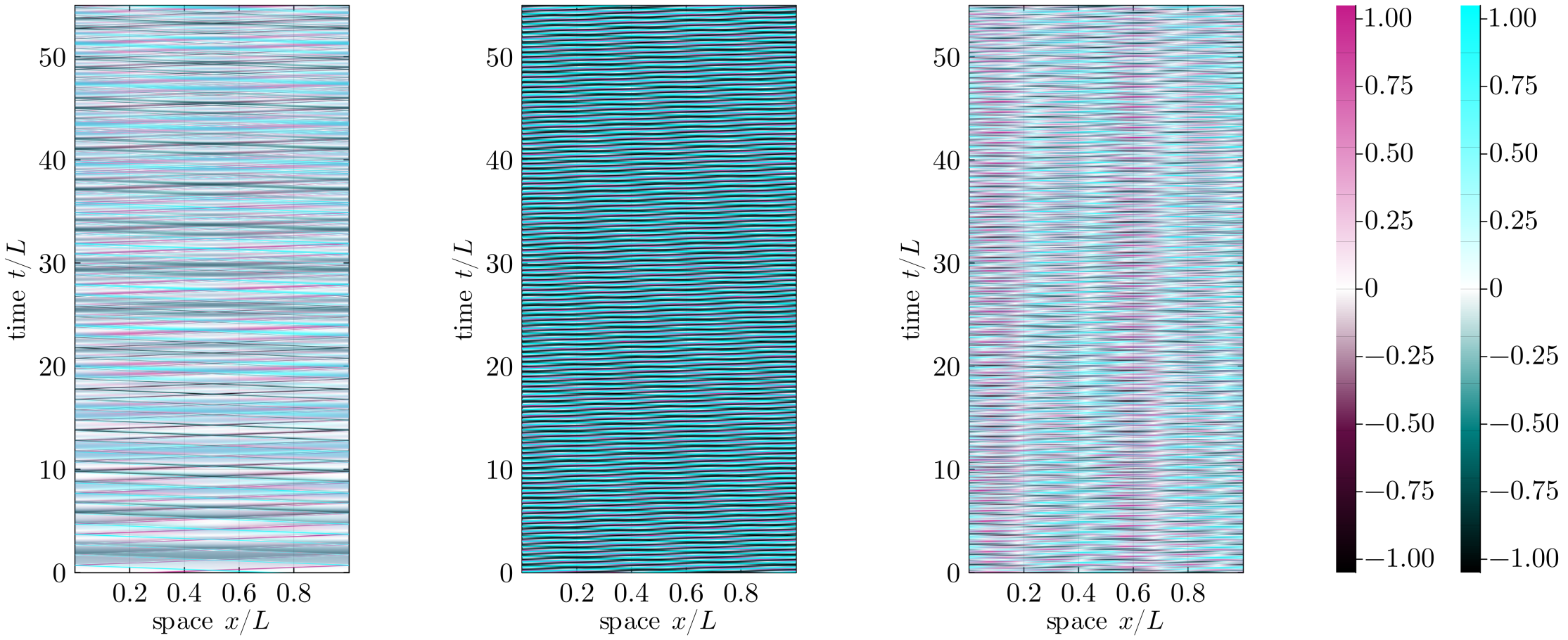}
    \caption{
        \label{fig:Liouville:trivial-vev:base-case}
        Evolution in the trivial vacuum (i.e., for $m_\phi^2\equiv m_\chi^2\equiv m^2$) of the polynomial Liouville potential, for the unit base case, i.e., $m\times L=1$, $\mathcal{C}_4\times L^2=1$ and initial data with characteristic wave number $k\times L/(2\pi)=1$ and amplitude $A=1$. From left to right, we show the evolution of the field values ($\phi$ in pink, see left legend; $\chi$ in cyan, see right legend) for the three initial-data families $\Phi^{(Gauss)}(k,\,A)$, $\Phi^{(wave)}(k,\,A)$, and $\Phi^{(rand)}(k,\,A)$ (see~\cref{sec:setup:ID}). 
        \href{https://zenodo.org/records/15209689}{Animations available online.}
    }
\end{figure*}

Depending on the relative values of the masses $m_\phi$, $m_\chi$ and the coupling $\mathcal{C}_4$, the point-particle potential can exhibit one or several locally (Lyapunov) stable equilibrium points, which are local minima of the integral of motion. 
In short, we can distinguish three different types of potentials:
\begin{itemize}
    \item 
    Potentials in which the origin is the only stable equilibrium point. 
    \item 
    Potentials in which the origin is supplemented by two further stable equilibrium points at which one field is trivial and the other non-trivial, i.e., $\phi=0$ and $\chi=\pm\chi_\text{vac} \ne 0$ and vice versa. 
    \item 
    Potentials in which the origin is supplemented by four further stable equilibrium point at which both fields are non-trivial, i.e., $\phi=\pm\phi_\text{vac} \ne 0$ and $\chi=\pm\chi_\text{vac} \ne 0$ (with all four sign combinations).
\end{itemize}
These equilibrium points apply to the point-particle model.
When generalising to field theory, we expect that these stable equilibrium points are good candidates for local field-theory vacua. We investigate these vacua and their longlivedness below.
\\

The relevant dimensionless ratios in the polynomial Liouville model are
\begin{align}
    m_{\phi/\chi}\times L
    \quad\text{and}\quad
    \mathcal{C}_4\times L^2\;.
\end{align}
These are supplemented by the characteristic initial wave number and amplitude of the initial-data families~$\Phi^{(Gauss)}(k,\,A)$, $\Phi^{(wave)}(k,\,A)$, $\Phi^{(rand)}(k,\,A)$ (see~\cref{sec:setup:ID}), i.e., in terms of dimensionless ratios,
\begin{align}
    k\times L/(2\pi)
    \quad\text{and}\quad
    A\;.
\end{align}
We recall that $L$ denotes the size of the simulated domain with periodic boundary conditions. The above four dimensionless ratios fully determine the respective physics. In summary, we find that, for all vacua, 
\begin{itemize}
    \item larger masses $m_{\phi/\chi}^2\times L^2$,
    \item smaller interaction coupling $\mathcal{C}_4\times L^2$,
    \item higher frequencies $k\times L/(2\pi)$,
    \item and smaller amplitudes $A$,
\end{itemize} 
have a stabilising effect and can lead to exponentially longlived motion in the respective limits. Vice versa, the opposite limits lead to an increasingly fast onset of the instability.

Finally, we note that decreasing the size of the computational domain $L$, is equivalent to a particular rescaling of the above dimensionless parameters which decreases the dimensionless ratios $m_{\phi/\chi}^2\times L^2$, $\mathcal{C}_4\times L^2$, and $k\times L/(2\pi)$. Overall, this effectively stabilises the motion. Vice versa, increasing the size of the computational domain has a destabilising effect.

\subsubsection{Longlived motion in the trivial vacuum}
\label{sec:longlived:Liouville:trivial-vev}

Here, we focus on the case in which the origin is the only stable equilibrium point of the point-particle potential. As one can see in~\cite[Fig.~9]{Deffayet:2023wdg}, this can be achieved by enforcing $m_\phi^2\equiv m_\chi^2\equiv m^2$ which we will investigate in this section.

As a first step, we start by setting all dimensionless ratios of the model and the initial data to unit value, i.e., $m\times L=1$, $\mathcal{C}_4\times L^2=1$, $k\times L/(2\pi)=1$, and $A=1$. The resulting evolution for all three initial data families, i.e., for $\Phi^{(Gauss)}(k,\,A)$, $\Phi^{(wave)}(k,\,A)$, and $\Phi^{(rand)}(k,\,A)$ (see~\cref{sec:setup:ID}), is shown in~\cref{fig:Liouville:trivial-vev:base-case} up to an evolution time of $t/L=50$. We refer to this as the unit base case. It turns out that the unit base case is quite longlived, i.e., the onset of the instability occurs at much larger times than $t/L=50$. In particular, the dimensionless stable evolution time is much larger than all of the dimensionless parameters of the unit base case. 

\begin{figure}
        \includegraphics[width=\linewidth]{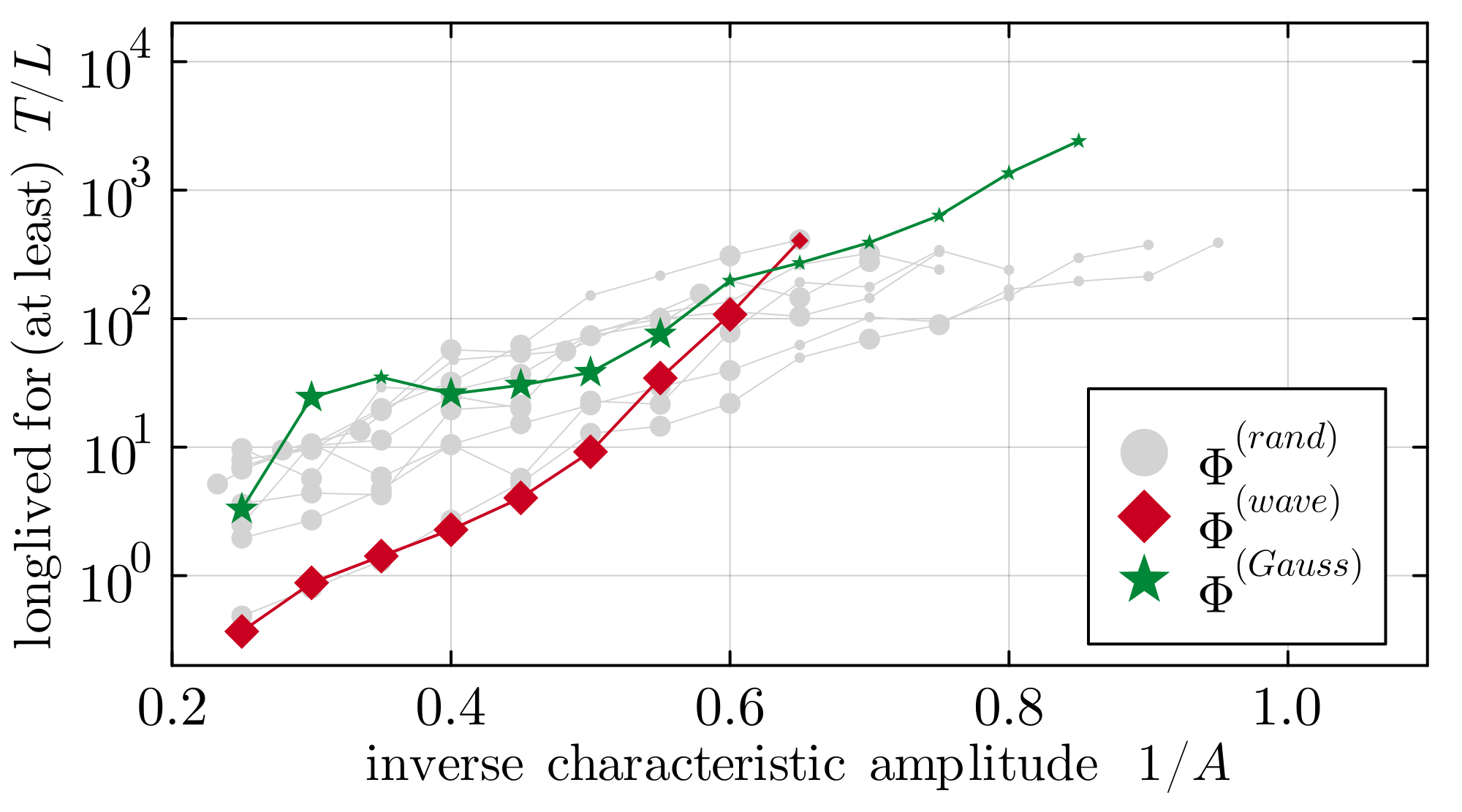}
        \includegraphics[width=\linewidth]{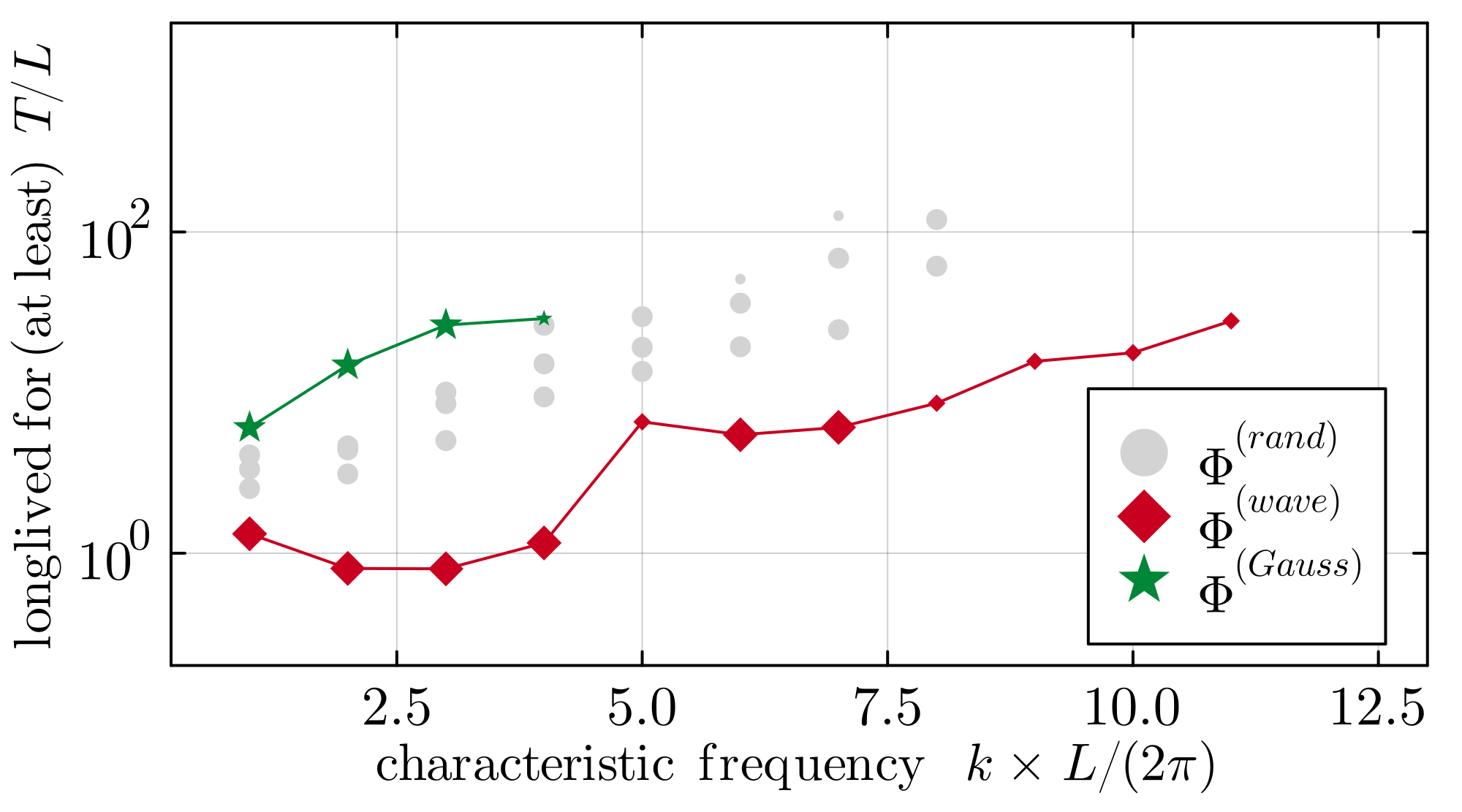}
    \caption{
        \label{fig:Liouville:scaling:ID}
        Stable evolution time $T/L$ in the trivial vacuum of the polynomial Liouville model, dependent on characteristic dimensionless parameters of the three initial-data families, i.e., on the inverse amplitude $1/A$ (upper panel and at $k\times L/(2\pi) = 1$) and on the wave number $k\times L/(2\pi)$ (lower panel and at $1/A=0.25$). The different symbols indicate different families of initial data, see legend, and convergence behaviour, see main text. In both panels, 
        $\mathcal{C}_4\times L^2 = 1$ and $m\times L=1$.
    }
\end{figure}

\begin{figure}
        \includegraphics[width=\linewidth]{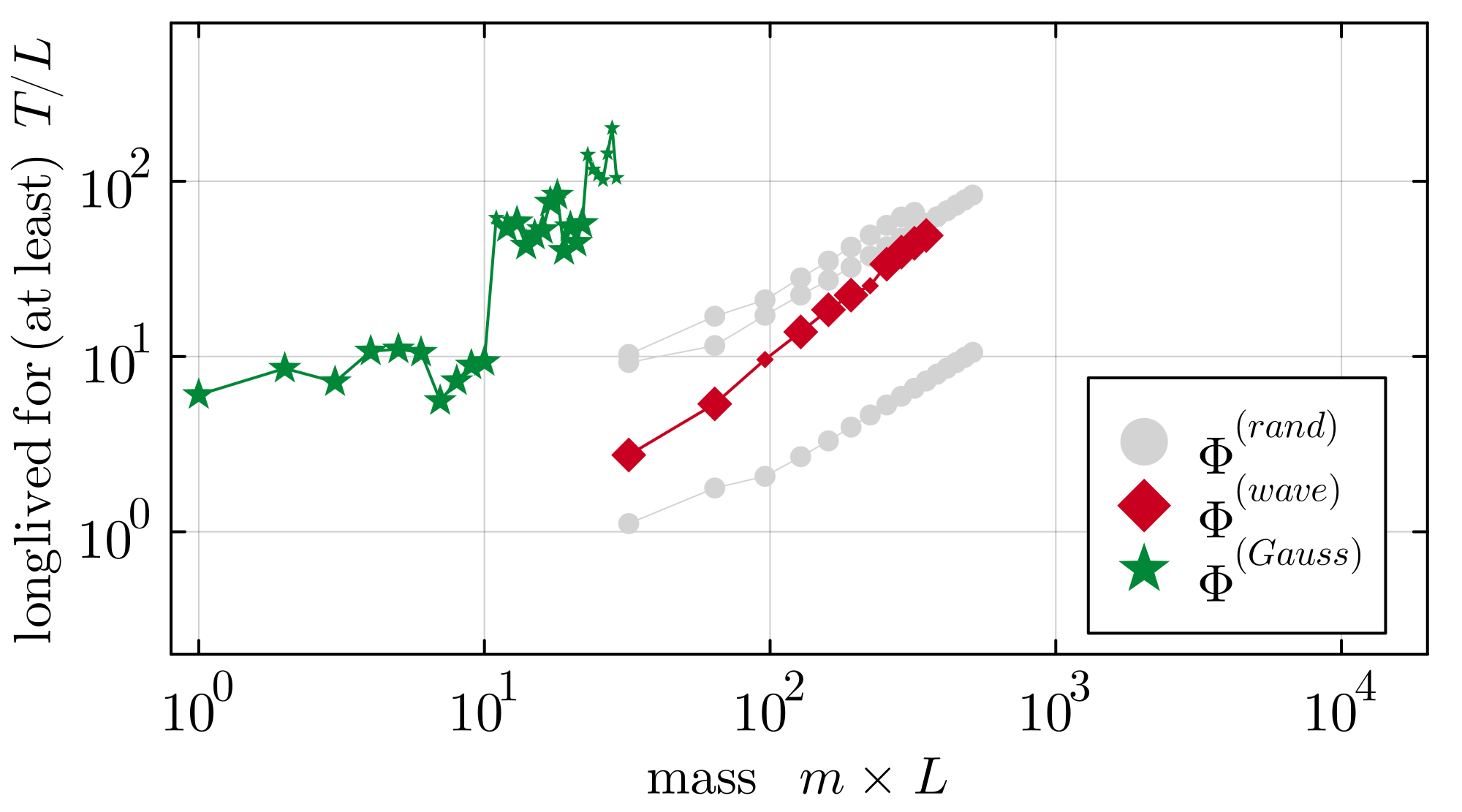}
        \includegraphics[width=\linewidth]{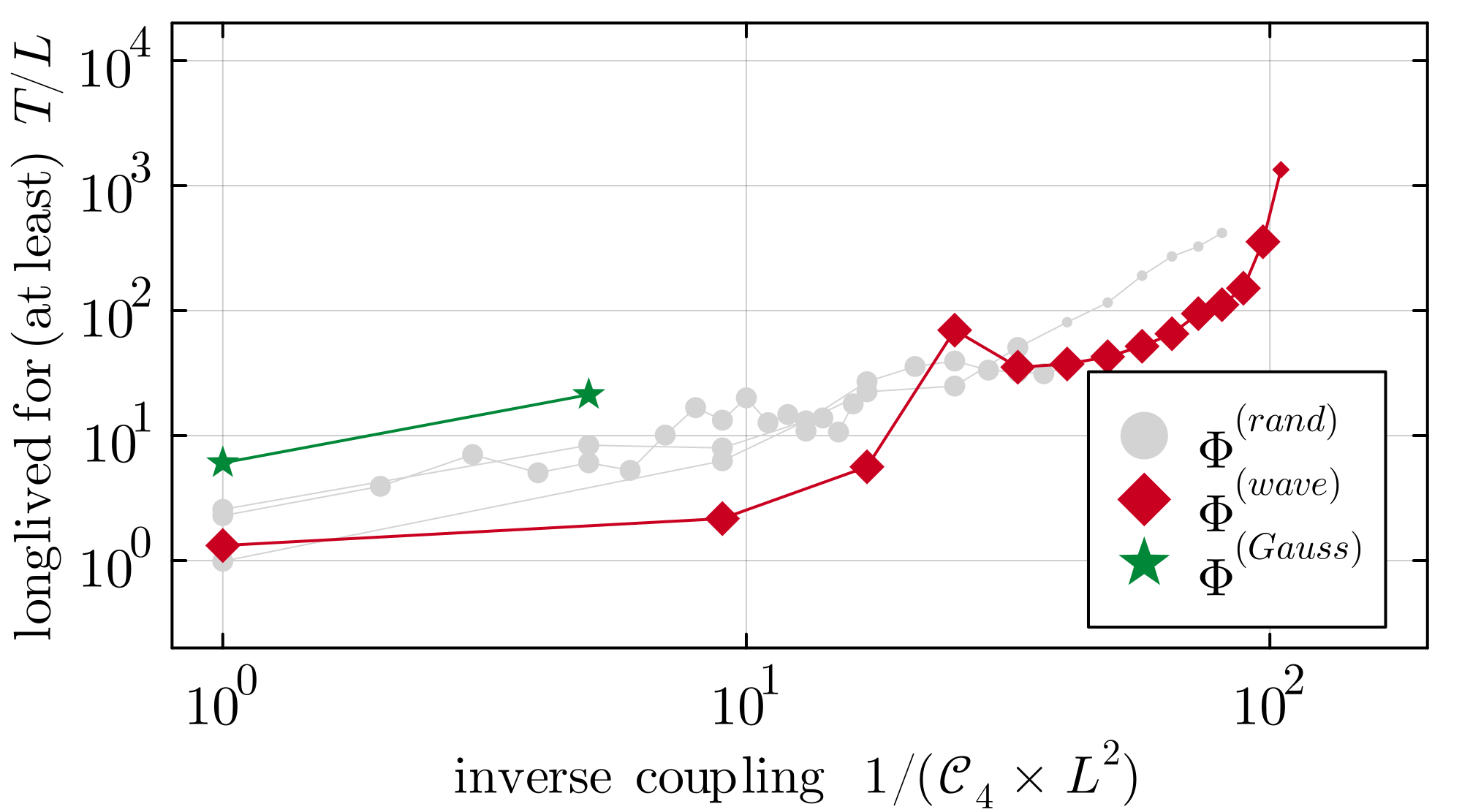}
    \caption{
        \label{fig:Liouville:scaling:model}
        Longlived stable evolution time $T/L$ in the in the trivial vacuum of the polynomial Liouville model, dependent on characteristic dimensionless parameters of the model, i.e., on dimensionless mass $m\times L$ (upper panel and at $\mathcal{C}_4\times L^2 = 1$) and on the inverse dimensionless coupling $\mathcal{C}_4\times L^2$ (lower panel and at $m\times L=1$). The different symbols indicate different families of initial data, see legend, and convergence behaviour, see main text.
        In both panels, 
        $1/A=0.25$ and $k\times L/(2\pi)=1$.
    }
\end{figure}

As a second step, we individually deform the dimensionless ratios away from their unit value to identify which deformations stabilise/destabilise the motion. For each such deformation, we measure the time at which the instability sets in, i.e., the time until the component energies grow $e$-fold in size (see above). For the four stabilising deformations of the unit base case, we show the respective scaling relations in~\cref{fig:Liouville:scaling:ID,fig:Liouville:scaling:model}. 
\cref{fig:Liouville:scaling:ID} shows the dependence on the dimensionless parameters of the initial data. \cref{fig:Liouville:scaling:model} shows the dependence on the dimensionless parameters of the model.
Each point corresponds to a converged numerical solution at the indicated model and initial data parameters. 
For the stochastic initial data family, we use stable random seeds (see~\cref{sec:setup:ID:stochastic}). This allows us to identify how each specific instance of (pseudo-)random initial data deforms with varying dimensionless mass $m\times L$, coupling $\mathcal{C}_4\times L^2$, and amplitude $A$. In~\cref{fig:Liouville:scaling:ID,fig:Liouville:scaling-nontrivial:ID}, we thus connect said instances of (pseudo-)random initial data with lines. In contrast, it is not meaningful to connect various instances of (pseudo-)random initial data evolution at varying wave number $k\times L/(2\pi)$ since this deformation is intrinsically random, even when employing a stable random seed.

For large points, the numerical precision is sufficient to confidently extract the time at which one of the component energies has grown $e$-fold in magnitude. For the small points, we cannot detect an $e$-fold increase in the component energies during the converged evolution time. For the latter, we thus plot the time at which convergence is lost. Hence, the small points indicate a confident lower bound on the stable evolution time. The interested reader can find visualisations for each respective point online.

Overall, our numerical results confidently support the conclusion that the evolution becomes increasingly longlived in the respective limits. 

%

\subsubsection{Longlived motion in a non-trivial vacuum}
\label{sec:longlived:Liouville:nontrivial-vev}

\begin{figure*}
        \includegraphics[width=\linewidth]{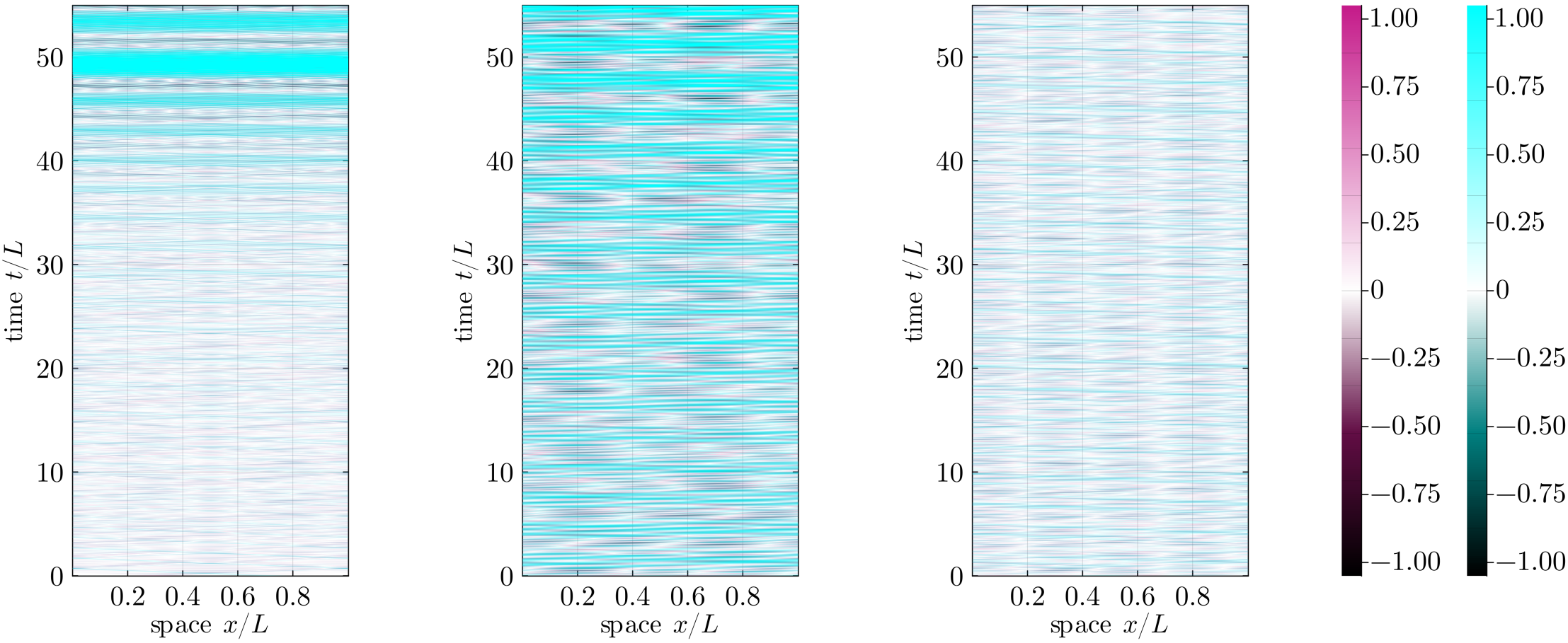}
    \caption{
        \label{fig:Liouville:nontrivial-vev:updated-base-case}
        Evolution in the nontrivial vacuum (i.e., for $m_\phi^2\equiv m_\chi^2/15\equiv m^2$) of the polynomial Liouville potential
        for the updated base case, i.e., for $m\times L=1$, $\mathcal{C}_4\times L^2=1$ and initial data with $k\times L/(2\pi)=1$ and $A=0.2$.
        From left to right, we show the evolution of the field values ($\phi$ in pink, see left legend; $\chi$ in cyan, see right legend) for the three initial-data families $\Phi^{(Gauss)}(k,\,A)$, $\Phi^{(wave)}(k,\,A)$, and $\Phi^{(rand)}(k,\,A)$ (see~\cref{sec:setup:ID}).
        \href{https://zenodo.org/records/15209689}{Animations available online.}
    }
\end{figure*}

In this section, we focus on the case in which the potential exhibits four non-trivial stable equilibrium points. 
This holds whenever the condition
\begin{align}
    \mathcal{C}_{4}<{\rm min}\left(\frac{\left(m_\phi^{2}-m_\chi^{2}\right)^{2}}{8\,m_\phi^{2}}\,,\frac{\left(m_\phi^{2}-m_\chi^{2}\right)^{2}}{8\,m_\chi^{2}}\right)\,,
    \label{eq:4-vev-condition}
\end{align}
is fulfilled, see~\cite{Deffayet:2023wdg}.
For instance, if we fix $m_\phi^2\equiv m_\chi^2/15\equiv m^2$, the above condition simplifies to
\begin{align}
    \mathcal{C}_{4}<\frac{49}{30}\,m^2\,,
    \label{eq:4-vev-condition-specific-case}
\end{align}
and, hence, the unit base case ($m\times L=1$ and $\mathcal{C}_4\times L^2=1$) exhibits the desired potential structure with four non-trivial stable equilibrium points. 
Throughout this section, we initialise both fields around their non-trivial vacuum value. For the point-particle system, the respective vacua are listed in the last line of~Tab.~I in~\cite{Deffayet:2023wdg}. We thus add the respective constant offset $\phi_\text{vac}$ and $\chi_\text{vac}$ to all initial data. In the following, whenever we refer to or visualise the field values of amplitudes, we refer to the fluctuation only, i.e., to the deviation of $\phi$ from $\phi_\text{vac}$ and of $\chi$ from $\chi_\text{vac}$, respectively.
The respective unit base case, i.e., for $m\times L=1$, $\mathcal{C}_4\times L^2=1$, $k\times L/(2\pi)=1$, and $A=1$.
For the unit base case, the amplitude is sufficiently large to decay the respective vacuum and subsequently the onset of the instability develops earlier than for the respective unit base case in the trivial vacuum. Our interpretation is that this occurs because the decay of the local vacuum frees up additional potential energy due to $\phi_\text{vac}$ and $\chi_\text{vac}$. Here, we are interested in the longlivedness of the motion in the nontrivial vacuum, i.e., to exemplify that this can be achieved, for instance, at lower amplitude. To do so, we thus decrease the amplitude to $A=0.2$. The resulting evolution of the updated base case, i.e., for $m\times L=1$, $\mathcal{C}_4\times L^2=1$, $k\times L/(2\pi)=1$, and $A=0.2$, is shown in~\cref{fig:Liouville:nontrivial-vev:updated-base-case}, once more for all three initial data families $\Phi^{(Gauss)}(k,\,A)$, $\Phi^{(wave)}(k,\,A)$, and $\Phi^{(rand)}(k,\,A)$ (see~\cref{sec:setup:ID}), from left to right.

As a second step, we again determine the time until we can detect an $e$-fold growth in any of the component energies when individually deforming the dimensionless ratios away from their unit value. 
We refrain from deforming the potential away from $m\times L=1$ and $\mathcal{C}_4\times L^2=1$, since these deformations deform the potential in a nontrivial manner and can thus expected to lead to nontrivial behaviour.

\begin{figure}
        \includegraphics[width=\linewidth]{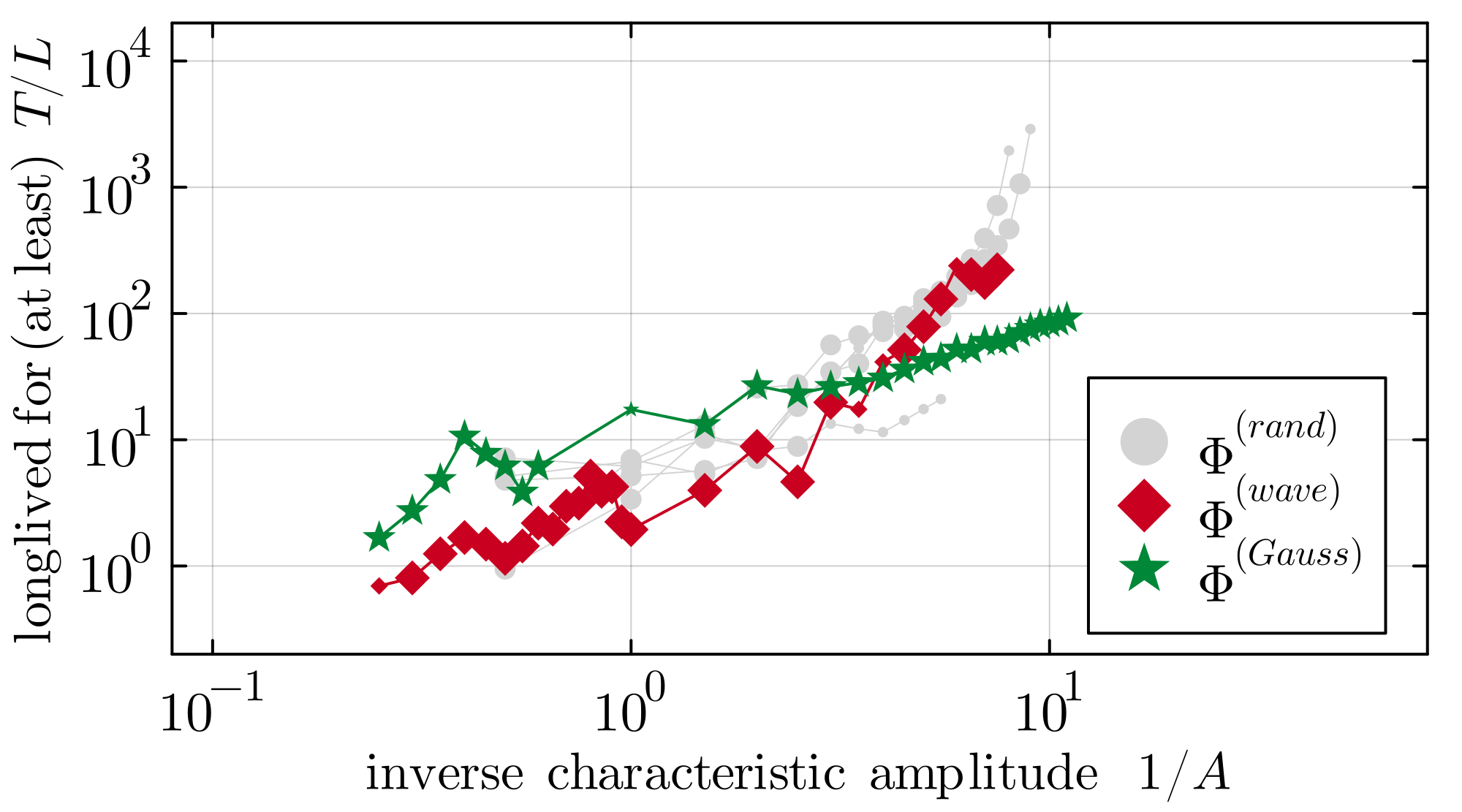}
        \includegraphics[width=\linewidth]{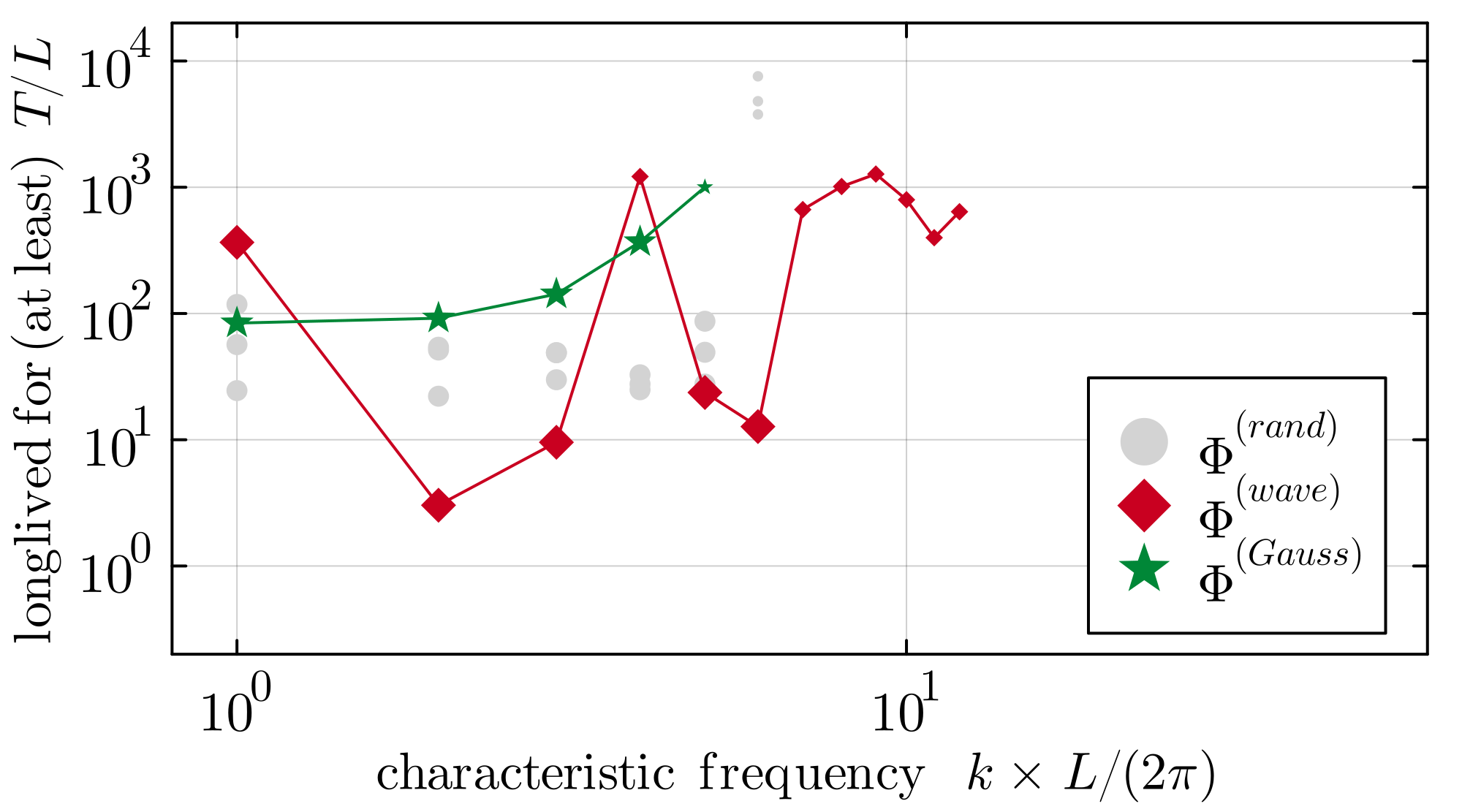}
    \caption{
        \label{fig:Liouville:scaling-nontrivial:ID}
        As in~\cref{fig:Liouville:scaling:ID} but for the nontrivial vacuum case (i.e., for $m_\phi^2\equiv m_\chi^2/15\equiv m^2$) with $m\times L=1$ and $\mathcal{C}_4\times L^2=1$. In the upper panel, we vary the amplitude at fixed frequency $k\times L/(2\pi) = 1$. In the lower panel, we vary the frequency at fixed amplitude $A=0.1$.
    }
\end{figure}

For the stabilising deformations of the characteristic initial data parameters, the respective extracted scaling relations are shown in~\cref{fig:Liouville:scaling-nontrivial:ID}. Once more, the results are compatible with increasingly longlived evolution in the respective limits. 

To summarise, we have presented numerical evidence that a sufficiently small-amplitude and/or sufficiently high-frequency (classical) fluctuations around a sufficiently deep nontrivial vacuum value can be exponentially longlived. This holds even if the nontrivial vacuum is induced by ghostly interactions. 

%
\subsection{Quenching catastrophic ghostly interactions}
\label{sec:longlived:catastrophic}

%
\begin{figure}
        \includegraphics[width=\linewidth]{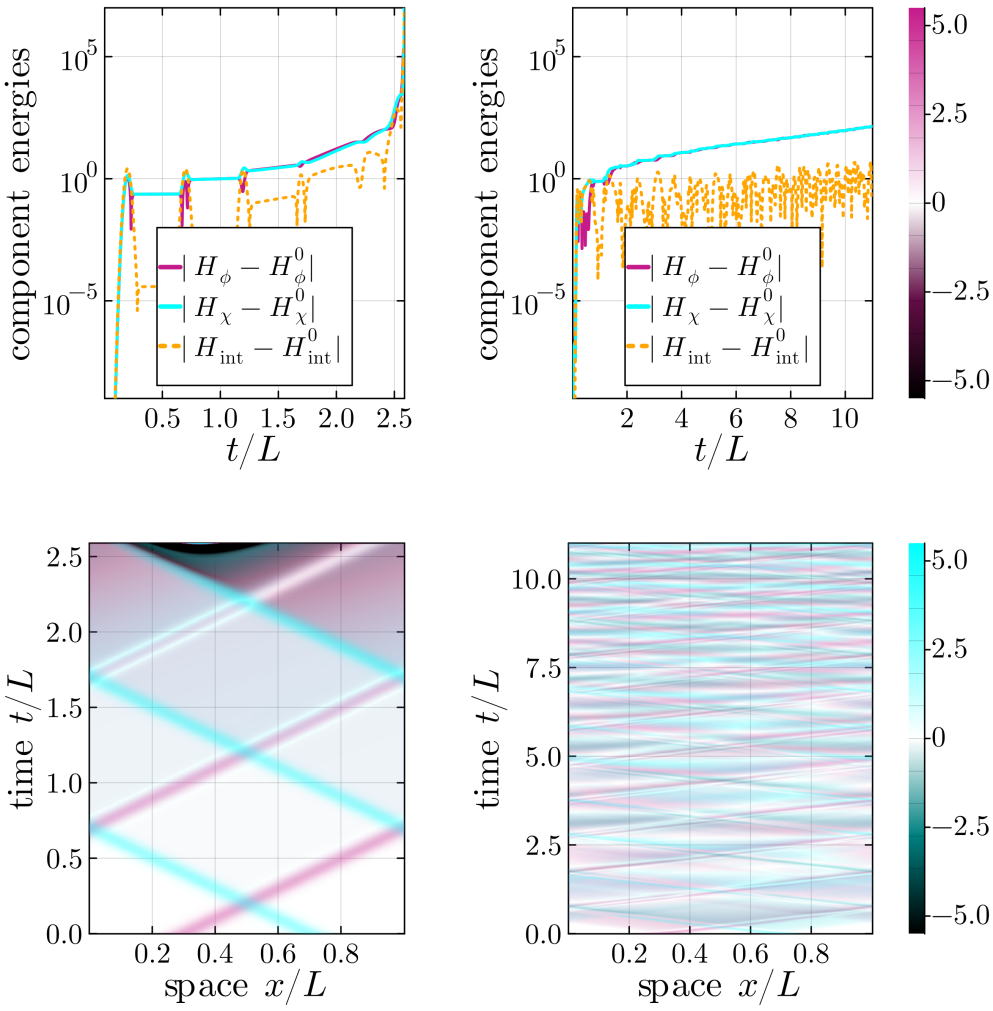}
    \caption{
        \label{fig:catastropic-quenched}
        We exemplify that sufficiently dominant self-interactions (e.g., $V_\phi = \lambda_{80}\,\phi^8$ and $V_\chi = \sigma\,\lambda_{08}\,\chi^8$, here with $\lambda_{80} \equiv \lambda_{08} \equiv \lambda_\text{self}$) can quench catastrophic runaways mediated by ghostly interactions (e.g., $V_\text{int}^{(33)}=\lambda_{33}\,\phi^3\chi^3$, here with $\lambda_{33}\times L^2=1$). To do so, we evolve the same Gaussian initial data $\Phi^{(Gauss)}(k,\,A)$ with $k\times L/(2\pi)=1$ and $A=2$, once for $\lambda_\text{self}\times L^2=0$ (left panels) and once for $\lambda_\text{self}\times L^2=1$ (right panels). The lower panels show the superposed evolution of both fields (see colour legends to the right of all figures; the upper legend specifies the $\phi$ field; the lower legend specifies the $\chi$ field).
        The upper panels show how the component energies evolve, respectively. 
        \href{https://zenodo.org/records/15209689}{Animations available online.}
    }
\end{figure}

In the previous section, we have demonstrated that trivial as well as non-trivial vacua in the ghostly polynomial Liouville model admit for increasingly longlived motion, for instance, with decreasing characteristic amplitude of the initial data.

As far as we are aware, the integrability of the point particle model (see~\cite{Deffayet:2023wdg}) does not extend to the field-theory case at hand. This suggests that integrability is not the crucial property that leads to longlived motion but rather that the generic structure of the potential is important. We thus expect that any vacuum of a potential with sufficiently dominant long-range self-interactions exhibits similar properties. In fact, we have confirmed that this is the case for the benign interactions investigated in~\cref{sec:characterising-instabilities:benign-vs-catastrophic:benign}. 

Moreover, as we will demonstrate in this section, even catastrophic ghostly interactions that -- if left unquenched -- lead to finite-time singularities, can be quenched with self-interactions. To do so, we revisit the ghostly interactions $V_\text{int}^{(33)}=\lambda_{33}\,\phi^3\chi^3$ investigated in~\cref{sec:characterising-instabilities:benign-vs-catastrophic:catastrophic} but now we add self-interactions $V_\phi = \lambda_{80}\,\phi^8$ and  $V_\chi = \sigma\,\lambda_{08}\,\chi^8$. For simplicity, we identify $\lambda_{80} \equiv \lambda_{08} \equiv \lambda_\text{self}$ in the following. 
In~\cref{fig:catastropic-quenched}, we compare the evolution of Gaussian initial data $\Phi^{(Gauss)}(k,\,A)$ with $k\times L/(2\pi)=1$ and $A=2$, once for $\lambda_{33}\times L^2=1$ and $\lambda_\text{self}\times L^2=0$ (unquenched), and once for $\lambda_{33}\times L^2=1$ and $\lambda_\text{self}\times L^2=1$ (quenched). While the unquenched case develops a finite-time singularity at $T/L\approx 2.587$, the quenched case shows no sign of a catastrophic runaway until (at least) $T/L=10$. 

We conclude that self-interactions seem to be able to quench a catastrophic runaway, or at least turn the former into a benign one.

\subsection{Longlived motion in a non-polynomial model}
\label{sec:longlived:PRL-model}

%
\begin{figure*}
        \includegraphics[width=\linewidth]{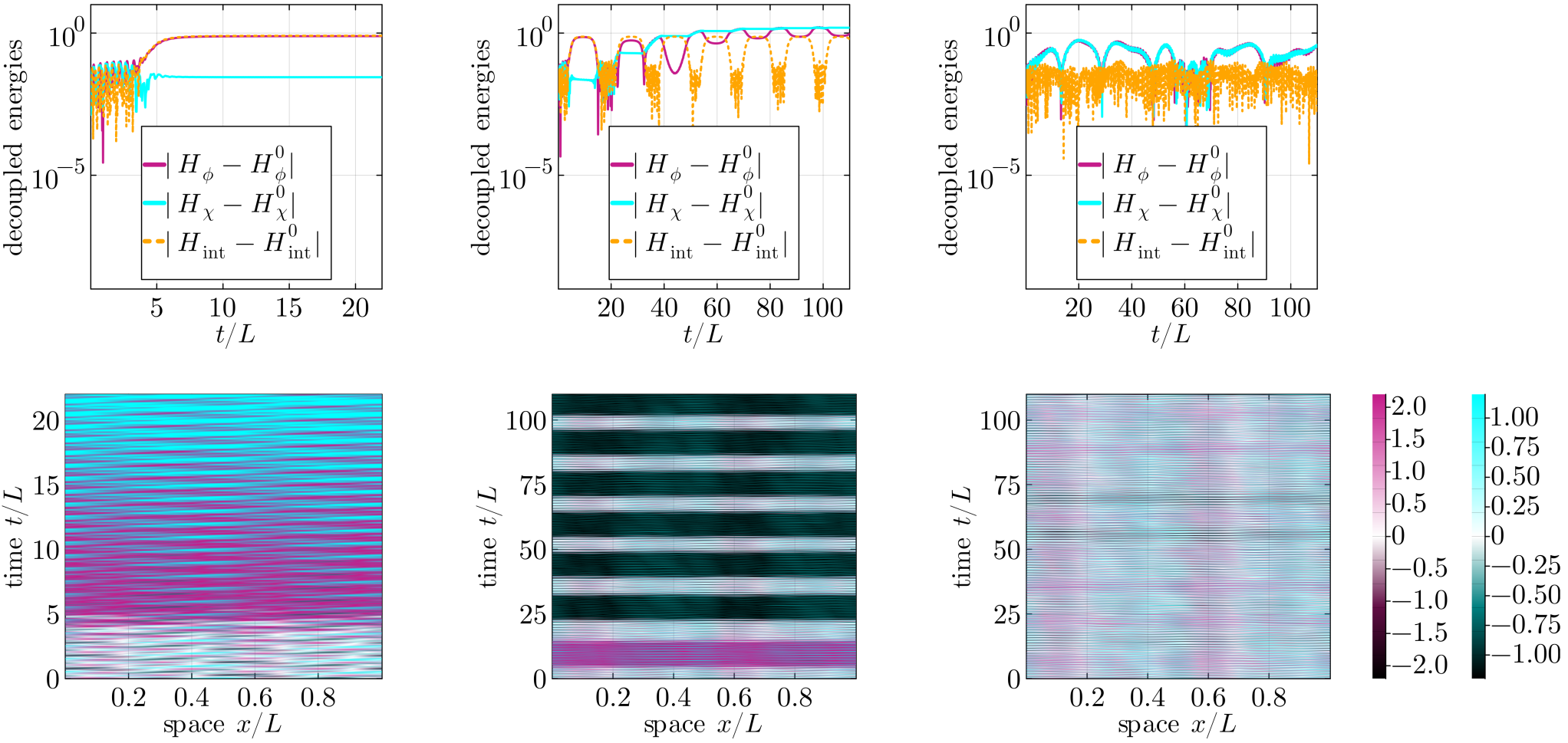}
    \caption{
        \label{fig:PRL}
        We exemplify the time evolution in the nonpolynomial model in~\cref{eq:PRL-potential} (for $\lambda\times L^2=1$) with varying mass terms. From left to right, we show the exemplary evolution of stochastic initial data $\Phi^{(rand)}(k,\,A)$ (with $k\times L/(2\pi)=1$ and $A=1$) for $m\times L=0$, $m\times L=0.1$, and $m\times L=1$. The upper panels show how the component energies (see~\cref{eq:component-energies}) evolve in time. The lower panels show the superposed evolution of $\phi$ and $\chi$. The colour scales are the same throughout all three panels and are indicated by the two legends on the right (left legend for $\phi$; right legend for $\chi$).  
        \href{https://zenodo.org/records/15209689}{Animations available online.}
    }
\end{figure*}

With the insights from the previous sections we turn to a non-polynomial potential first investigated in~\cite{Deffayet:2021nnt} for the case of point particles. (Once more the potential leads to integrable point-particle motion but we do not expect this to be of particular significance for the field theory at hand.)

The respective ghostly potential is given by
\begin{equation}
    V_{\text{int}}=\frac{\lambda}{\sqrt{1+\left(\phi^{2}-\chi^{2}\right)^{2}+2\left(\phi^{2}+\chi^{2}\right)}}
    \label{eq:PRL-potential}\,,
\end{equation}
where $\lambda$ is a coupling constant and we recall that the fields are dimensionless in $1+1$ dimensions. We note that the ghostly potential asymptotes to zero for large $\phi\gg 1$ or large $\chi\gg 1$.
For simplicity, we identify the two masses $m_\phi\equiv m_\chi\equiv m$ and exemplify the dynamics starting from stochastic initial data $\Phi^{(rand)}(k,\,A)$ with $k\times L/(2\pi)=1$ and $A=1$. In~\cref{fig:PRL}, we fix $\lambda\times L^2=1$ and demonstrate how the time evolution changes with growing mass terms, i.e., for $m\times L=0$, $m\times L=0.1$, and $m\times L=1$, from left to right. In the latter case, we cannot confidently identify any remaining growth.
In fact, our numerics suggest that generic initial data may lead to some growth in the decoupled energies, however, the latter eventually asymptote to constant values. 
In the absence of analytical proof, it remains an open question whether such models are completely stable, i.e., whether the decoupled energies remain bounded for all initial data and for all future time.

\section{Conclusions}
\label{sec:conclusions}

We numerically solve the initial value problem in $(1+1)$ dimensional Lorentz-invariant scalar field theories with opposite-sign kinetic terms and non-derivative interactions. Our results dispel expectations about an inevitable catastrophic instability (see~\cref{sec:characterising-instabilities}), suggest that heavy high-frequency ghost fields effectively decouple (see~\cref{sec:massive}), and clarify that the respective classical motion can be longlived (see~\cref{sec:longlived}).

We view these results as a crucial first step to more rigorously scrutinise the dynamics of ghosts in classical field theory. 
Generalisations to higher spatial dimensions as well as other types of fields should be addressed in future studies. In light of our results they certainly seem promising to pursue. Further, it is desirable to support our physical conclusions by analytical arguments, see~\cite{Gross:2020tph} for a study relating perturbative analytical approximations and numerical tendencies.

By nature, numerical solutions are limited to finite evolution times and specific initial data and, as such, cannot provide proof of global stability. They nevertheless allow us to reliably demonstrate several significant physical conclusions and study the onset of the instability $T$. We operationally define this time $T$ by an $e$-fold growth of the individual decoupled energies. We emphasize that we confirm self-convergence rates throughout. All of our results are thus expected to hold in the classical continuum field theory.
\\

Our first set of results (see~\cref{sec:characterising-instabilities}) characterises the unquenched instability mediated by a single polynomial ghostly interaction $V_\text{int}^{(nm)}[\phi,\chi]=\lambda_{nm}\,\phi^n\chi^m$ with $2\leqslant n,m\in\mathbb{N}$.
We confirm the expectation that the unquenched ghost instability drives the individual component (``kinetic'') energies to grow with opposite sign. 

In~\cref{sec:characterising-instabilities:benign-vs-catastrophic}, we distinguish catastrophic instabilities, for which the component (``kinetic'') energies diverge at finite evolution time, from benign instabilities where such finite-time singularities are absent. We find that interactions with odd $2\leqslant n,m\in\mathbb{N}$ lead to catastrophic instabilities while interactions with even $2\leqslant n,m\in\mathbb{N}$ seem to remain benign. We can show that the distinction is related to the behaviour of effective mass terms: In the benign case, one of the effective mass terms is manifestly positive, hence non-tachyonic. In the catastrophic case, both effective mass terms can become tachyonic. 

The above (and all following) results firmly establish that a catastrophic instability in classical field theory is not inevitable and certainly not instantaneous. In fact, the notion of instantaneous decay is directly disproved by local well-posedness of the respective initial value formulation. The latter is trivial in the present case and has previously been established for successively more general classes of gravitational higher-derivative theories which, indeed, provides one of the motivations for the present work. In~\cref{sec:characterising-instabilities:frequency-dependence}, we highlight that a physical explanation for well-posedness and, hence, for the absence of instantaneous decay can be given in terms of frequency-dependence. While the classical instability can populate high-frequency excitations, it is not driven by high-frequency fluctuations: At fixed amplitude (or kinetic energy) shorter wavelength modes are more stable not less stable. We also highlight that a transfer of energy to high-frequency modes -- a so-called direct cascade -- is not a phenomenon limited to non-linear systems with ghosts. In fact, direct cascades are typical for reheating with canonical scalar fields, see e.g. \cite{Micha:2002ey,Micha:2004bv,Lozanov:2017hjm}.
\\

Our second set of results (see~\cref{sec:massive}) concerns mass terms. In~\cref{sec:massive:longlived}, we demonstrate that increasingly heavy mass terms can efficiently quench the instability, see~\cref{fig:benign:mass-quenching}. To be specific, all of our results are consistent with a polynomial relation between the mass $m$ and the onset time of the instability $T$, i.e., $T/L\sim m^2L^2$, see the left-hand panel in~\cref{fig:benign:mass-quenching}.
In~\cref{sec:massive:decoupling}, we provide an example which explicitly demonstrates that a sufficiently heavy (and sufficiently high-frequency) ghost field effectively decouples from the remaining dynamics and can effectively be integrated out, see~\cref{fig:decoupling}. In the context of effective field theory, we thus see no reason to expect that ghost fields violate the decoupling theorem, see~\cite{Appelquist:1974tg} for a formulation of the theorem in the context of quantum field theory. Of course, as for all of our results, the effect of quantisation remains to be clarified.  
\\

Our third set of results (see~\cref{sec:longlived}) studies the impact of self-interactions. This analysis is motivated by recent proofs of local and global stability for point-particle systems~\cite{Deffayet:2021nnt,Deffayet:2023wdg}. 
Here, we investigate field theories with the same potential interactions.

In a first part (see~\cref{sec:longlived:Liouville}), we demonstrate that benign ghostly instabilities can be delayed by sufficiently dominant non-derivative self-interactions. To do so, we work with the field-theory generalisation of an integrable polynomial point-particle system, see~\cite{Deffayet:2023wdg}. 
We identify the respective dimensionless limits in which the motion becomes arbitrarily longlived. Concerning initial data, we find that (sufficiently) small amplitude and/or high frequency can lead to (arbitrarily) longlived motion. Longlivedness seems to hold not just for fluctuations around the trivial vacuum but also for fluctuations around nontrivial vacua. For the trivial vacuum, our numerical results are consistent with exponential longlivedness as a function of inverse amplitude and of frequency, see~\cref{fig:Liouville:scaling:ID}. For the nontrivial vacuum,
we can identify longlivedness in the same dimensionless limits but statements about the respective scaling are less conclusive, see~\cref{fig:Liouville:scaling-nontrivial:ID}.
For the trivial vacuum, we also investigate the dependence on the dimensionless model parameters and demonstrate that (sufficiently) large mass and/or weak interaction coupling can also lead to (arbitrarily) longlived motion, see~\cref{fig:Liouville:scaling:model}.

In a second part (see~\cref{sec:longlived:catastrophic}), we demonstrate that even catastrophic ghostly interactions can be quenched by sufficiently strong self-interactions, see~\cref{fig:catastropic-quenched}. In particular, for suitable initial data in a compact phase-space region, the remaining instability is benign and the resulting motion can, once more, become longlived. 

Third (see~\cref{sec:longlived:PRL-model}), we demonstrate that ghostly interactions can become even more stable if the ghostly interactions are localised to small field values. To do so, we numerically solve the field-theory equivalent of the non-polynomial model introduced in~\cite{Deffayet:2021nnt}, see~\cref{fig:PRL}. In contrast to all previous polynomial models, the respective ghostly interaction potential vanishes for large field values. Hence, even mass terms seem sufficient to completely quench the instability. In fact, we find that, even in the massless model, the component energies seem to not diverge at all but rather asymptote to finite values. For increasing mass, the transition to this asymptotic behaviour occurs at later and later times and a non-trivial interaction remains. The latter seems to be stable, at least for all practical purposes. 
\\

It is interesting to note that unbounded from below gradient energies can appear in usual, linearly stable systems which are free of ghosts. This happens in strongly anisotropic cases when energies are calculated with respect to a supersonic observer \cite{Babichev:2017lmw,Babichev:2024uro,Sawicki:2024ryt}. These negative energies may cause a physical instability which is, if not identical \cite{Babichev:2024uro}, then at least similar~\cite{Sawicki:2024ryt}, to the ghostly one investigated here. This supersonic instability corresponds to the usual Cherenkov radiation which is clearly not instantaneous and describes an observed physical process. 

To conclude, we think that the question of stability for interacting field theories with ghosts may hide a lot of surprises, including interesting physical phenomena.

\subsection*{Acknowledgements}
\label{sec:Acknowledgements}
We thank Mustafa Amin, Ramiro Cayuso, Hyun Lim, Luis Lehner, Frans Pretorius, Toby Wiseman and Richard P.~Woodard for discussions.
The work of S.M. was supported in part by Japan Society for the Promotion of Science (JSPS) Grants-in-Aid for Scientific Research No.~24K07017 and the World Premier International Research Center Initiative (WPI), MEXT, Japan. 
The work of A.~V. was supported by European Structural and Investment Funds and the Czech Ministry of Education, Youth and Sports (Project FORTE CZ.02.01.01/00/22 008/0004632).
Numerical computations with up to $N=2^{11}=2048$ can be obtained on a standard laptop. All numerical computations which require higher resolution have been performed on a single node (16 cores at 2.8 GHz each and up to 1 TB of RAM) of the Thanos cluster at LPENS.

\appendix

\section{Numerical evolution}
\label{app:numerics}

In this appendix, we provide further instructive details on the numerics described in~\cref{sec:setup:numerics}. 
We re-iterate that we have implemented standard periodic boundary conditions, discretize the spatial domain by means of central 4th-order finite differencing, and chose to evolve in time via a 4th-order accurate Runge-Kutta (\texttt{RK4}) scheme~\cite{runge1895numerische, kutta1901beitrag}, see also~\cite{press1988numerical}). As is common practice, we fix the temporal resolution $\Delta t$ to the spatial resolution $\Delta x$ by the so-called Courant–Friedrichs–Lewy (CFL) factor~\cite{Courant:1967} $C=\Delta t/\Delta x$. In all numerical evolutions, we pick $C=0.25$. In particular, we find that the latter choice is sufficient to avoid stiffness.

For all further details of the implementation, the interested reader may additionally consult the \texttt{git}-repository in which we make the entire source code publicly available. All results contained in this paper can be reproduced by downloading this source code, installing all dependencies and, in particular, the \texttt{julia} computing infrastructure, and executing the respective source files on a suitable machine. More details on dependencies and how to run the code can be found in the \texttt{git}-repository as well.

Finally, we discuss two important points in the remainder of this appendix: First, we briefly review the concept of self-convergence rate and present an example of the automated convergence tests which have been implemented. Second, we exemplify that different time stepping schemes reproduce the same continuum solution.

\subsection{Convergence}
\label{app:numerics:convergence}

%
\begin{figure}
    \begin{centering}
        \includegraphics[height=0.6\linewidth]{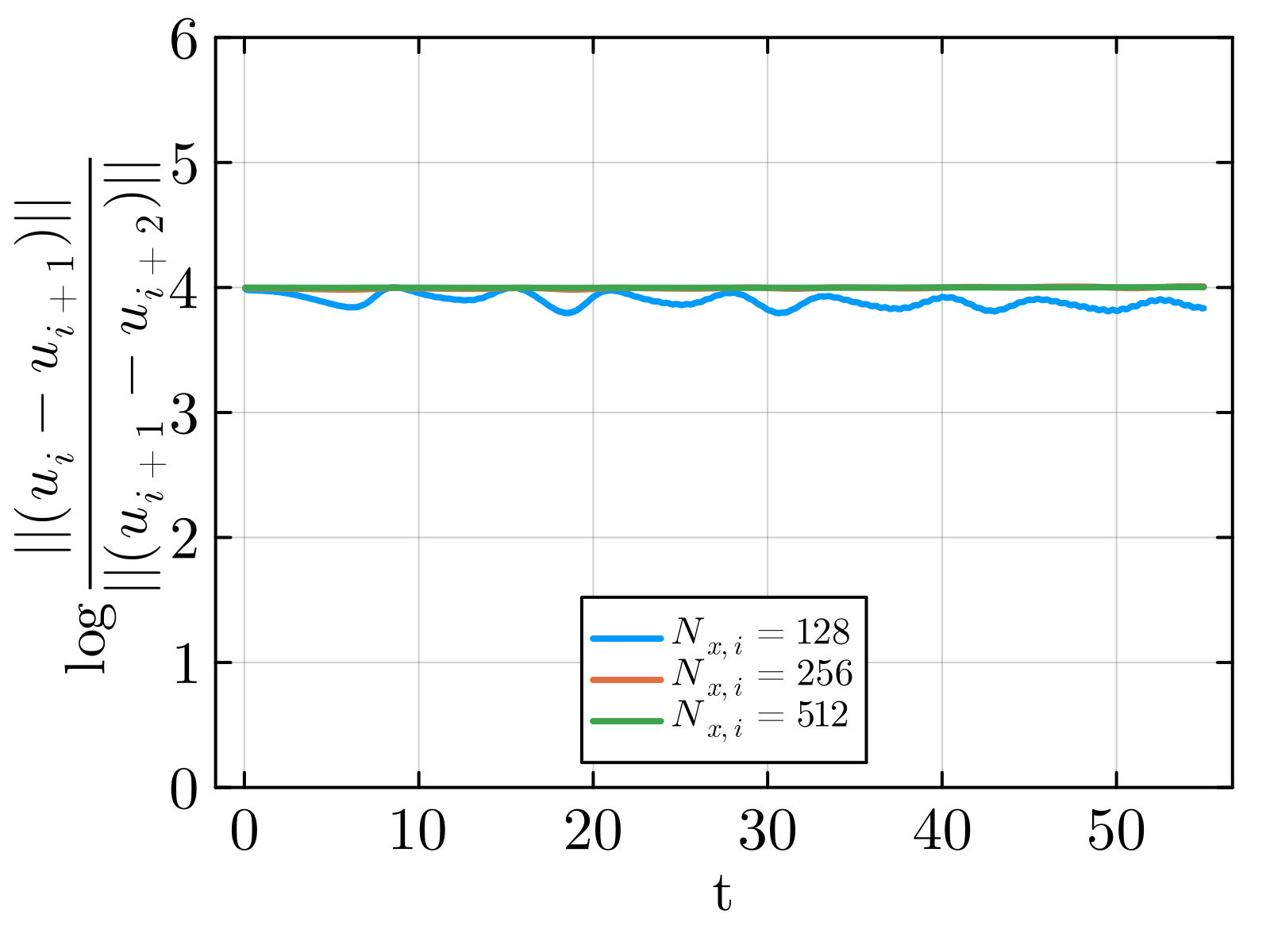}
        \includegraphics[height=0.63\linewidth]{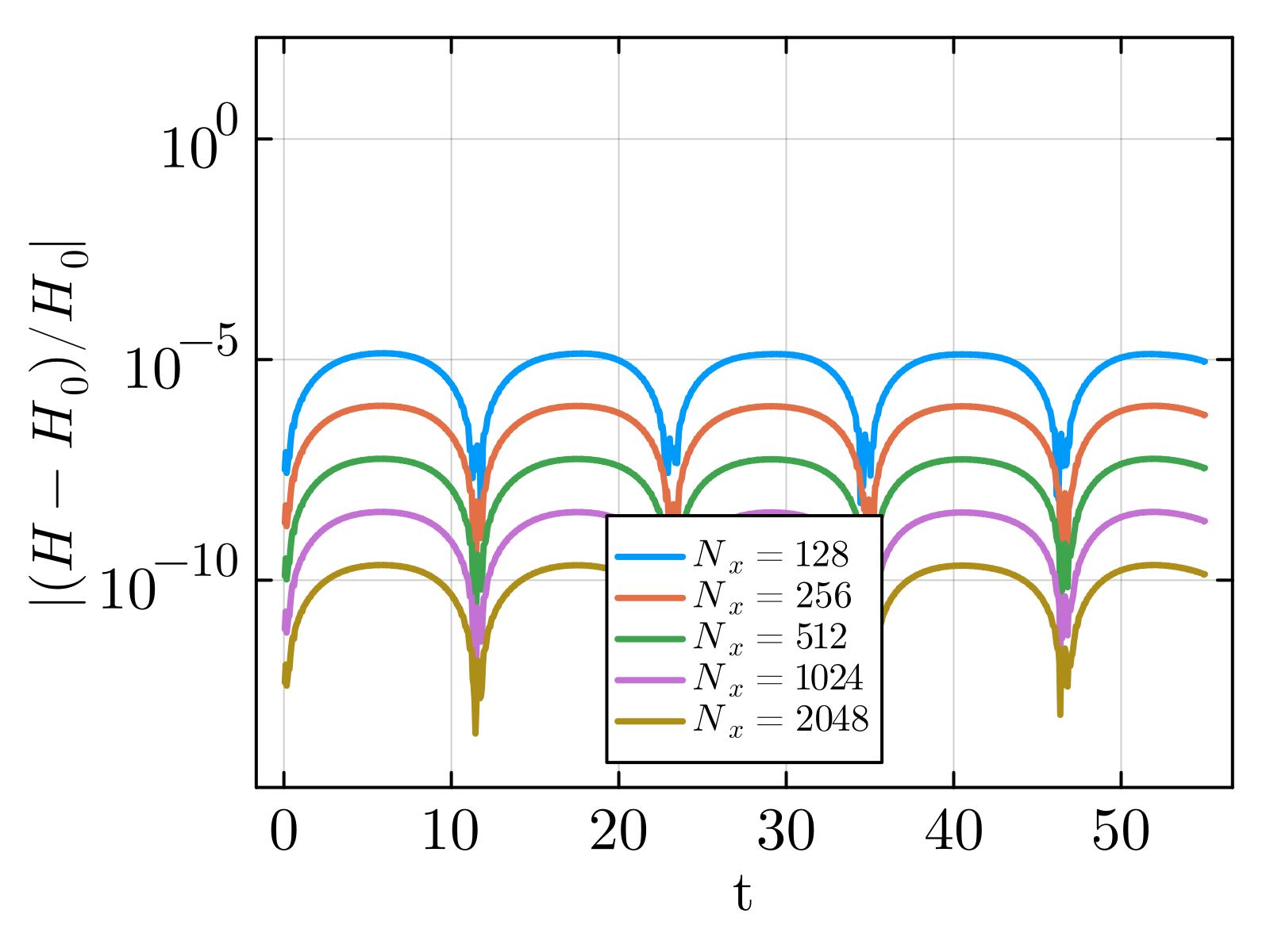}
    \end{centering}
    \caption{\label{fig:exemplary_convergence_test}
        As an example, we show a self-convergence test on the state vector $u$ (upper panel) and the relative error of the Hamiltonian constraint $H$ (lower panel) for the evolution shown in the centre panel of~\cref{fig:Liouville:trivial-vev:base-case}. 
    }
\end{figure}

In the \texttt{julia} code, we have implemented an automated self-convergence test. The code will always run three neighbouring resolutions, labelled here by $i=1,2,3$, such that $N_i = N_{i+1}/2$.

From this, the code calculates the self-convergence rate
\begin{align}
    \mathcal{C} = \log\left(\frac{||u_i - u_{i+1}||}{||u_{i+1} - u_{i+2}||}\right)\;.
\end{align}
Herein, $u_i$ denotes the state vector of all evolution variables, i.e., a collection of all four 1st-order fields evaluated on all spatial points within the discretization. Moreover, $||\cdot||$ denotes a suitable norm. To be explicit, we determine self-convergence with respect to the standard discretised $L^2$ norm. To calculate this norm, the state vectors obtained with higher resolutions have to be scaled down to the lowest resolution. We note that it is crucial to ensure, both in this downscaling and in the initial data, that the respective points correspond to the same physical point in the continuum. Otherwise, the discretisation of the initial data can dominate the convergence rate. Since we are working with a 4th-order accurate finite-difference scheme and a 4th-order accurate time-evolution scheme (\texttt{RK4}), the expected convergence rate for sufficiently high resolution is $\mathcal{C}=4$.

In the upper panel of~\cref{fig:exemplary_convergence_test}, we show convergence plots for the exemplary evolution case corresponding to the centre panel of~\cref{fig:Liouville:trivial-vev:base-case}. 
As described above, our \texttt{julia} code automatically computes the convergence rate on \emph{every} evolution and ensures that all downstream analysis only concerns the portion of the evolution on which the convergence rate is maintained, i.e., $3<\mathcal{C}<5$, and, hence, convergence to the continuum solution is thereby established. While an example of the convergence rate is presented in~\cref{fig:exemplary_convergence_test}, the interested reader can consult the \texttt{git}-repository for the respective convergence plots for \emph{all} other results of this paper. 
\\

Alternatively, convergence is often verified by calculating $H^{(N)} - H_0$, where $H^{(N)}$ is the total Hamiltonian at the respective time step and resolution $N$ and $H_0$ is its value at the initial time. As the total Hamiltonian is conserved, the above Hamiltonian constraint violations should decrease when the resolution is increased. While this has a direct physical interpretation, the specific rate of convergence may vary with initial data, nonlinearities in the Hamiltonian, and also depends on the discrete integration routine which is used to calculate the spatial integral of the Hamiltonian density. We nevertheless present the convergence of the Hamiltonian constraint in the lower panel of~\cref{fig:exemplary_convergence_test}. Once more the \texttt{git}-repository contains the respective convergence plots for \emph{all} other results of this paper.

\subsection{Time evolution schemes}

All of the results presented in this paper have been obtained with a 4th-order Runge-Kutta (\texttt{RK4}) scheme~\cite{runge1895numerische, kutta1901beitrag}, see also~\cite{press1988numerical}).
The \texttt{DifferentialEquations.jl}~\cite{rackauckas2017differentialequations} package however provides a large variety of other evolution schemes which are readily available. We emphasize that the specific choice of time stepping scheme is not of any relevance as long as suitable convergence to the continuum solution is rigorously established. 
We have tested various different 4th-order schemes and have verified their convergence rates on exemplary evolutions. The interested reader can reproduce these crosschecks by running the respective scripts available in the \texttt{git} repository~\cite{ghostlyPDE_1D} published alongside this article. We thus are confident that all of the presented results are independent of the choice of 4th-order time-stepping routine. The interested reader can straightforwardly adapt the respective line in the \texttt{julia} code~\cite{ghostlyPDE_1D} and verify specific results themselves.

\bibliographystyle{apsrev4-1}
\bibliography{stableGhosts}

\begin{thebibliography}{73}%
\makeatletter
\providecommand \@ifxundefined [1]{%
 \@ifx{#1\undefined}
}%
\providecommand \@ifnum [1]{%
 \ifnum #1\expandafter \@firstoftwo
 \else \expandafter \@secondoftwo
 \fi
}%
\providecommand \@ifx [1]{%
 \ifx #1\expandafter \@firstoftwo
 \else \expandafter \@secondoftwo
 \fi
}%
\providecommand \natexlab [1]{#1}%
\providecommand \enquote  [1]{``#1''}%
\providecommand \bibnamefont  [1]{#1}%
\providecommand \bibfnamefont [1]{#1}%
\providecommand \citenamefont [1]{#1}%
\providecommand \href@noop [0]{\@secondoftwo}%
\providecommand \href [0]{\begingroup \@sanitize@url \@href}%
\providecommand \@href[1]{\@@startlink{#1}\@@href}%
\providecommand \@@href[1]{\endgroup#1\@@endlink}%
\providecommand \@sanitize@url [0]{\catcode `\\12\catcode `\$12\catcode
  `\&12\catcode `\#12\catcode `\^12\catcode `\_12\catcode `\%12\relax}%
\providecommand \@@startlink[1]{}%
\providecommand \@@endlink[0]{}%
\providecommand \url  [0]{\begingroup\@sanitize@url \@url }%
\providecommand \@url [1]{\endgroup\@href {#1}{\urlprefix }}%
\providecommand \urlprefix  [0]{URL }%
\providecommand \Eprint [0]{\href }%
\providecommand \doibase [0]{http://dx.doi.org/}%
\providecommand \selectlanguage [0]{\@gobble}%
\providecommand \bibinfo  [0]{\@secondoftwo}%
\providecommand \bibfield  [0]{\@secondoftwo}%
\providecommand \translation [1]{[#1]}%
\providecommand \BibitemOpen [0]{}%
\providecommand \bibitemStop [0]{}%
\providecommand \bibitemNoStop [0]{.\EOS\space}%
\providecommand \EOS [0]{\spacefactor3000\relax}%
\providecommand \BibitemShut  [1]{\csname bibitem#1\endcsname}%
\let\auto@bib@innerbib\@empty
\bibitem [{\citenamefont {Abdul~Karim}\ \emph {et~al.}(2025)\citenamefont
  {Abdul~Karim} \emph {et~al.}}]{DESI:2025zgx}%
  \BibitemOpen
  \bibfield  {author} {\bibinfo {author} {\bibfnamefont {M.}~\bibnamefont
  {Abdul~Karim}} \emph {et~al.} (\bibinfo {collaboration} {DESI}),\ }\href@noop
  {} {\  (\bibinfo {year} {2025})},\ \Eprint {http://arxiv.org/abs/2503.14738}
  {arXiv:2503.14738 [astro-ph.CO]} \BibitemShut {NoStop}%
\bibitem [{\citenamefont {Lodha}\ \emph {et~al.}(2025)\citenamefont {Lodha}
  \emph {et~al.}}]{Lodha:2025qbg}%
  \BibitemOpen
  \bibfield  {author} {\bibinfo {author} {\bibfnamefont {K.}~\bibnamefont
  {Lodha}} \emph {et~al.},\ }\href@noop {} {\  (\bibinfo {year} {2025})},\
  \Eprint {http://arxiv.org/abs/2503.14743} {arXiv:2503.14743 [astro-ph.CO]}
  \BibitemShut {NoStop}%
\bibitem [{\citenamefont {Caldwell}(2002)}]{Caldwell:1999ew}%
  \BibitemOpen
  \bibfield  {author} {\bibinfo {author} {\bibfnamefont {R.~R.}\ \bibnamefont
  {Caldwell}},\ }\href {\doibase 10.1016/S0370-2693(02)02589-3} {\bibfield
  {journal} {\bibinfo  {journal} {Phys. Lett. B}\ }\textbf {\bibinfo {volume}
  {545}},\ \bibinfo {pages} {23} (\bibinfo {year} {2002})},\ \Eprint
  {http://arxiv.org/abs/astro-ph/9908168} {arXiv:astro-ph/9908168} \BibitemShut
  {NoStop}%
\bibitem [{\citenamefont {Vikman}(2005)}]{Vikman:2004dc}%
  \BibitemOpen
  \bibfield  {author} {\bibinfo {author} {\bibfnamefont {A.}~\bibnamefont
  {Vikman}},\ }\href {\doibase 10.1103/PhysRevD.71.023515} {\bibfield
  {journal} {\bibinfo  {journal} {Phys. Rev. D}\ }\textbf {\bibinfo {volume}
  {71}},\ \bibinfo {pages} {023515} (\bibinfo {year} {2005})},\ \Eprint
  {http://arxiv.org/abs/astro-ph/0407107} {arXiv:astro-ph/0407107} \BibitemShut
  {NoStop}%
\bibitem [{\citenamefont {Hu}(2005)}]{Hu:2004kh}%
  \BibitemOpen
  \bibfield  {author} {\bibinfo {author} {\bibfnamefont {W.}~\bibnamefont
  {Hu}},\ }\href {\doibase 10.1103/PhysRevD.71.047301} {\bibfield  {journal}
  {\bibinfo  {journal} {Phys. Rev. D}\ }\textbf {\bibinfo {volume} {71}},\
  \bibinfo {pages} {047301} (\bibinfo {year} {2005})},\ \Eprint
  {http://arxiv.org/abs/astro-ph/0410680} {arXiv:astro-ph/0410680} \BibitemShut
  {NoStop}%
\bibitem [{\citenamefont {Caldwell}\ and\ \citenamefont
  {Doran}(2005)}]{Caldwell:2005ai}%
  \BibitemOpen
  \bibfield  {author} {\bibinfo {author} {\bibfnamefont {R.~R.}\ \bibnamefont
  {Caldwell}}\ and\ \bibinfo {author} {\bibfnamefont {M.}~\bibnamefont
  {Doran}},\ }\href {\doibase 10.1103/PhysRevD.72.043527} {\bibfield  {journal}
  {\bibinfo  {journal} {Phys. Rev. D}\ }\textbf {\bibinfo {volume} {72}},\
  \bibinfo {pages} {043527} (\bibinfo {year} {2005})},\ \Eprint
  {http://arxiv.org/abs/astro-ph/0501104} {arXiv:astro-ph/0501104} \BibitemShut
  {NoStop}%
\bibitem [{\citenamefont {Kunz}\ and\ \citenamefont
  {Sapone}(2006)}]{Kunz:2006wc}%
  \BibitemOpen
  \bibfield  {author} {\bibinfo {author} {\bibfnamefont {M.}~\bibnamefont
  {Kunz}}\ and\ \bibinfo {author} {\bibfnamefont {D.}~\bibnamefont {Sapone}},\
  }\href {\doibase 10.1103/PhysRevD.74.123503} {\bibfield  {journal} {\bibinfo
  {journal} {Phys. Rev. D}\ }\textbf {\bibinfo {volume} {74}},\ \bibinfo
  {pages} {123503} (\bibinfo {year} {2006})},\ \Eprint
  {http://arxiv.org/abs/astro-ph/0609040} {arXiv:astro-ph/0609040} \BibitemShut
  {NoStop}%
\bibitem [{\citenamefont {Nesseris}\ and\ \citenamefont
  {Perivolaropoulos}(2007)}]{Nesseris:2006er}%
  \BibitemOpen
  \bibfield  {author} {\bibinfo {author} {\bibfnamefont {S.}~\bibnamefont
  {Nesseris}}\ and\ \bibinfo {author} {\bibfnamefont {L.}~\bibnamefont
  {Perivolaropoulos}},\ }\href {\doibase 10.1088/1475-7516/2007/01/018}
  {\bibfield  {journal} {\bibinfo  {journal} {JCAP}\ }\textbf {\bibinfo
  {volume} {01}},\ \bibinfo {pages} {018} (\bibinfo {year} {2007})},\ \Eprint
  {http://arxiv.org/abs/astro-ph/0610092} {arXiv:astro-ph/0610092} \BibitemShut
  {NoStop}%
\bibitem [{\citenamefont {Guo}\ \emph {et~al.}(2005)\citenamefont {Guo},
  \citenamefont {Piao}, \citenamefont {Zhang},\ and\ \citenamefont
  {Zhang}}]{Guo:2004fq}%
  \BibitemOpen
  \bibfield  {author} {\bibinfo {author} {\bibfnamefont {Z.-K.}\ \bibnamefont
  {Guo}}, \bibinfo {author} {\bibfnamefont {Y.-S.}\ \bibnamefont {Piao}},
  \bibinfo {author} {\bibfnamefont {X.-M.}\ \bibnamefont {Zhang}}, \ and\
  \bibinfo {author} {\bibfnamefont {Y.-Z.}\ \bibnamefont {Zhang}},\ }\href
  {\doibase 10.1016/j.physletb.2005.01.017} {\bibfield  {journal} {\bibinfo
  {journal} {Phys. Lett. B}\ }\textbf {\bibinfo {volume} {608}},\ \bibinfo
  {pages} {177} (\bibinfo {year} {2005})},\ \Eprint
  {http://arxiv.org/abs/astro-ph/0410654} {arXiv:astro-ph/0410654} \BibitemShut
  {NoStop}%
\bibitem [{\citenamefont {Cai}\ \emph {et~al.}(2010)\citenamefont {Cai},
  \citenamefont {Saridakis}, \citenamefont {Setare},\ and\ \citenamefont
  {Xia}}]{Cai:2009zp}%
  \BibitemOpen
  \bibfield  {author} {\bibinfo {author} {\bibfnamefont {Y.-F.}\ \bibnamefont
  {Cai}}, \bibinfo {author} {\bibfnamefont {E.~N.}\ \bibnamefont {Saridakis}},
  \bibinfo {author} {\bibfnamefont {M.~R.}\ \bibnamefont {Setare}}, \ and\
  \bibinfo {author} {\bibfnamefont {J.-Q.}\ \bibnamefont {Xia}},\ }\href
  {\doibase 10.1016/j.physrep.2010.04.001} {\bibfield  {journal} {\bibinfo
  {journal} {Phys. Rept.}\ }\textbf {\bibinfo {volume} {493}},\ \bibinfo
  {pages} {1} (\bibinfo {year} {2010})},\ \Eprint
  {http://arxiv.org/abs/0909.2776} {arXiv:0909.2776 [hep-th]} \BibitemShut
  {NoStop}%
\bibitem [{\citenamefont {Deffayet}\ \emph {et~al.}(2010)\citenamefont
  {Deffayet}, \citenamefont {Pujolas}, \citenamefont {Sawicki},\ and\
  \citenamefont {Vikman}}]{Deffayet:2010qz}%
  \BibitemOpen
  \bibfield  {author} {\bibinfo {author} {\bibfnamefont {C.}~\bibnamefont
  {Deffayet}}, \bibinfo {author} {\bibfnamefont {O.}~\bibnamefont {Pujolas}},
  \bibinfo {author} {\bibfnamefont {I.}~\bibnamefont {Sawicki}}, \ and\
  \bibinfo {author} {\bibfnamefont {A.}~\bibnamefont {Vikman}},\ }\href
  {\doibase 10.1088/1475-7516/2010/10/026} {\bibfield  {journal} {\bibinfo
  {journal} {JCAP}\ }\textbf {\bibinfo {volume} {10}},\ \bibinfo {pages} {026}
  (\bibinfo {year} {2010})},\ \Eprint {http://arxiv.org/abs/1008.0048}
  {arXiv:1008.0048 [hep-th]} \BibitemShut {NoStop}%
\bibitem [{\citenamefont {Ostrogradsky}(1850)}]{Ostrogradsky:1850fid}%
  \BibitemOpen
  \bibfield  {author} {\bibinfo {author} {\bibfnamefont {M.}~\bibnamefont
  {Ostrogradsky}},\ }\href@noop {} {\bibfield  {journal} {\bibinfo  {journal}
  {Mem. Acad. St. Petersbourg}\ }\textbf {\bibinfo {volume} {6}},\ \bibinfo
  {pages} {385} (\bibinfo {year} {1850})}\BibitemShut {NoStop}%
\bibitem [{\citenamefont {Woodard}(2015)}]{Woodard:2015zca}%
  \BibitemOpen
  \bibfield  {author} {\bibinfo {author} {\bibfnamefont {R.~P.}\ \bibnamefont
  {Woodard}},\ }\href {\doibase 10.4249/scholarpedia.32243} {\bibfield
  {journal} {\bibinfo  {journal} {Scholarpedia}\ }\textbf {\bibinfo {volume}
  {10}},\ \bibinfo {pages} {32243} (\bibinfo {year} {2015})},\ \Eprint
  {http://arxiv.org/abs/1506.02210} {arXiv:1506.02210 [hep-th]} \BibitemShut
  {NoStop}%
\bibitem [{\citenamefont {Pais}\ and\ \citenamefont
  {Uhlenbeck}(1950)}]{Pais:1950za}%
  \BibitemOpen
  \bibfield  {author} {\bibinfo {author} {\bibfnamefont {A.}~\bibnamefont
  {Pais}}\ and\ \bibinfo {author} {\bibfnamefont {G.~E.}\ \bibnamefont
  {Uhlenbeck}},\ }\href {\doibase 10.1103/PhysRev.79.145} {\bibfield  {journal}
  {\bibinfo  {journal} {Phys. Rev.}\ }\textbf {\bibinfo {volume} {79}},\
  \bibinfo {pages} {145} (\bibinfo {year} {1950})}\BibitemShut {NoStop}%
\bibitem [{\citenamefont {de~Urries}\ and\ \citenamefont
  {Julve}(1998)}]{deUrries:1998obu}%
  \BibitemOpen
  \bibfield  {author} {\bibinfo {author} {\bibfnamefont {F.~J.}\ \bibnamefont
  {de~Urries}}\ and\ \bibinfo {author} {\bibfnamefont {J.}~\bibnamefont
  {Julve}},\ }\href {\doibase 10.1088/0305-4470/31/33/006} {\bibfield
  {journal} {\bibinfo  {journal} {J. Phys. A}\ }\textbf {\bibinfo {volume}
  {31}},\ \bibinfo {pages} {6949} (\bibinfo {year} {1998})},\ \Eprint
  {http://arxiv.org/abs/hep-th/9802115} {arXiv:hep-th/9802115} \BibitemShut
  {NoStop}%
\bibitem [{\citenamefont {Lee}\ and\ \citenamefont {Wick}(1970)}]{Lee:1970iw}%
  \BibitemOpen
  \bibfield  {author} {\bibinfo {author} {\bibfnamefont {T.~D.}\ \bibnamefont
  {Lee}}\ and\ \bibinfo {author} {\bibfnamefont {G.~C.}\ \bibnamefont {Wick}},\
  }\href {\doibase 10.1103/PhysRevD.2.1033} {\bibfield  {journal} {\bibinfo
  {journal} {Phys. Rev. D}\ }\textbf {\bibinfo {volume} {2}},\ \bibinfo {pages}
  {1033} (\bibinfo {year} {1970})}\BibitemShut {NoStop}%
\bibitem [{\citenamefont {Garriga}\ and\ \citenamefont
  {Vilenkin}(2013)}]{Garriga:2012pk}%
  \BibitemOpen
  \bibfield  {author} {\bibinfo {author} {\bibfnamefont {J.}~\bibnamefont
  {Garriga}}\ and\ \bibinfo {author} {\bibfnamefont {A.}~\bibnamefont
  {Vilenkin}},\ }\href {\doibase 10.1088/1475-7516/2013/01/036} {\bibfield
  {journal} {\bibinfo  {journal} {JCAP}\ }\textbf {\bibinfo {volume} {01}},\
  \bibinfo {pages} {036} (\bibinfo {year} {2013})},\ \Eprint
  {http://arxiv.org/abs/1202.1239} {arXiv:1202.1239 [hep-th]} \BibitemShut
  {NoStop}%
\bibitem [{\citenamefont {Lovelock}(1971)}]{Lovelock:1971yv}%
  \BibitemOpen
  \bibfield  {author} {\bibinfo {author} {\bibfnamefont {D.}~\bibnamefont
  {Lovelock}},\ }\href {\doibase 10.1063/1.1665613} {\bibfield  {journal}
  {\bibinfo  {journal} {J. Math. Phys.}\ }\textbf {\bibinfo {volume} {12}},\
  \bibinfo {pages} {498} (\bibinfo {year} {1971})}\BibitemShut {NoStop}%
\bibitem [{\citenamefont {Stelle}(1977)}]{Stelle:1976gc}%
  \BibitemOpen
  \bibfield  {author} {\bibinfo {author} {\bibfnamefont {K.~S.}\ \bibnamefont
  {Stelle}},\ }\href {\doibase 10.1103/PhysRevD.16.953} {\bibfield  {journal}
  {\bibinfo  {journal} {Phys. Rev. D}\ }\textbf {\bibinfo {volume} {16}},\
  \bibinfo {pages} {953} (\bibinfo {year} {1977})}\BibitemShut {NoStop}%
\bibitem [{\citenamefont {Hawking}\ and\ \citenamefont
  {Hertog}(2002)}]{Hawking:2001yt}%
  \BibitemOpen
  \bibfield  {author} {\bibinfo {author} {\bibfnamefont {S.~W.}\ \bibnamefont
  {Hawking}}\ and\ \bibinfo {author} {\bibfnamefont {T.}~\bibnamefont
  {Hertog}},\ }\href {\doibase 10.1103/PhysRevD.65.103515} {\bibfield
  {journal} {\bibinfo  {journal} {Phys. Rev. D}\ }\textbf {\bibinfo {volume}
  {65}},\ \bibinfo {pages} {103515} (\bibinfo {year} {2002})},\ \Eprint
  {http://arxiv.org/abs/hep-th/0107088} {arXiv:hep-th/0107088} \BibitemShut
  {NoStop}%
\bibitem [{\citenamefont {Linde}(1988)}]{Linde:1988ws}%
  \BibitemOpen
  \bibfield  {author} {\bibinfo {author} {\bibfnamefont {A.~D.}\ \bibnamefont
  {Linde}},\ }\href {\doibase 10.1016/0370-2693(88)90770-8} {\bibfield
  {journal} {\bibinfo  {journal} {Phys. Lett. B}\ }\textbf {\bibinfo {volume}
  {200}},\ \bibinfo {pages} {272} (\bibinfo {year} {1988})}\BibitemShut
  {NoStop}%
\bibitem [{\citenamefont {Kaplan}\ and\ \citenamefont
  {Sundrum}(2006)}]{Kaplan:2005rr}%
  \BibitemOpen
  \bibfield  {author} {\bibinfo {author} {\bibfnamefont {D.~E.}\ \bibnamefont
  {Kaplan}}\ and\ \bibinfo {author} {\bibfnamefont {R.}~\bibnamefont
  {Sundrum}},\ }\href {\doibase 10.1088/1126-6708/2006/07/042} {\bibfield
  {journal} {\bibinfo  {journal} {JHEP}\ }\textbf {\bibinfo {volume} {07}},\
  \bibinfo {pages} {042} (\bibinfo {year} {2006})},\ \Eprint
  {http://arxiv.org/abs/hep-th/0505265} {arXiv:hep-th/0505265} \BibitemShut
  {NoStop}%
\bibitem [{\citenamefont {Cline}\ \emph {et~al.}(2004)\citenamefont {Cline},
  \citenamefont {Jeon},\ and\ \citenamefont {Moore}}]{Cline:2003gs}%
  \BibitemOpen
  \bibfield  {author} {\bibinfo {author} {\bibfnamefont {J.~M.}\ \bibnamefont
  {Cline}}, \bibinfo {author} {\bibfnamefont {S.}~\bibnamefont {Jeon}}, \ and\
  \bibinfo {author} {\bibfnamefont {G.~D.}\ \bibnamefont {Moore}},\ }\href
  {\doibase 10.1103/PhysRevD.70.043543} {\bibfield  {journal} {\bibinfo
  {journal} {Phys. Rev. D}\ }\textbf {\bibinfo {volume} {70}},\ \bibinfo
  {pages} {043543} (\bibinfo {year} {2004})},\ \Eprint
  {http://arxiv.org/abs/hep-ph/0311312} {arXiv:hep-ph/0311312} \BibitemShut
  {NoStop}%
\bibitem [{\citenamefont {Cline}\ \emph {et~al.}(2023)\citenamefont {Cline},
  \citenamefont {Puel}, \citenamefont {Toma},\ and\ \citenamefont
  {Wang}}]{Cline:2023cwm}%
  \BibitemOpen
  \bibfield  {author} {\bibinfo {author} {\bibfnamefont {J.~M.}\ \bibnamefont
  {Cline}}, \bibinfo {author} {\bibfnamefont {M.}~\bibnamefont {Puel}},
  \bibinfo {author} {\bibfnamefont {T.}~\bibnamefont {Toma}}, \ and\ \bibinfo
  {author} {\bibfnamefont {Q.~S.}\ \bibnamefont {Wang}},\ }\href {\doibase
  10.1103/PhysRevD.108.095033} {\bibfield  {journal} {\bibinfo  {journal}
  {Phys. Rev. D}\ }\textbf {\bibinfo {volume} {108}},\ \bibinfo {pages}
  {095033} (\bibinfo {year} {2023})},\ \Eprint
  {http://arxiv.org/abs/2308.12989} {arXiv:2308.12989 [hep-ph]} \BibitemShut
  {NoStop}%
\bibitem [{\citenamefont {Cline}\ \emph {et~al.}(2024)\citenamefont {Cline},
  \citenamefont {Puel},\ and\ \citenamefont {Toma}}]{Cline:2023hfw}%
  \BibitemOpen
  \bibfield  {author} {\bibinfo {author} {\bibfnamefont {J.~M.}\ \bibnamefont
  {Cline}}, \bibinfo {author} {\bibfnamefont {M.}~\bibnamefont {Puel}}, \ and\
  \bibinfo {author} {\bibfnamefont {T.}~\bibnamefont {Toma}},\ }\href {\doibase
  10.1016/j.physletb.2023.138377} {\bibfield  {journal} {\bibinfo  {journal}
  {Phys. Lett. B}\ }\textbf {\bibinfo {volume} {848}},\ \bibinfo {pages}
  {138377} (\bibinfo {year} {2024})},\ \Eprint
  {http://arxiv.org/abs/2308.01333} {arXiv:2308.01333 [hep-ph]} \BibitemShut
  {NoStop}%
\bibitem [{\citenamefont {Brandenberger}\ and\ \citenamefont
  {Peter}(2017)}]{Brandenberger:2016vhg}%
  \BibitemOpen
  \bibfield  {author} {\bibinfo {author} {\bibfnamefont {R.}~\bibnamefont
  {Brandenberger}}\ and\ \bibinfo {author} {\bibfnamefont {P.}~\bibnamefont
  {Peter}},\ }\href {\doibase 10.1007/s10701-016-0057-0} {\bibfield  {journal}
  {\bibinfo  {journal} {Found. Phys.}\ }\textbf {\bibinfo {volume} {47}},\
  \bibinfo {pages} {797} (\bibinfo {year} {2017})},\ \Eprint
  {http://arxiv.org/abs/1603.05834} {arXiv:1603.05834 [hep-th]} \BibitemShut
  {NoStop}%
\bibitem [{\citenamefont {Woodard}(2007)}]{Woodard:2006nt}%
  \BibitemOpen
  \bibfield  {author} {\bibinfo {author} {\bibfnamefont {R.~P.}\ \bibnamefont
  {Woodard}},\ }\href {\doibase 10.1007/978-3-540-71013-4_14} {\bibfield
  {journal} {\bibinfo  {journal} {Lect. Notes Phys.}\ }\textbf {\bibinfo
  {volume} {720}},\ \bibinfo {pages} {403} (\bibinfo {year} {2007})},\ \Eprint
  {http://arxiv.org/abs/astro-ph/0601672} {arXiv:astro-ph/0601672} \BibitemShut
  {NoStop}%
\bibitem [{\citenamefont {Copeland}\ \emph {et~al.}(2006)\citenamefont
  {Copeland}, \citenamefont {Sami},\ and\ \citenamefont
  {Tsujikawa}}]{Copeland:2006wr}%
  \BibitemOpen
  \bibfield  {author} {\bibinfo {author} {\bibfnamefont {E.~J.}\ \bibnamefont
  {Copeland}}, \bibinfo {author} {\bibfnamefont {M.}~\bibnamefont {Sami}}, \
  and\ \bibinfo {author} {\bibfnamefont {S.}~\bibnamefont {Tsujikawa}},\ }\href
  {\doibase 10.1142/S021827180600942X} {\bibfield  {journal} {\bibinfo
  {journal} {Int. J. Mod. Phys. D}\ }\textbf {\bibinfo {volume} {15}},\
  \bibinfo {pages} {1753} (\bibinfo {year} {2006})},\ \Eprint
  {http://arxiv.org/abs/hep-th/0603057} {arXiv:hep-th/0603057} \BibitemShut
  {NoStop}%
\bibitem [{\citenamefont {Sotiriou}\ and\ \citenamefont
  {Faraoni}(2010)}]{Sotiriou:2008rp}%
  \BibitemOpen
  \bibfield  {author} {\bibinfo {author} {\bibfnamefont {T.~P.}\ \bibnamefont
  {Sotiriou}}\ and\ \bibinfo {author} {\bibfnamefont {V.}~\bibnamefont
  {Faraoni}},\ }\href {\doibase 10.1103/RevModPhys.82.451} {\bibfield
  {journal} {\bibinfo  {journal} {Rev. Mod. Phys.}\ }\textbf {\bibinfo {volume}
  {82}},\ \bibinfo {pages} {451} (\bibinfo {year} {2010})},\ \Eprint
  {http://arxiv.org/abs/0805.1726} {arXiv:0805.1726 [gr-qc]} \BibitemShut
  {NoStop}%
\bibitem [{\citenamefont {Clifton}\ \emph {et~al.}(2012)\citenamefont
  {Clifton}, \citenamefont {Ferreira}, \citenamefont {Padilla},\ and\
  \citenamefont {Skordis}}]{Clifton:2011jh}%
  \BibitemOpen
  \bibfield  {author} {\bibinfo {author} {\bibfnamefont {T.}~\bibnamefont
  {Clifton}}, \bibinfo {author} {\bibfnamefont {P.~G.}\ \bibnamefont
  {Ferreira}}, \bibinfo {author} {\bibfnamefont {A.}~\bibnamefont {Padilla}}, \
  and\ \bibinfo {author} {\bibfnamefont {C.}~\bibnamefont {Skordis}},\ }\href
  {\doibase 10.1016/j.physrep.2012.01.001} {\bibfield  {journal} {\bibinfo
  {journal} {Phys. Rept.}\ }\textbf {\bibinfo {volume} {513}},\ \bibinfo
  {pages} {1} (\bibinfo {year} {2012})},\ \Eprint
  {http://arxiv.org/abs/1106.2476} {arXiv:1106.2476 [astro-ph.CO]} \BibitemShut
  {NoStop}%
\bibitem [{\citenamefont {Joyce}\ \emph {et~al.}(2015)\citenamefont {Joyce},
  \citenamefont {Jain}, \citenamefont {Khoury},\ and\ \citenamefont
  {Trodden}}]{Joyce:2014kja}%
  \BibitemOpen
  \bibfield  {author} {\bibinfo {author} {\bibfnamefont {A.}~\bibnamefont
  {Joyce}}, \bibinfo {author} {\bibfnamefont {B.}~\bibnamefont {Jain}},
  \bibinfo {author} {\bibfnamefont {J.}~\bibnamefont {Khoury}}, \ and\ \bibinfo
  {author} {\bibfnamefont {M.}~\bibnamefont {Trodden}},\ }\href {\doibase
  10.1016/j.physrep.2014.12.002} {\bibfield  {journal} {\bibinfo  {journal}
  {Phys. Rept.}\ }\textbf {\bibinfo {volume} {568}},\ \bibinfo {pages} {1}
  (\bibinfo {year} {2015})},\ \Eprint {http://arxiv.org/abs/1407.0059}
  {arXiv:1407.0059 [astro-ph.CO]} \BibitemShut {NoStop}%
\bibitem [{\citenamefont {de~Rham}(2014)}]{deRham:2014zqa}%
  \BibitemOpen
  \bibfield  {author} {\bibinfo {author} {\bibfnamefont {C.}~\bibnamefont
  {de~Rham}},\ }\href {\doibase 10.12942/lrr-2014-7} {\bibfield  {journal}
  {\bibinfo  {journal} {Living Rev. Rel.}\ }\textbf {\bibinfo {volume} {17}},\
  \bibinfo {pages} {7} (\bibinfo {year} {2014})},\ \Eprint
  {http://arxiv.org/abs/1401.4173} {arXiv:1401.4173 [hep-th]} \BibitemShut
  {NoStop}%
\bibitem [{\citenamefont {Berti}\ \emph {et~al.}(2015)\citenamefont {Berti}
  \emph {et~al.}}]{Berti:2015itd}%
  \BibitemOpen
  \bibfield  {author} {\bibinfo {author} {\bibfnamefont {E.}~\bibnamefont
  {Berti}} \emph {et~al.},\ }\href {\doibase 10.1088/0264-9381/32/24/243001}
  {\bibfield  {journal} {\bibinfo  {journal} {Class. Quant. Grav.}\ }\textbf
  {\bibinfo {volume} {32}},\ \bibinfo {pages} {243001} (\bibinfo {year}
  {2015})},\ \Eprint {http://arxiv.org/abs/1501.07274} {arXiv:1501.07274
  [gr-qc]} \BibitemShut {NoStop}%
\bibitem [{\citenamefont {Errasti~D\'\i{}ez}\ \emph {et~al.}(2024)\citenamefont
  {Errasti~D\'\i{}ez}, \citenamefont {Gaset~Rif\`a},\ and\ \citenamefont
  {Staudt}}]{ErrastiDiez:2024hfq}%
  \BibitemOpen
  \bibfield  {author} {\bibinfo {author} {\bibfnamefont {V.}~\bibnamefont
  {Errasti~D\'\i{}ez}}, \bibinfo {author} {\bibfnamefont {J.}~\bibnamefont
  {Gaset~Rif\`a}}, \ and\ \bibinfo {author} {\bibfnamefont {G.}~\bibnamefont
  {Staudt}},\ }\href {\doibase 10.1002/prop.202400268} {\  (\bibinfo {year}
  {2024}),\ 10.1002/prop.202400268},\ \Eprint {http://arxiv.org/abs/2408.16832}
  {arXiv:2408.16832 [hep-th]} \BibitemShut {NoStop}%
\bibitem [{\citenamefont {Deffayet}\ \emph {et~al.}(2022)\citenamefont
  {Deffayet}, \citenamefont {Mukohyama},\ and\ \citenamefont
  {Vikman}}]{Deffayet:2021nnt}%
  \BibitemOpen
  \bibfield  {author} {\bibinfo {author} {\bibfnamefont {C.}~\bibnamefont
  {Deffayet}}, \bibinfo {author} {\bibfnamefont {S.}~\bibnamefont {Mukohyama}},
  \ and\ \bibinfo {author} {\bibfnamefont {A.}~\bibnamefont {Vikman}},\ }\href
  {\doibase 10.1103/PhysRevLett.128.041301} {\bibfield  {journal} {\bibinfo
  {journal} {Phys. Rev. Lett.}\ }\textbf {\bibinfo {volume} {128}},\ \bibinfo
  {pages} {041301} (\bibinfo {year} {2022})},\ \Eprint
  {http://arxiv.org/abs/2108.06294} {arXiv:2108.06294 [gr-qc]} \BibitemShut
  {NoStop}%
\bibitem [{\citenamefont {Deffayet}\ \emph {et~al.}(2023)\citenamefont
  {Deffayet}, \citenamefont {Held}, \citenamefont {Mukohyama},\ and\
  \citenamefont {Vikman}}]{Deffayet:2023wdg}%
  \BibitemOpen
  \bibfield  {author} {\bibinfo {author} {\bibfnamefont {C.}~\bibnamefont
  {Deffayet}}, \bibinfo {author} {\bibfnamefont {A.}~\bibnamefont {Held}},
  \bibinfo {author} {\bibfnamefont {S.}~\bibnamefont {Mukohyama}}, \ and\
  \bibinfo {author} {\bibfnamefont {A.}~\bibnamefont {Vikman}},\ }\href
  {\doibase 10.1088/1475-7516/2023/11/031} {\bibfield  {journal} {\bibinfo
  {journal} {JCAP}\ }\textbf {\bibinfo {volume} {11}},\ \bibinfo {pages} {031}
  (\bibinfo {year} {2023})},\ \Eprint {http://arxiv.org/abs/2305.09631}
  {arXiv:2305.09631 [gr-qc]} \BibitemShut {NoStop}%
\bibitem [{\citenamefont {Kaparulin}\ \emph {et~al.}(2014)\citenamefont
  {Kaparulin}, \citenamefont {Lyakhovich},\ and\ \citenamefont
  {Sharapov}}]{Kaparulin:2014vpa}%
  \BibitemOpen
  \bibfield  {author} {\bibinfo {author} {\bibfnamefont {D.~S.}\ \bibnamefont
  {Kaparulin}}, \bibinfo {author} {\bibfnamefont {S.~L.}\ \bibnamefont
  {Lyakhovich}}, \ and\ \bibinfo {author} {\bibfnamefont {A.~A.}\ \bibnamefont
  {Sharapov}},\ }\href {\doibase 10.1140/epjc/s10052-014-3072-3} {\bibfield
  {journal} {\bibinfo  {journal} {Eur. Phys. J. C}\ }\textbf {\bibinfo {volume}
  {74}},\ \bibinfo {pages} {3072} (\bibinfo {year} {2014})},\ \Eprint
  {http://arxiv.org/abs/1407.8481} {arXiv:1407.8481 [hep-th]} \BibitemShut
  {NoStop}%
\bibitem [{\citenamefont {Pagani}\ \emph {et~al.}(1987)\citenamefont {Pagani},
  \citenamefont {Tecchiolli},\ and\ \citenamefont {Zerbini}}]{Pagani:1987ue}%
  \BibitemOpen
  \bibfield  {author} {\bibinfo {author} {\bibfnamefont {E.}~\bibnamefont
  {Pagani}}, \bibinfo {author} {\bibfnamefont {G.}~\bibnamefont {Tecchiolli}},
  \ and\ \bibinfo {author} {\bibfnamefont {S.}~\bibnamefont {Zerbini}},\ }\href
  {\doibase 10.1007/BF00402140} {\bibfield  {journal} {\bibinfo  {journal}
  {Lett. Math. Phys.}\ }\textbf {\bibinfo {volume} {14}},\ \bibinfo {pages}
  {311} (\bibinfo {year} {1987})}\BibitemShut {NoStop}%
\bibitem [{\citenamefont {Smilga}(2005)}]{Smilga:2004cy}%
  \BibitemOpen
  \bibfield  {author} {\bibinfo {author} {\bibfnamefont {A.~V.}\ \bibnamefont
  {Smilga}},\ }\href {\doibase 10.1016/j.nuclphysb.2004.10.037} {\bibfield
  {journal} {\bibinfo  {journal} {Nucl. Phys. B}\ }\textbf {\bibinfo {volume}
  {706}},\ \bibinfo {pages} {598} (\bibinfo {year} {2005})},\ \Eprint
  {http://arxiv.org/abs/hep-th/0407231} {arXiv:hep-th/0407231} \BibitemShut
  {NoStop}%
\bibitem [{\citenamefont {Carroll}\ \emph {et~al.}(2003)\citenamefont
  {Carroll}, \citenamefont {Hoffman},\ and\ \citenamefont
  {Trodden}}]{Carroll:2003st}%
  \BibitemOpen
  \bibfield  {author} {\bibinfo {author} {\bibfnamefont {S.~M.}\ \bibnamefont
  {Carroll}}, \bibinfo {author} {\bibfnamefont {M.}~\bibnamefont {Hoffman}}, \
  and\ \bibinfo {author} {\bibfnamefont {M.}~\bibnamefont {Trodden}},\ }\href
  {\doibase 10.1103/PhysRevD.68.023509} {\bibfield  {journal} {\bibinfo
  {journal} {Phys. Rev. D}\ }\textbf {\bibinfo {volume} {68}},\ \bibinfo
  {pages} {023509} (\bibinfo {year} {2003})},\ \Eprint
  {http://arxiv.org/abs/astro-ph/0301273} {arXiv:astro-ph/0301273} \BibitemShut
  {NoStop}%
\bibitem [{\citenamefont {Ilhan}\ and\ \citenamefont
  {Kovner}(2013)}]{Ilhan:2013xe}%
  \BibitemOpen
  \bibfield  {author} {\bibinfo {author} {\bibfnamefont {I.~B.}\ \bibnamefont
  {Ilhan}}\ and\ \bibinfo {author} {\bibfnamefont {A.}~\bibnamefont {Kovner}},\
  }\href {\doibase 10.1103/PhysRevD.88.044045} {\bibfield  {journal} {\bibinfo
  {journal} {Phys. Rev. D}\ }\textbf {\bibinfo {volume} {88}},\ \bibinfo
  {pages} {044045} (\bibinfo {year} {2013})},\ \Eprint
  {http://arxiv.org/abs/1301.4879} {arXiv:1301.4879 [hep-th]} \BibitemShut
  {NoStop}%
\bibitem [{\citenamefont {Pav\v{s}i\v{c}}(2016)}]{Pavsic:2016ykq}%
  \BibitemOpen
  \bibfield  {author} {\bibinfo {author} {\bibfnamefont {M.}~\bibnamefont
  {Pav\v{s}i\v{c}}},\ }\href {\doibase 10.1142/S0219887816300154} {\bibfield
  {journal} {\bibinfo  {journal} {Int. J. Geom. Meth. Mod. Phys.}\ }\textbf
  {\bibinfo {volume} {13}},\ \bibinfo {pages} {1630015} (\bibinfo {year}
  {2016})},\ \Eprint {http://arxiv.org/abs/1607.06589} {arXiv:1607.06589
  [gr-qc]} \BibitemShut {NoStop}%
\bibitem [{\citenamefont {Smilga}(2017)}]{Smilga:2017arl}%
  \BibitemOpen
  \bibfield  {author} {\bibinfo {author} {\bibfnamefont {A.}~\bibnamefont
  {Smilga}},\ }\href {\doibase 10.1142/S0217751X17300253} {\bibfield  {journal}
  {\bibinfo  {journal} {Int. J. Mod. Phys. A}\ }\textbf {\bibinfo {volume}
  {32}},\ \bibinfo {pages} {1730025} (\bibinfo {year} {2017})},\ \Eprint
  {http://arxiv.org/abs/1710.11538} {arXiv:1710.11538 [hep-th]} \BibitemShut
  {NoStop}%
\bibitem [{\citenamefont {Pav\v{s}i\v{c}}(2013)}]{Pavsic:2013noa}%
  \BibitemOpen
  \bibfield  {author} {\bibinfo {author} {\bibfnamefont {M.}~\bibnamefont
  {Pav\v{s}i\v{c}}},\ }\href {\doibase 10.1142/S0217732313501654} {\bibfield
  {journal} {\bibinfo  {journal} {Mod. Phys. Lett. A}\ }\textbf {\bibinfo
  {volume} {28}},\ \bibinfo {pages} {1350165} (\bibinfo {year} {2013})},\
  \Eprint {http://arxiv.org/abs/1302.5257} {arXiv:1302.5257 [gr-qc]}
  \BibitemShut {NoStop}%
\bibitem [{\citenamefont {Pav\v{s}i\v{c}}(2020)}]{Pavsic:2020aqi}%
  \BibitemOpen
  \bibfield  {author} {\bibinfo {author} {\bibfnamefont {M.}~\bibnamefont
  {Pav\v{s}i\v{c}}},\ }\href {\doibase 10.1142/S0217751X20300203} {\bibfield
  {journal} {\bibinfo  {journal} {Int. J. Mod. Phys. A}\ }\textbf {\bibinfo
  {volume} {35}},\ \bibinfo {pages} {2030020} (\bibinfo {year} {2020})},\
  \Eprint {http://arxiv.org/abs/2012.04976} {arXiv:2012.04976 [hep-th]}
  \BibitemShut {NoStop}%
\bibitem [{\citenamefont {Boulanger}\ \emph {et~al.}(2019)\citenamefont
  {Boulanger}, \citenamefont {Buisseret}, \citenamefont {Dierick},\ and\
  \citenamefont {White}}]{Boulanger:2018tue}%
  \BibitemOpen
  \bibfield  {author} {\bibinfo {author} {\bibfnamefont {N.}~\bibnamefont
  {Boulanger}}, \bibinfo {author} {\bibfnamefont {F.}~\bibnamefont
  {Buisseret}}, \bibinfo {author} {\bibfnamefont {F.}~\bibnamefont {Dierick}},
  \ and\ \bibinfo {author} {\bibfnamefont {O.}~\bibnamefont {White}},\ }\href
  {\doibase 10.1140/epjc/s10052-019-6569-y} {\bibfield  {journal} {\bibinfo
  {journal} {Eur. Phys. J. C}\ }\textbf {\bibinfo {volume} {79}},\ \bibinfo
  {pages} {60} (\bibinfo {year} {2019})},\ \Eprint
  {http://arxiv.org/abs/1811.07733} {arXiv:1811.07733 [physics.class-ph]}
  \BibitemShut {NoStop}%
\bibitem [{\citenamefont {Damour}\ and\ \citenamefont
  {Smilga}(2022)}]{Damour:2021fva}%
  \BibitemOpen
  \bibfield  {author} {\bibinfo {author} {\bibfnamefont {T.}~\bibnamefont
  {Damour}}\ and\ \bibinfo {author} {\bibfnamefont {A.}~\bibnamefont
  {Smilga}},\ }\href {\doibase 10.1103/PhysRevD.105.045018} {\bibfield
  {journal} {\bibinfo  {journal} {Phys. Rev. D}\ }\textbf {\bibinfo {volume}
  {105}},\ \bibinfo {pages} {045018} (\bibinfo {year} {2022})},\ \Eprint
  {http://arxiv.org/abs/2110.11175} {arXiv:2110.11175 [hep-th]} \BibitemShut
  {NoStop}%
\bibitem [{\citenamefont {Figueras}\ \emph {et~al.}(2024)\citenamefont
  {Figueras}, \citenamefont {Held},\ and\ \citenamefont
  {Kov\'acs}}]{Figueras:2024bba}%
  \BibitemOpen
  \bibfield  {author} {\bibinfo {author} {\bibfnamefont {P.}~\bibnamefont
  {Figueras}}, \bibinfo {author} {\bibfnamefont {A.}~\bibnamefont {Held}}, \
  and\ \bibinfo {author} {\bibfnamefont {A.~D.}\ \bibnamefont {Kov\'acs}},\
  }\href@noop {} {\  (\bibinfo {year} {2024})},\ \Eprint
  {http://arxiv.org/abs/2407.08775} {arXiv:2407.08775 [gr-qc]} \BibitemShut
  {NoStop}%
\bibitem [{\citenamefont {{Noakes}}(1983)}]{Noakes:1983}%
  \BibitemOpen
  \bibfield  {author} {\bibinfo {author} {\bibfnamefont {D.~R.}\ \bibnamefont
  {{Noakes}}},\ }\href {\doibase 10.1063/1.525906} {\bibfield  {journal}
  {\bibinfo  {journal} {Journal of Mathematical Physics}\ }\textbf {\bibinfo
  {volume} {24}},\ \bibinfo {pages} {1846} (\bibinfo {year}
  {1983})}\BibitemShut {NoStop}%
\bibitem [{\citenamefont {Held}\ and\ \citenamefont
  {Lim}(2021)}]{Held:2021pht}%
  \BibitemOpen
  \bibfield  {author} {\bibinfo {author} {\bibfnamefont {A.}~\bibnamefont
  {Held}}\ and\ \bibinfo {author} {\bibfnamefont {H.}~\bibnamefont {Lim}},\
  }\href {\doibase 10.1103/PhysRevD.104.084075} {\bibfield  {journal} {\bibinfo
   {journal} {Phys. Rev. D}\ }\textbf {\bibinfo {volume} {104}},\ \bibinfo
  {pages} {084075} (\bibinfo {year} {2021})},\ \Eprint
  {http://arxiv.org/abs/2104.04010} {arXiv:2104.04010 [gr-qc]} \BibitemShut
  {NoStop}%
\bibitem [{\citenamefont {Held}\ and\ \citenamefont
  {Lim}(2023)}]{Held:2023aap}%
  \BibitemOpen
  \bibfield  {author} {\bibinfo {author} {\bibfnamefont {A.}~\bibnamefont
  {Held}}\ and\ \bibinfo {author} {\bibfnamefont {H.}~\bibnamefont {Lim}},\
  }\href {\doibase 10.1103/PhysRevD.108.104025} {\bibfield  {journal} {\bibinfo
   {journal} {Phys. Rev. D}\ }\textbf {\bibinfo {volume} {108}},\ \bibinfo
  {pages} {104025} (\bibinfo {year} {2023})},\ \Eprint
  {http://arxiv.org/abs/2306.04725} {arXiv:2306.04725 [gr-qc]} \BibitemShut
  {NoStop}%
\bibitem [{\citenamefont {Held}\ and\ \citenamefont
  {Lim}(2025)}]{Held:2025ckb}%
  \BibitemOpen
  \bibfield  {author} {\bibinfo {author} {\bibfnamefont {A.}~\bibnamefont
  {Held}}\ and\ \bibinfo {author} {\bibfnamefont {H.}~\bibnamefont {Lim}},\
  }\href@noop {} {\  (\bibinfo {year} {2025})},\ \Eprint
  {http://arxiv.org/abs/2503.13428} {arXiv:2503.13428 [gr-qc]} \BibitemShut
  {NoStop}%
\bibitem [{\citenamefont {Gross}\ \emph {et~al.}(2021)\citenamefont {Gross},
  \citenamefont {Strumia}, \citenamefont {Teresi},\ and\ \citenamefont
  {Zirilli}}]{Gross:2020tph}%
  \BibitemOpen
  \bibfield  {author} {\bibinfo {author} {\bibfnamefont {C.}~\bibnamefont
  {Gross}}, \bibinfo {author} {\bibfnamefont {A.}~\bibnamefont {Strumia}},
  \bibinfo {author} {\bibfnamefont {D.}~\bibnamefont {Teresi}}, \ and\ \bibinfo
  {author} {\bibfnamefont {M.}~\bibnamefont {Zirilli}},\ }\href {\doibase
  10.1103/PhysRevD.103.115025} {\bibfield  {journal} {\bibinfo  {journal}
  {Phys. Rev. D}\ }\textbf {\bibinfo {volume} {103}},\ \bibinfo {pages}
  {115025} (\bibinfo {year} {2021})},\ \Eprint
  {http://arxiv.org/abs/2007.05541} {arXiv:2007.05541 [hep-th]} \BibitemShut
  {NoStop}%
\bibitem [{\citenamefont {Fring}\ \emph
  {et~al.}(2024{\natexlab{a}})\citenamefont {Fring}, \citenamefont {Taira},\
  and\ \citenamefont {Turner}}]{Fring:2024xhd}%
  \BibitemOpen
  \bibfield  {author} {\bibinfo {author} {\bibfnamefont {A.}~\bibnamefont
  {Fring}}, \bibinfo {author} {\bibfnamefont {T.}~\bibnamefont {Taira}}, \ and\
  \bibinfo {author} {\bibfnamefont {B.}~\bibnamefont {Turner}},\ }\href
  {\doibase 10.1007/JHEP09(2024)199} {\bibfield  {journal} {\bibinfo  {journal}
  {JHEP}\ }\textbf {\bibinfo {volume} {09}},\ \bibinfo {pages} {199} (\bibinfo
  {year} {2024}{\natexlab{a}})},\ \Eprint {http://arxiv.org/abs/2406.18255}
  {arXiv:2406.18255 [nlin.SI]} \BibitemShut {NoStop}%
\bibitem [{\citenamefont {Fring}\ and\ \citenamefont
  {Turner}(2023{\natexlab{a}})}]{Fring:2023pww}%
  \BibitemOpen
  \bibfield  {author} {\bibinfo {author} {\bibfnamefont {A.}~\bibnamefont
  {Fring}}\ and\ \bibinfo {author} {\bibfnamefont {B.}~\bibnamefont {Turner}},\
  }\href {\doibase 10.1088/1751-8121/ace0e6} {\bibfield  {journal} {\bibinfo
  {journal} {J. Phys. A}\ }\textbf {\bibinfo {volume} {56}},\ \bibinfo {pages}
  {295203} (\bibinfo {year} {2023}{\natexlab{a}})},\ \Eprint
  {http://arxiv.org/abs/2301.11317} {arXiv:2301.11317 [hep-th]} \BibitemShut
  {NoStop}%
\bibitem [{\citenamefont {Fring}\ \emph
  {et~al.}(2024{\natexlab{b}})\citenamefont {Fring}, \citenamefont {Taira},\
  and\ \citenamefont {Turner}}]{Fring:2024brg}%
  \BibitemOpen
  \bibfield  {author} {\bibinfo {author} {\bibfnamefont {A.}~\bibnamefont
  {Fring}}, \bibinfo {author} {\bibfnamefont {T.}~\bibnamefont {Taira}}, \ and\
  \bibinfo {author} {\bibfnamefont {B.}~\bibnamefont {Turner}},\ }\href
  {\doibase 10.3390/universe10050198} {\bibfield  {journal} {\bibinfo
  {journal} {Universe}\ }\textbf {\bibinfo {volume} {10}},\ \bibinfo {pages}
  {198} (\bibinfo {year} {2024}{\natexlab{b}})},\ \Eprint
  {http://arxiv.org/abs/2403.11949} {arXiv:2403.11949 [hep-th]} \BibitemShut
  {NoStop}%
\bibitem [{\citenamefont {Fring}\ and\ \citenamefont
  {Turner}(2023{\natexlab{b}})}]{Fring:2023ijk}%
  \BibitemOpen
  \bibfield  {author} {\bibinfo {author} {\bibfnamefont {A.}~\bibnamefont
  {Fring}}\ and\ \bibinfo {author} {\bibfnamefont {B.}~\bibnamefont {Turner}},\
  }\href {\doibase 10.1140/epjp/s13360-023-04726-3} {\bibfield  {journal}
  {\bibinfo  {journal} {Eur. Phys. J. Plus}\ }\textbf {\bibinfo {volume}
  {138}},\ \bibinfo {pages} {1136} (\bibinfo {year} {2023}{\natexlab{b}})},\
  \Eprint {http://arxiv.org/abs/2307.15210} {arXiv:2307.15210 [hep-th]}
  \BibitemShut {NoStop}%
\bibitem [{\citenamefont {Sarbach}\ and\ \citenamefont
  {Tiglio}(2012)}]{Sarbach:2012pr}%
  \BibitemOpen
  \bibfield  {author} {\bibinfo {author} {\bibfnamefont {O.}~\bibnamefont
  {Sarbach}}\ and\ \bibinfo {author} {\bibfnamefont {M.}~\bibnamefont
  {Tiglio}},\ }\href {\doibase 10.12942/lrr-2012-9} {\bibfield  {journal}
  {\bibinfo  {journal} {Living Rev. Rel.}\ }\textbf {\bibinfo {volume} {15}},\
  \bibinfo {pages} {9} (\bibinfo {year} {2012})},\ \Eprint
  {http://arxiv.org/abs/1203.6443} {arXiv:1203.6443 [gr-qc]} \BibitemShut
  {NoStop}%
\bibitem [{\citenamefont {Rackauckas}\ and\ \citenamefont
  {Nie}(2017)}]{rackauckas2017differentialequations}%
  \BibitemOpen
  \bibfield  {author} {\bibinfo {author} {\bibfnamefont {C.}~\bibnamefont
  {Rackauckas}}\ and\ \bibinfo {author} {\bibfnamefont {Q.}~\bibnamefont
  {Nie}},\ }\href@noop {} {\bibfield  {journal} {\bibinfo  {journal} {Journal
  of Open Research Software}\ }\textbf {\bibinfo {volume} {5}} (\bibinfo {year}
  {2017})}\BibitemShut {NoStop}%
\bibitem [{\citenamefont {Runge}(1895)}]{runge1895numerische}%
  \BibitemOpen
  \bibfield  {author} {\bibinfo {author} {\bibfnamefont {C.}~\bibnamefont
  {Runge}},\ }\href@noop {} {\bibfield  {journal} {\bibinfo  {journal}
  {Mathematische Annalen}\ }\textbf {\bibinfo {volume} {46}},\ \bibinfo {pages}
  {167} (\bibinfo {year} {1895})}\BibitemShut {NoStop}%
\bibitem [{\citenamefont {Kutta}(1901)}]{kutta1901beitrag}%
  \BibitemOpen
  \bibfield  {author} {\bibinfo {author} {\bibfnamefont {W.}~\bibnamefont
  {Kutta}},\ }\href@noop {} {\emph {\bibinfo {title} {Beitrag zur
  n{\"a}herungsweisen Integration totaler Differentialgleichungen}}}\ (\bibinfo
   {publisher} {Teubner},\ \bibinfo {year} {1901})\BibitemShut {NoStop}%
\bibitem [{\citenamefont {Press}\ \emph {et~al.}(1988)\citenamefont {Press},
  \citenamefont {Vetterling}, \citenamefont {Teukolsky},\ and\ \citenamefont
  {Flannery}}]{press1988numerical}%
  \BibitemOpen
  \bibfield  {author} {\bibinfo {author} {\bibfnamefont {W.~H.}\ \bibnamefont
  {Press}}, \bibinfo {author} {\bibfnamefont {W.~T.}\ \bibnamefont
  {Vetterling}}, \bibinfo {author} {\bibfnamefont {S.~A.}\ \bibnamefont
  {Teukolsky}}, \ and\ \bibinfo {author} {\bibfnamefont {B.~P.}\ \bibnamefont
  {Flannery}},\ }\href@noop {} {\emph {\bibinfo {title} {Numerical recipes}}}\
  (\bibinfo  {publisher} {Citeseer},\ \bibinfo {year} {1988})\BibitemShut
  {NoStop}%
\bibitem [{\citenamefont {Tsitouras}(2011)}]{tsitouras2011runge}%
  \BibitemOpen
  \bibfield  {author} {\bibinfo {author} {\bibfnamefont {C.}~\bibnamefont
  {Tsitouras}},\ }\href@noop {} {\bibfield  {journal} {\bibinfo  {journal}
  {Computers \& Mathematics with Applications}\ }\textbf {\bibinfo {volume}
  {62}},\ \bibinfo {pages} {770} (\bibinfo {year} {2011})}\BibitemShut
  {NoStop}%
\bibitem [{\citenamefont {Robert}\ and\ \citenamefont
  {Smilga}(2008)}]{Robert:2006nj}%
  \BibitemOpen
  \bibfield  {author} {\bibinfo {author} {\bibfnamefont {D.}~\bibnamefont
  {Robert}}\ and\ \bibinfo {author} {\bibfnamefont {A.~V.}\ \bibnamefont
  {Smilga}},\ }\href {\doibase 10.1063/1.2904474} {\bibfield  {journal}
  {\bibinfo  {journal} {J. Math. Phys.}\ }\textbf {\bibinfo {volume} {49}},\
  \bibinfo {pages} {042104} (\bibinfo {year} {2008})},\ \Eprint
  {http://arxiv.org/abs/math-ph/0611023} {arXiv:math-ph/0611023} \BibitemShut
  {NoStop}%
\bibitem [{\citenamefont {Micha}\ and\ \citenamefont
  {Tkachev}(2003)}]{Micha:2002ey}%
  \BibitemOpen
  \bibfield  {author} {\bibinfo {author} {\bibfnamefont {R.}~\bibnamefont
  {Micha}}\ and\ \bibinfo {author} {\bibfnamefont {I.~I.}\ \bibnamefont
  {Tkachev}},\ }\href {\doibase 10.1103/PhysRevLett.90.121301} {\bibfield
  {journal} {\bibinfo  {journal} {Phys. Rev. Lett.}\ }\textbf {\bibinfo
  {volume} {90}},\ \bibinfo {pages} {121301} (\bibinfo {year} {2003})},\
  \Eprint {http://arxiv.org/abs/hep-ph/0210202} {arXiv:hep-ph/0210202}
  \BibitemShut {NoStop}%
\bibitem [{\citenamefont {Micha}\ and\ \citenamefont
  {Tkachev}(2004)}]{Micha:2004bv}%
  \BibitemOpen
  \bibfield  {author} {\bibinfo {author} {\bibfnamefont {R.}~\bibnamefont
  {Micha}}\ and\ \bibinfo {author} {\bibfnamefont {I.~I.}\ \bibnamefont
  {Tkachev}},\ }\href {\doibase 10.1103/PhysRevD.70.043538} {\bibfield
  {journal} {\bibinfo  {journal} {Phys. Rev. D}\ }\textbf {\bibinfo {volume}
  {70}},\ \bibinfo {pages} {043538} (\bibinfo {year} {2004})},\ \Eprint
  {http://arxiv.org/abs/hep-ph/0403101} {arXiv:hep-ph/0403101} \BibitemShut
  {NoStop}%
\bibitem [{\citenamefont {Lozanov}\ and\ \citenamefont
  {Amin}(2018)}]{Lozanov:2017hjm}%
  \BibitemOpen
  \bibfield  {author} {\bibinfo {author} {\bibfnamefont {K.~D.}\ \bibnamefont
  {Lozanov}}\ and\ \bibinfo {author} {\bibfnamefont {M.~A.}\ \bibnamefont
  {Amin}},\ }\href {\doibase 10.1103/PhysRevD.97.023533} {\bibfield  {journal}
  {\bibinfo  {journal} {Phys. Rev. D}\ }\textbf {\bibinfo {volume} {97}},\
  \bibinfo {pages} {023533} (\bibinfo {year} {2018})},\ \Eprint
  {http://arxiv.org/abs/1710.06851} {arXiv:1710.06851 [astro-ph.CO]}
  \BibitemShut {NoStop}%
\bibitem [{\citenamefont {Appelquist}\ and\ \citenamefont
  {Carazzone}(1975)}]{Appelquist:1974tg}%
  \BibitemOpen
  \bibfield  {author} {\bibinfo {author} {\bibfnamefont {T.}~\bibnamefont
  {Appelquist}}\ and\ \bibinfo {author} {\bibfnamefont {J.}~\bibnamefont
  {Carazzone}},\ }\href {\doibase 10.1103/PhysRevD.11.2856} {\bibfield
  {journal} {\bibinfo  {journal} {Phys. Rev. D}\ }\textbf {\bibinfo {volume}
  {11}},\ \bibinfo {pages} {2856} (\bibinfo {year} {1975})}\BibitemShut
  {NoStop}%
\bibitem [{\citenamefont {Babichev}\ \emph {et~al.}(2018)\citenamefont
  {Babichev}, \citenamefont {Charmousis}, \citenamefont {Esposito-Far\`ese},\
  and\ \citenamefont {Leh\'ebel}}]{Babichev:2017lmw}%
  \BibitemOpen
  \bibfield  {author} {\bibinfo {author} {\bibfnamefont {E.}~\bibnamefont
  {Babichev}}, \bibinfo {author} {\bibfnamefont {C.}~\bibnamefont
  {Charmousis}}, \bibinfo {author} {\bibfnamefont {G.}~\bibnamefont
  {Esposito-Far\`ese}}, \ and\ \bibinfo {author} {\bibfnamefont
  {A.}~\bibnamefont {Leh\'ebel}},\ }\href {\doibase
  10.1103/PhysRevLett.120.241101} {\bibfield  {journal} {\bibinfo  {journal}
  {Phys. Rev. Lett.}\ }\textbf {\bibinfo {volume} {120}},\ \bibinfo {pages}
  {241101} (\bibinfo {year} {2018})},\ \Eprint
  {http://arxiv.org/abs/1712.04398} {arXiv:1712.04398 [gr-qc]} \BibitemShut
  {NoStop}%
\bibitem [{\citenamefont {Babichev}(2024)}]{Babichev:2024uro}%
  \BibitemOpen
  \bibfield  {author} {\bibinfo {author} {\bibfnamefont {E.}~\bibnamefont
  {Babichev}},\ }\href@noop {} {\  (\bibinfo {year} {2024})},\ \Eprint
  {http://arxiv.org/abs/2412.20093} {arXiv:2412.20093 [gr-qc]} \BibitemShut
  {NoStop}%
\bibitem [{\citenamefont {Sawicki}\ \emph {et~al.}(2024)\citenamefont
  {Sawicki}, \citenamefont {Trenkler},\ and\ \citenamefont
  {Vikman}}]{Sawicki:2024ryt}%
  \BibitemOpen
  \bibfield  {author} {\bibinfo {author} {\bibfnamefont {I.}~\bibnamefont
  {Sawicki}}, \bibinfo {author} {\bibfnamefont {G.}~\bibnamefont {Trenkler}}, \
  and\ \bibinfo {author} {\bibfnamefont {A.}~\bibnamefont {Vikman}},\
  }\href@noop {} {\  (\bibinfo {year} {2024})},\ \Eprint
  {http://arxiv.org/abs/2412.21169} {arXiv:2412.21169 [gr-qc]} \BibitemShut
  {NoStop}%
\bibitem [{\citenamefont {{Courant}}\ \emph {et~al.}(1967)\citenamefont
  {{Courant}}, \citenamefont {{Friedrichs}},\ and\ \citenamefont
  {{Lewy}}}]{Courant:1967}%
  \BibitemOpen
  \bibfield  {author} {\bibinfo {author} {\bibfnamefont {R.}~\bibnamefont
  {{Courant}}}, \bibinfo {author} {\bibfnamefont {K.}~\bibnamefont
  {{Friedrichs}}}, \ and\ \bibinfo {author} {\bibfnamefont {H.}~\bibnamefont
  {{Lewy}}},\ }\href {\doibase 10.1147/rd.112.0215} {\bibfield  {journal}
  {\bibinfo  {journal} {IBM Journal of Research and Development}\ }\textbf
  {\bibinfo {volume} {11}},\ \bibinfo {pages} {215} (\bibinfo {year}
  {1967})}\BibitemShut {NoStop}%
\bibitem [{\citenamefont {Held}(2025)}]{ghostlyPDE_1D}%
  \BibitemOpen
  \bibfield  {author} {\bibinfo {author} {\bibfnamefont {A.}~\bibnamefont
  {Held}},\ }\href@noop {} {\enquote {\bibinfo {title} {Solving (1+1)d ghostly
  pdes with differentialequations.jl},}\ }\bibinfo {howpublished}
  {\url{https://github.com/aaron-hd/ghostlyPDE_1D}} (\bibinfo {year}
  {2025})\BibitemShut {NoStop}%
\end{thebibliography}%

\end{document}